\documentclass[article]{JHEP}
\usepackage{amsmath,epsfig}
\usepackage{amssymb,amsfonts}
\usepackage{feynmf}
\relax
\renewcommand{\theequation}{\arabic{section}.\arabic{subsection}.\arabic{equation}}
\def\be{\begin{equation}}
\def\ee{\end{equation}}
\def\bea{\begin{eqnarray}}
\def\eea{\end{eqnarray}}
\def\be{\begin{equation}}
\def\ee{\end{equation}}
\newcommand{\een}{\end{subequations}}
\newcommand{\ben}{\begin{subequations}}
\newcommand{\beq}{\begin{eqalignno}}
\newcommand{\eeq}{\end{eqalignno}}

\renewcommand{\arraystretch}{1.1}

\renewcommand{\b}[1]{\textbf{#1}}
\newcommand{\bb}[1]{\bar{\textbf{#1}}}
\renewcommand{\u}[1]{{U(#1)}}
\newcommand{\su}[1]{{SU(#1)}}

\newcommand{\PLB}[3]{\emph{ Phys.~Lett.} \textbf{B#1} (#2) #3}
\newcommand\fverb{\setbox\pippobox=\hbox\bgroup\verb}
\newcommand\fverbdo{\egroup\medskip\noindent%
                        \fbox{\unhbox\pippobox}\ }
\newcommand\fverbit{\egroup\item[\fbox{\unhbox\pippobox}]}
\newbox\pippobox
\def\sla{\raise.15ex\hbox{$/$}\kern-.57em}

\def\dr{\dot r}
\def\dt{\dot\varphi}
\def\bbox{\nabla^2}

\def\mt{{\tilde m}}

\def\e{\epsilon}

\def\m{\mu}
\def\n{\nu}
\def\r{\rho}
\def\s{\sigma}
\def\t{\theta}
\def\sp{\;\;\;,\;\;\;}

\def\p{\partial}

\def\a{\alpha}

\def\l{\lambda}

\def\c{\chi}
\def\ch{\hat \chi}
\def\rc{r_c}
\def\mpl{{M_{\rm P}}}

\def\gn{G_{N}}
\def\ba{\begin{eqnarray}}

\def\ea{\end{eqnarray}}
\def\rct{{\tilde r}_c}

\newcommand{\ii}{i}
\newcommand{\jj}{j}

\def\z{\zeta}
\def \lta {\mathrel{\vcenter
     {\hbox{$<$}\nointerlineskip\hbox{$\sim$}}}}
\def \gta {\mathrel{\vcenter
     {\hbox{$>$}\nointerlineskip\hbox{$\sim$}}}}

\newcommand{\Zint}{\mathbb{Z}}

\newcommand{\nn}{\nonumber}
\def\simlt{\mathrel{\lower2.5pt\vbox{\lineskip=0pt\baselineskip=0pt
           \hbox{$<$}\hbox{$\sim$}}}}
\def\simgt{\mathrel{\lower2.5pt\vbox{\lineskip=0pt\baselineskip=0pt
           \hbox{$>$}\hbox{$\sim$}}}}

\def\hre#1#2{\href{http://arxiv.org/abs/#1/#2}{[arXiv:#1/#2]}}
\def\hspi#1#2{\href{http://www.slac.stanford.edu/spires/find/hep/www?irn=#1}{#2}}
\newcommand{\da}{\dot{a}}
\newcommand{\db}{\dot{b}}
\newcommand{\dn}{\dot{n}}
\newcommand{\dda}{\ddot{a}}
\newcommand{\ddb}{\ddot{b}}

\newcommand{\pa}{a^{\prime}}
\newcommand{\pb}{b^{\prime}}
\newcommand{\pn}{n^{\prime}}
\newcommand{\ppa}{a^{\prime \prime}}

\newcommand{\ppn}{n^{\prime \prime}}
\newcommand{\fda}{\frac{\da}{a}}
\newcommand{\fdb}{\frac{\db}{b}}
\newcommand{\fdn}{\frac{\dn}{n}}
\newcommand{\fdda}{\frac{\dda}{a}}
\newcommand{\fddb}{\frac{\ddb}{b}}

\newcommand{\fpa}{\frac{\pa}{a}}
\newcommand{\fpb}{\frac{\pb}{b}}
\newcommand{\fpn}{\frac{\pn}{n}}
\newcommand{\fppa}{\frac{\ppa}{a}}

\newcommand{\fppn}{\frac{\ppn}{n}}

\newcommand{\drho}{\delta \rho}

\newcommand{\dchi}{\delta \chi}

\title{D-branes in Standard Model building, Gravity and Cosmology}

\author{
~~~~~Elias Kiritsis\\
~\\
CPHT, UMR du CNRS 7644, Ecole Polytechnique,\\
 91128, Palaiseau, FRANCE\\
~~~~~~~~~~~~~~~~~~~~~~~and\\
Department of Physics, University of Crete\\
PO Box 2208, 71003 Heraklion, GREECE\\
~\\
{\tt E-mail: kiritsis@cpht.polytechnique.fr}
}

\preprint{\hepth{0310001}\\CPHT-RR 090.0903}% OR:
\abstract{D-branes are by now an integral part of our toolbox towards
understanding nature. In this review we will describe recent progress in
their use to realize fundamental
interactions. The realization of the Standard Model and relevant physics
and problems will be detailed.
New ideas on realizing 4-dimensional gravity use the brane idea in an
important way.
Such approaches will be reviewed and compared to the standard paradigm of
compactification.
Branes can play a pivotal role both in early- and late-universe cosmology
mainly via the brane-universe paradigm. Brane realizations of various
cosmological ideas (early inflation, sources for dark matter and dark
energy, massive gravity etc) will be also reviewed.}

\dedicated{\tt To appear in Physics Reports}

\begin{document}

\maketitle %%%%%%%%%% THIS IS IGNORED %%%%%%%%%%%

\section{Introduction and perspective}

The basic problem in particle physics is to extend our knowledge
of fundamental physics beyond the energies so far explored by
accelerators and other experiments. The Standard Model of  physics, the
culmination of twentieth century research in the high energy
frontier, has been confirmed to a great degree of accuracy in many
experiments. Despite the fact that the Higgs particle remains
experimentally elusive, few doubt that there may be major
surprises in this direction. It is fair to say that it remains to
be decided whether the Higgs is a fundamental scalar or a bound
state, and what some of its parameters are.

In a related direction, there is concrete experimental evidence
 consistently indicating that neutrinos have (tiny)
masses and mixings and the SM should be extended to accommodate
this. Several ideas on how this can be done, have been forwarded
many years ago and we are currently awaiting for experiment to decide.

On the other hand there are two theoretical issues that since the
late seventies have made physicists believe that the SM, although
extremely successful in describing current experimental data,
cannot be the final story.

${\bf (A)}$ (Quantum) gravity is not incorporated.

${\bf (B)}$ The SM suffers from the hierarchy problem, namely,
that some low energy scales like the mass of the Higgs are
technically unnatural if the SM is extended to energies much
beyond the present energy frontier.

There may be further questions concerning the explanation of the
values of the parameters of the SM but, although important,  they are not as
central
as the ones above.

To these issues we should add some novel questions coming from the
spectacular progress in the last twenty years of observational
cosmology, which can no more be ignored:

${\bf (C)}$ What and how  has made the universe inflate in its
early stages?

${\bf (D)}$ What comprises the dark matter of the universe?

${\bf (E)}$ Why the present day cosmological constant scale (or dark energy)
is of the same order of magnitude as
the Humble constant and many orders of magnitude smaller than the
natural scale of four-dimensional gravity, $M_P$ or other fundamental particle physics scales?

It is felt by many physicists, that the previous questions are most probably interrelated.

The adventure of grand unification in the late seventies/early
eighties followed a path, extending the successful ideas
of non-abelian gauge symmetry that led to the SM description of
the fundamental interactions to its natural conclusion.
In this context, the natural scale of the unified theory
turned out to be $M_U\sim {\cal O}(10^{16-17})$ GeV which is
 not far from the Planck mass $M_P\sim 10^{19}$ GeV where
quantum gravitational effects could no more be ignored.
Moreover, a new question surfaced:

${\bf (B')}$ Why $M_U$ is so close to $M_P$?

In addition, the issue of question (B), became acute: the theories
are technically useless due to the instability induced by the
hierarchy problem.

Several ideas were put forward to deal with the large hierarchy of
scales. The minimal one, technicolor, despite many attempts so far
has failed to produce satisfactory models, mostly because it
invokes strong coupling physics of non-abelian gauge theories that
makes such theories complicated and not very predictive.

An independent idea, involving a new symmetry, supersymmetry, had
more success, at the expense of introducing new degrees of
freedom. Since the low energy world is not supersymmetric, if
supersymmetry is responsible for the stability of the fundamental
theory, it must be spontaneously broken at low energy.
Mechanisms for this are known, but it is fair to say that no
complete and successful model exists so far.

However, physicists today more than ever believe that
supersymmetry in some form holds the key to the successful
resolution of the hierarchy problem and if so, its avatars should
be visible in the next round of accelerator experiments.
Moreover, this belief was strengthened by string theory which
eventually come into the arena of fundamental interactions.

It soon became obvious that spontaneously broken global
supersymmetry, although technically  enough to stabilize grand
unified theories, is too constrained to describe the low energy
data. The advent of local supersymmetry allows such constraints
to be weakened but brings gravity back in
the game. Gravity is an integral part of a locally supersymmetric
theory (supergravity) and this opens  the new pandora's box of
quantum gravity: the theory is non-renormalizable and the issue of
stability (quadratic divergences) resurfaces.

String theory, after a short lived stint as a theory of hadrons,
regained popularity, because it was (and still is) the only theory
that provides a workable theory of quantum gravity (at least in
perturbation theory and for energies below the Planck scale).
It provided a new theoretical argument for supersymmetry: the
presence of space-time fermions in the theory implies an underlying
supersymmetry.

In particular, heterotic string theory, provided a framework that
incorporated the successes of grand unification, supersymmetry,
and a controllable treatment of quantum gravity and justifiably monopolized
the attention of high-energy physicists for a decade.

The heterotic string approach to the fundamental interactions
\cite{het},
provided an answer to the question ${\bf (B')}$. It predicts that
the unification scale is close to the four-dimensional Planck
scale.

Model building in the context of the heterotic string has given
several models which at low energy come close to the standard
model \cite{hetmod1,hetmod2}. However, although the setup is very appealing, there are
still unresolved problems in this context.

$\bullet$ Semi-realistic heterotic ground-states contain particles with
fractional electric charges (other than the quarks).
Strong coupling dynamics in the hidden sector is advocated in
order to bypass this problem.

$\bullet$ N=1 supersymmetric vacua have an involved
structure and the solutions to the
D- and F-flatness conditions are complicated.

$\bullet$ The mechanism of supersymmetry breaking is not well understood.
There are two candidate mechanisms: Gaugino condensation can be
argued to happen in semi-realistic heterotic vacua, breaking
supersymmetry. It involves, however, strong coupling dynamics
and because of this, it has so far failed to have a quantitative
description. Moreover, it is not known (modulo some partial recent
progress using non-perturbative dualities) how to implement it
beyond the effective field theory level.

An alternative mechanism of supersymmetry breaking which is
geometrical and can be implemented at the full string theory level is
the Scherk-Schwarz mechanism, which was introduced earlier in higher dimensional
supergravity, \cite{ss}, and was subsequently implemented in string
theory \cite{sss1}-\cite{sss4}. The supersymmetry breaking scale is proportional
to an inverse compactification length.
However, in the context of simple (orbifold) compactifications of the heterotic string
all compactification scales are of the order of the string scale
$M_s$ which is of the same order as the four-dimensional Planck
scale. It is thus difficult to generate a supersymmetry breaking
scale of a few TeV, necessary for the successful resolution of the
hierarchy problem.

$\bullet$ The hierarchy of SM fermion masses is not easy to
generate. There are ideas in this direction using higher-order
terms in the tree-level potential and strong coupling dynamics in
the hidden sector, but no complete model exists.

$\bullet$ There is a generic axion problem associated with many
semi-realistic heterotic vacua.

Despite these shortcomings, the heterotic string is an elegant
candidate for describing nature and may provide further
breakthroughs in this respect.

In the last decade, further progress in the understanding of string
theory indicated that other ten-dimensional supersymmetric string
theories (type-IIA/B closed strings and type-I closed and open
strings) should be treated on the same footing as the heterotic theory.
The other string theories till then had been handicapped because IIA/B string theory
could not accommodate the SM fields in the perturbative spectrum \cite{iib},
and the type-I theory  had an intricate and little-understood
structure \cite{sagn1}-\cite{sagn6}.
It was successfully argued \cite{ht,w} that the five distinct ten-dimensional
supersymmetric string theories can be viewed as vacua of a larger
(unique?) theory, coined M-theory.
Moreover, non-perturbative dualities relate the strong coupling
behavior of a given theory to another weakly coupled string theory
or a still elusive eleven-dimensional theory.

A key element in these non-perturbative dualities was the
recognition that extended solitonic-like  objects (D-branes and
NS5-branes) form an integral part of the theory.
Moreover, D-branes were shown to have a simple weak-coupling
description as defects where closed strings can open up \cite{pol}.
A detailed understanding of their fluctuations ensued, which
geometrized gauge theories and provided new vistas in the
relationship between  gauge theory and gravity/string theory
\cite{mald1,mald2} with a interesting reconsideration of the strong coupling
dynamics of gauge theories.
An important byproduct was that D-brane degrees of freedom could
explain the macroscopic entropy of black holes \cite{bh1,bh2} and made plausible
that a unitary description of their physics may be at hand.

These developments enlarged substantially the arena for the search
of SM-like vacua in string theory.
In novel contexts, D-branes play a central role, and the purpose
of the present lectures is to discuss their role in new avenues for SM
building, as well as to new realizations of four-dimensional
gravity.
Moreover, D-branes turn out to provide new contexts and new intuition
for older mechanisms (like inflation) deemed necessary in
cosmology, and to indicate new approaches to the problems of dark
matter and the nature of the initial cosmological singularity.

In the search for new string vacua, an exciting new possibility
emerged, namely that a compactification scale \cite{low10}-\cite{low12} or the string scale maybe much lower than the
four-dimensional Planck scale \cite{low1}-\cite{low5} and in fact as low as a few
TeV, without grossly conflicting with available experimental data.
Such ground states, namely orientifolds, can be considered as
generalized compactifications of type-I string theory. They
contain D-branes whose (localized) fluctuations should describe
the SM fields. This gives flesh and blood to earlier ideas of a
brane-universe \cite{shap1}-\cite{shap3}, gives a new perspective of the hierarchy
problem \cite{rs1},
borrowing on ideas from the gauge-theory/gravity correspondence and provides new realizations
of four-dimensional gravity \cite{rs2}.
Moreover, it provides novel contexts for the early-universe
cosmology \cite{betal1}-\cite{alex2}.

\renewcommand{\theequation}{\arabic{section}.\arabic{equation}}
\section{A survey of various string theory
compactifications}
\setcounter{equation}{0}

Different string theories have distinct ways of realizing the
gauge interactions that are responsible for the SM forces.
Ten-dimensional gravity is always an ingredient, coming from the
closed string sector. The simplest way to convert it to four
dimensional gravity is via compactification and this is what we will assume here.
In  section \ref{novel} we will describe other ways of turning higher-dimensional
gravity to four-dimensional, but the implementation of
such ideas in string theory is still in its infancy.

Upon compactification to four dimensions on a six-dimensional
manifold of volume $V_6~M_s^{-6}$ the four-dimensional Planck
scale is given at tree level by
\be
M_P^2={V_6\over g_s^2}M_s^2
\label{hetp}\ee
where $g_s$ is the string coupling constant and the volume $V_6$
is measured in string units. We have dropped numerical
factors of order one.

\renewcommand{\theequation}{\arabic{section}.\arabic{subsection}.\arabic{equation}}
\subsection{The heterotic string}
\setcounter{equation}{0}

The ten-dimensional theory, apart from the gravitational
super-multiplet, contains also a (super) Yang-Mills (sYM) sector with gauge
group $E_8\times E_8$ or $SO(32)$.

Here, the gauge fields descent directly from ten
dimensions, and we obtain for the  four-dimensional gauge coupling
constants
\be
{1\over g^2_{YM}}={V_6\over g_s^2}
\label{hetym}\ee
Such tree-level relations are corrected in perturbation theory and
the couplings run with energy. The tree-level couplings correspond
to their values at the string (unification) scale. In a stable and
reliable perturbation theory, such corrections are small and
this indicates that in order to comply with experimental data
$g_{YM}\sim {\cal O}(1)$ and (\ref{hetp},\ref{hetym}) imply that
\be
M_P^2={M_s^2\over g^2_{YM}}~~~~~~\Rightarrow ~~~~~~   M_P\sim M_s
\ee
Thus, we obtain that for realistic perturbative vacua, the unification scale is tied
to the four-dimensional Planck scale. This is a remarkable
postdiction of the heterotic string that answers question {\bf
B'}.

The issue of supersymmetry breaking is of crucial importance
in order to eventually make contact with the low-energy dynamics
of the Standard Model.

There are two alternatives here, gaugino condensation (dynamical) and
Scherk-Schwarz (geometrical) supersymmetry breaking.

The first possibility can be implemented in the heterotic string,
however it involves non-perturbative dynamics and consequently it
is not well-controllable in perturbation theory. Moreover, we do
not know how to describe this dynamics at the string level.

If  supersymmetry is broken \`a la Scherk-Schwarz, then, the
supersymmetry breaking scale is related to the size $R$ of  an internal
compact direction as
\be
M_{susy}\sim {1\over R}
\ee
The successful resolution of the hierarchy problem requires that $M_{susy}\sim $ a few TeV so that
$M_{susy}/M_P<<1$.
This implies, $R>>M_s^{-1}$ and from (\ref{hetym}) $g_s>>1$ in order to
keep $g_{YM}\sim {\cal O}(1)$.
Thus, we are pushed in the non-perturbative regime.
To obtain information, on the strong coupling dynamics of the
heterotic string we must use non-perturbative dualities.
Under such dualities, the heterotic string is mapped to type-I
string theory \cite{w}, type II theory \cite{ht} or heterotic M-theory \cite{hw}.
A review of such dualities and their nontrivial tests can be found
in \cite{trieste,instanton}.

\subsection{The type-I string}
\setcounter{equation}{0}

In ten dimensions, the strong coupling limit of the heterotic
SO(32)
string is the weakly-coupled type-I string. There are many non-trivial
tests of this duality in ten or less dimensions \cite{dualI1}-\cite{dualI4}.

In the Einstein frame, this duality implies
\be
G^{I}_{\m\n}=G_{\m\n}^{het}\sp A_{\m}^{I}=A_{\m}^{het}\sp
g_s^{het}={1\over g_s^I}\sp M_P^{het}=M_P^{I}
\label{het-I}\ee
Since the ten-dimensional Planck scale, $\hat M_P$, is given by
$\hat M_{P}^8=M_s^8/g_s^2$ in both theories we obtain from
(\ref{het-I})
\be
M_s^{I}={M_{s}^{het}\over \sqrt{g_s^{het}}}
\ee

Thus, at strong heterotic coupling, the dual type-I string scale
is much smaller that the heterotic one: dual type-I strings have a
 much bigger size than the heterotic ones.

 We are, however, interested in four-dimensional compactifications.
 It can be shown \cite{ap} that upon toroidal compactification to
 four-dimensions the strongly coupled heterotic string is dual to
 a weakly coupled type-I string when the compactification torus
 has 4,5 or 6 large dimensions.

 In the type-I vacua, gauge symmetries can arise from D$_p$-branes that
 stretch along the four Minkowski directions and wrap their extra
 $p-3$ dimensions in a submanifold of the compactification
 manifold.
 Let us denote by $V_{||}$ the volume of such a submanifold in
 string units.

 The relation of the four-dimensional Planck scale to the
 string scale is the same as in (\ref{hetp}) since gravity comes
 from the closed string sector. However, the four-dimensional YM
 coupling of the D-brane gauge fields now
become
\be
{1\over g_{YM}^2}={V_{||}\over g_s}
\label{ym}\ee
and
\be
{M_P^2\over M_s^2}={V_6\over g_s~V_{||}}\;\;\;.
\ee
where $M_P$ is the four-dimensional Planck scale.
 Now $M_s$ can be much smaller than $M_P$ while keeping the theory
perturbative,
$g_s<1$, by having the volume of the space transverse to the D$_p$ branes ${V_6\over
V_{||}}>>1$.
Thus, in this context, the string scale $M_s$ can be anywhere between the four-dimensional
Planck scale and a few TeV
without any obvious experimental objection \cite{low1}-\cite{low5}.
The possibility of perturbative string model building with a
very
low string scale is intriguing and inherently interesting due to
several reasons

$\bullet$ If $M_s$ is a few TeV, string effects will be visible at
experiments around that energy scale, namely accelerator
experiments in the near future. If nature turns out to work that
way, the experimental signals will be forthcoming. In the other
extreme case $M_s\sim M_P$, there is little chance to see telltale
signals of the string at TeV-scale experiments.

$\bullet$ Supersymmetry can be broken directly at the string scale
without the need for fancy supersymmetry breaking mechanisms
(for example by direct orbifolding).
Past the string scale there is no hierarchy problem since there is
no field theoretic running of couplings.

It seems that there is thus no-hierarchy problem to solve in this
case. It is not difficult to realize, however, that here the
essence of the problem is hidden elsewhere.
Some of the internal dimensions in this context must be much
larger than the string scale. The radii are expectation values of
 associated scalar fields that will generically have a potential
 in the absence of supersymmetry. The hierarchy problem now reads:
 why the minimum of the potential is at $R>>1$?
 Although we can imagine potentials with logs providing large hierarchies of their
 minima and computations in semi-realistic models indicate that
 this is possible \cite{lau1,lau2},
 it  has not yet been achieved in realistic models in string
 theory.

\subsection{Type II string theory}
\setcounter{equation}{0}

Another non-perturbative duality relates the heterotic string
compactified on $T^4$ to type-IIA theory compactified on $K3$
\cite{ht,w}.
There is by now direct \cite{kop} and indirect \cite{kv} non-trivial
evidence for this.

This duality keeps the Einstein metrics and Planck scales of the
two theories fixed, and inverts the six-dimensional string
coupling $\tilde g_s$ defined as
\be
{1\over \tilde g_s^2}={V_4\over g_s^2}\sp \tilde g^{het}_s={1\over
\tilde g_s^{IIA}}
\ee
Moreover, it dualizes the six-dimensional two-index antisymmetric
tensor. In this way, the heterotic string is a magnetic string
soliton of the electric type IIA string and vice versa.

It can be shown \cite{ap} that this duality maps a strongly
coupled heterotic string to a weakly coupled type II string when
we have 1,2,3,4,6 large dimensions.

In type II compactifications, the four-dimensional matter gauge
fields come from the RR sector and generate abelian symmetries.
There are no charged massive gauge fields in the perturbative
spectrum to enhance the gauge symmetry to a non-abelian one.
However, such enhancement can happen non-perturbatively. When
two-cycles of the compactification manifold shrink to zero volume,
states produced by D$_2$ branes wrapping these cycles generate
massless charged gauge bosons that enhance the symmetry to a
non-abelian group.

We thus consider a compactification on $K_3\times T^2$ of volume
$V_{K3}$ and $V_2$ respectively in string units.

As in the heterotic case, the relation between the
four-dimensional Planck scale and the string scale remains the
same
\be
M_P^2={V_{K3}V_2\over g_s^2}M_s^2
\ee
On the other hand, since the non-abelian gauge symmetry comes from
the RR sector there is no $V_{K3}$ nor $g_s$ dependence in the four-dimensional YM
coupling
\be
{1\over g_{YM}^2}=V_2~~~~~\Rightarrow~~~~~{M_P^2\over
M_s^2}={V_{K3}\over g_s^2}{1\over g_{YM}^2}
\ee
We can consequently generate a large hierarchy between $M_P$ and $M_s$ by
choosing $V_{K3}>>1$ or $g_s<<1$.

The second case with $V_{K3}\sim 1$ and $g_s\sim 10^{-14}$ was
explored in \cite{adg} and new interesting features of the
 dynamics uncovered.
It is however, hard to construct calculable models since the
physics at the singularity of the internal space
is related to Little String Theory
(LST) \cite{lst} which is not very well understood.

\section{Orientifold vacua, chirality and supersymmetry}

An efficient way of constructing type-I vacua is in terms of
orientifolds \cite{sagn1}-\cite{sagn6}, an upgrade to the notion of orbifold
in closed string theory.

One starts from a type IIB (flat) vacuum described by a CFT invariant
under world-sheet parity $\Omega$.
One then orbifolds by a group of discrete symmetries that also
includes $\Omega$. The structure of the orientifold group is then
$G=G_1+\Omega~G_2$ where $G_{1,2}$ are symmetry groups of the CFT.
At generic values of the toroidal moduli, such symmetries are
(left and/or right) translations and rotations preserving the lattice.

The one-loop amplitude implementing the $\Omega$ projection is
interpreted as a Klein-bottle amplitude \cite{carlo,dab} and has
potential ultraviolet divergences (tadpoles).

Such tadpoles can be interpreted as sources in space-time
introduced by the orientifold ($\Omega$) projection. They couple
to the massless IIB fields, in particular the metric (so they
have energy or tension), the dilaton and the RR-forms (under which
they are minimally coupled).
Such sources are localized in sub-manifolds of space-time,
typically hyper-planes and are known as orientifold planes, $O_p$.
They are BPS-like ($|$tension$|$=$|$charge$|$) and there are four
different kinds in each odd dimension characterized by positive vs
negative tension and positive vs negative RR charge.
The most common orientifold planes have negative tension, a fact that
comes handy in the search of a flat vacuum.

Flatness and stability will be assured if D-branes are introduced
in a way that guarantees the cancellation of tadpoles  (energy and
charge).
They can be thought of as the twisted sector of the $\Omega$
projection.

Thus, an orientifold vacuum can be described as a collection of
orientifold planes and D-branes in various configurations.
RR-tadpole cancellation is necessary for consistency, since it is
equivalent to the vanishing of gauge charge in a compact space.
NS-tadpole cancellation is equivalent to the vanishing of forces
in the D-brane/O-plane vacuum configuration.
A vacuum with un-canceled NS tadpoles is not in equilibrium and
this shows up at (open) one-loop. In particular, there will be UV divergences
that signal this lack of equilibrium.
From the closed string point of view this is an IR instability.

Orientifold planes, unlike D-branes do not have fluctuations. This
is as well since in the case of negative tension such fluctuations
would have negative kinetic terms.
However, acting as sources, they do couple to the massless sector fields of the closed
sector.

\subsection{D-brane mechanics}

D$_p$-branes are defined as sub-manifolds of space-time where the closed strings can open up
and attach.
In flat space-time, such open strings have Neumann boundary conditions along the
$p+1$ longitudinal directions of a D$_p$-brane and Dirichlet for the rest.

The massless spectrum of fluctuations of a single D-brane,
according to the Polchinski prescription \cite{pol}, is that of a
vibrating open superstring with its two endpoints forced to be
attached to the brane.
For the flat type II theory, this gives a vector $A_{\mu}(x^{\n})$, $9-p$
scalars $\Phi_{I}(x^{\n})$ and a  16-component fermion.
This is the dimensional reduction of the D=10 N=1 gauge multiplet
to $p+1$ dimensions. The expectation values of the scalars
$\Phi_I$ are the transverse coordinates of the position of the
brane.

N distinct parallel branes are described by the $N^2$ possible
strings stretched between them, generating the spectrum (and
interactions) of a U(N) sYM theory.
Moving the branes apart, the gauge symmetry is Higgsed to $U(1)^N$ in
the generic case.

In general, for a string coordinate with Neumann  boundary conditions on
both endpoints (NN), the mode expansion contains a center of mass
momentum, no winding and integral oscillator modes. There are zero modes in
the Ramond sector.

For a string coordinate with  Dirichlet  boundary conditions on
both endpoints (DD), the mode expansion contains a winding number,
no momentum and
 integral oscillator modes. There are again zero modes in
the R-sector.

Finally, for a string coordinate with  Dirichlet  boundary conditions on
one endpoint and  Neumann on the other, (DN), the mode expansion contains no
winding or momentum
and half-integral oscillator modes.
There are fermionic zero modes in
the NS-sector.

T-duality acts on D-branes by interchanging N and D boundary
condition. Thus, a T-duality in a direction transverse to a
D$_p$
brane transforms it into a D$_{p+1}$ brane while T-duality
in a direction longitudinal  to a D$_p$
brane transforms it into a D$_{p-1}$ brane.
T-duality splits O-planes and fractionalizes their charge.

An open string with one endpoint attached on a stack of N coincident parallel D-branes
has a spectrum that transforms as the fundamental
(or anti-fundamental depending on the orientation) representation
of the U(N) gauge symmetry.

One important property that is crucial for model building is the
issue of the reduction of the supersymmetry of the world-volume gauge
theory. This is a prerequisite in order to obtain chiral
four-dimensional fermions, an crucial ingredient of the standard
model.

As we will expand below, there are two such methods available in
the context of orientifolds: internal world-volume magnetic fields
\cite{bachas,bdl,b1,b3,poly}
and branes stuck at orbifold singularities \cite{dm,bsing}.
The second option is an extreme case of the generic situation of branes wrapping
non-trivial cycles of the compactification manifold.

\subsection{D-branes  at orbifold singularities}

We will describe here the example of D$_3$ and D$_7$ branes stuck
at orbifold singularities following \cite{bsing}.

We thus consider $n$ D$_3$ branes transverse to a $Z_N$ singularity.
Since the brane description is local in this case, all that matters
is the local structure of the singularity and this is similar to
$R^6/Z_N$.
The $Z_N$ rotation acts on the SO(6) R-symmetry quantum
numbers of the massless brane fields. The vectors $A_{\mu}$
transform in the singlet, the fermions in the spinor and the
scalars in the vector.

Complexifying the scalars by pairs, the $Z_N$ rotation acts on
them as
\be
R_{\theta}={\rm diag}\left(~e^{2\pi i {b_1\over N}}~,~e^{-2\pi i {b_1\over N}}~,~ e^{2\pi i {b_2\over
N}}~,~ e^{-2\pi i {b_2\over
N}}~,~ e^{2\pi i {b_3\over N}}~,~ e^{-2\pi i {b_3\over N}}~\right)
\ee
while on the spinor it acts as
\be
S_{\theta}={\rm diag}\left(~e^{2\pi i {a_1\over N}}~,~ e^{2\pi i {a_2\over
N}}~,~ e^{2\pi i {a_3\over N}}~,e^{2\pi i {a_4\over N}}~\right)
\ee
with
\be
a_1={b_2+b_3-b_1\over 2}\sp a_2={b_1-b_2+b_3\over 2}\sp
a_3={b_1+b_2-b_3\over 2}\sp a_4=-{b_1+b_2+b_3\over 2}
\ee

Moreover we can parameterize the action of the rotation on the
Chan-Paton (CP) indices without loss of generality using the
matrices
\be
 \gamma_{3,\theta}={\rm diag}\left({\bf 1}_{n_0},\theta~{\bf
1}_{n_1},\cdots, \theta^{N-1}~{\bf 1}_{n_{N-1}}\right)
\ee
where $\theta =e^{2\pi i\over N}$, $n=\sum_{i=0}^{N-1}n_i$ and ${\bf 1}_{n}$ is the unit
$n\times n$ matrix.

The orbifold action on the gauge boson state is
\be
A_{\m}\sim \psi^{\mu}_{-{1\over 2}}|~\lambda ~\rangle \to \psi^{\mu}_{-{1\over 2}}|~\gamma_{3,\theta}~\lambda
~\gamma^{-1}_{3,\theta}~\rangle
\ee
where the matrix $\lambda$ keeps track of the CP indices.

Thus, the invariant gauge bosons must satisfy $\lambda
=~\gamma_{3,\theta}~\lambda~
\gamma^{-1}_{3,\theta}$.
The solutions to this equation are $n_i\times n_i$ block diagonal
matrices, and thus the invariant gauge bosons are in the adjoint
of $\prod_{i=0}^{N-1}~U(n_i)$.

The three complex scalars $\Phi_k$ obtained from the complexification of the six real scalars, transform as
\be
\Phi_k\sim \psi^{k}_{-{1\over 2}}|~\lambda ~\rangle \to e^{-2\pi i {b_k\over N}}~
\psi^{k}_{-{1\over 2}}|~\gamma_{3,\theta}~\lambda
~\gamma^{-1}_{3,\theta}~\rangle
\ee
and the invariant scalars must satisfy $\lambda =e^{2\pi i {b_k\over
N}}~~\gamma_{3,\theta}~\lambda~
\gamma^{-1}_{3,\theta}$. The solution to this equation
gives scalars in the following representation of the gauge group
$${\rm scalars}~~~~~~~~~~\to~~~~~~\sum_{k=1}^3\sum_{i=0}^{N-1}\left(n_i,\bar
n_{i-b_k}\right)$$

Finally the fermions are labelled as
\be
\psi_a\sim |\lambda~;s_1,s_2,s_3,s_4\rangle
\ee
where $s_i=\pm {1\over 2}$, with $\sum_{i=1}^4~s_i=odd$ (GSO projection). The
states with $s_4=-{1\over 2}$ correspond to left, four-dimensional
Weyl fermions while $s_4={1\over 2}$ corresponds to right, four-dimensional
Weyl fermions.
The $s_{1,2,3}$ spinor quantum numbers are R-symmetry spinor
quantum numbers. We can thus label the 8 on-shell fermion states
as $|\lambda;\a,s_4\rangle$ where $\alpha=1,2,3,4$ is the R-spinor
quantum number.
The fermions  transform as
\be
|\lambda;\alpha,s_4\rangle\to e^{2\pi i{a_{\a}\over N}}~|~\gamma_{3,\theta}~\lambda
~\gamma^{-1}_{3,\theta}~;\a,s_4~\rangle
\ee
and the invariant fermions satisfy $\lambda =e^{2\pi i {a_{\a}\over
N}}~~\gamma_{3,\theta}~\lambda~
\gamma^{-1}_{3,\theta}$
The solution to this equation
gives left moving fermions in the following representation of the gauge group
$${\rm left ~~Weyl ~~fermions}~~~~~~~~~~\to~~~~~~\sum_{\a=1}^4\sum_{i=0}^{N-1}\left(n_i,\bar
n_{i+a_{\a}}\right)$$
a representation that is generically chiral.

When $\sum_{i=1}^3~b_i=0$ so that $a_4=0$ we have an N=1 supersymmetric
configuration (the rotation $\in SU(3)\subset SO(6)$) and this is
an N=1 orbifold singularity.
The $a_4$ fermions become the gaugini while the $a_{1,2,3}$
fermions are the N=1 supersymmetric partners of the scalars.

To cancel the twisted tadpoles we must introduce also m D$_7$
branes (that we take to be transverse to the last complex
coordinate, the third plane).

For the 77 strings the story is similar with a new CP matrix
\begin{equation}
\gamma_{7,\theta}=\left\{ \begin{array}{lll}
\displaystyle {\rm diag}\left({\bf 1}_{m_0},\theta~{\bf
1}_{m_1},\cdots, \theta^{N-1}~{\bf 1}_{m_{N-1}}\right),&\phantom{aa} &b_3~~{\rm even},\\ \\
{\rm diag}\left(\theta~{\bf 1}_{m_0},\theta^3~{\bf
1}_{m_1},\cdots, \theta^{2N-1}~{\bf
1}_{m_{N-1}}\right),&\phantom{aa}&b_3~~{\rm odd}
\end{array}\right.
\nn
\end{equation}
The extra fields that are localized on the D$_3$ brane
world-volume come from the 37 and 73 strings.
For such strings, there are 4 ND directions which provide 4 zero
modes in the NS sector (directions 4,5,6,7) while from the NN and
DD directions we have zero modes in the R sector ( directions
2,3,8,9)

 The invariant (complex) scalars (NS sector) must satisfy
 \be
\lambda_{37}=e^{-i\pi{b_1+b_2\over
N}}~~\gamma_{3,\theta}~\lambda_{37}
~\gamma^{-1}_{7,\theta}~
\ee
\be
\lambda_{73}=e^{-i\pi{b_1+b_2\over
N}}~~\gamma_{7,\theta}~\lambda_{73}
~\gamma^{-1}_{3,\theta}~
\ee
with spectrum

\begin{equation}
\begin{array}{lll}
\displaystyle \sum_{i=0}^{N-1}\left[\left(n_i,\bar m_{i-{b_1+b_2\over 2}}\right)+
\left(m_i,\bar n_{i-{b_1+b_2\over 2}}\right)\right]
,&\phantom{aa} &b_3~~{\rm even},\\ \\
\displaystyle\sum_{i=0}^{N-1}\left[\left(n_i,\bar m_{i-{b_1+b_2+1\over 2}}\right)+\left(m_i,
\bar n_{i-{b_1+b_2-1\over 2}}\right)\right],&\phantom{aa}&b_3~~{\rm odd}
\end{array}
\nn
\end{equation}

The invariant fermions coming from the R-sector must satisfy
 \be
\lambda_{37}=e^{i\pi{b_3\over
N}}~~\gamma_{3,\theta}~\lambda_{37}
~\gamma^{-1}_{7,\theta}~
\ee
\be
\lambda_{73}=e^{i\pi{b_3\over
N}}~~\gamma_{7,\theta}~\lambda_{73}
~\gamma^{-1}_{3,\theta}~
\ee
with spectrum
\begin{equation}
\begin{array}{lll}
\displaystyle \sum_{i=0}^{N-1}\left[\left(n_i,\bar m_{i+{b_3\over 2}}\right)+
\left(m_i,\bar n_{i+{b_3\over 2}}\right)\right]
,&\phantom{aa} &b_3~~{\rm even},\\ \\
\displaystyle\sum_{i=0}^{N-1}\left[\left(n_i,\bar m_{i+{b_3-1\over 2}}\right)+\left(m_i,
\bar n_{i+{b_3+1\over 2}}\right)\right],&\phantom{aa}&b_3~~{\rm odd}
\end{array}
\nn
\end{equation}

A similar analysis holds for D$_7$ branes stretched transverse to
the other two planes.

Finally, the twisted tadpole cancelation conditions in the sector labeled by $k=1,2,\cdots, N-1$ are
\be
Tr\left(\gamma_{3,\theta}^k\right)\prod_{i=1}^3\left[2\sin {\pi k
b_i\over N}\right]+Tr\left(\gamma_{7,\theta}^k\right)\sum_{i=1}^3
2\sin {\pi k
b_i\over N}=0
\ee
For viable models twisted tadpoles have to be canceled   and there
are many solutions to this.

Thus, such configurations provide a generically chiral spectrum of
four-dimensional fermions.

\subsection{Internal magnetic fields and branes intersecting at angles}
\setcounter{equation}{0}

It is well known that magnetic fields can generate chirality. The
reason is that an (internal)  magnetic field splits the masses of
spinors according to their spin quantum numbers, and if
appropriately adjusted, it can lead to chiral pieces of a non-chiral
spinor remaining massless while the other parts picking up a mass.
This mechanism can be implemented also in open string theory
\cite{bachas,bdl}.
What we will show here is that there is a T-dual version of this
mechanism for chirality generation that has a pure geometrical
interpretation, namely strings stretched between intersecting
branes.

Consider an open string that has Neumann boundary conditions on a
Euclidean 2-plane with coordinates $x^{1,2}$.

\be
\left.\p_{\s}x^1\right|_{\s=0}=0\sp
\left.\p_{\s}x^2\right|_{\s=0}=0\sp
\left.\p_{\s}x^1\right|_{\s=\pi}=0\sp \left.\p_{\s}x^2\right|_{\s=\pi}=0
\ee

We will turn on a constant magnetic field for the gauge field
associated with the $\s=\pi$ end-point of the string.
The world-sheet action is
\be
S={1\over 2\pi \a'}\left[\int
d\tau~d\s~{1\over 2}\left(\p_{\tau}x^i\p_{\tau}x^i+\p_{\s}x^i\p_{\s}x^i\right)
+\left.\int d\tau~A_{i}\p_{\tau}x^i\right|_{\s=\pi}\right]
\ee
$$
={1\over 4\pi \a'}\left[\int
d\tau~d\s\left(\p_{\tau}x^i\p_{\tau}x^i+\p_{\s}x^i\p_{\s}x^i\right)
-\left.\int d\tau~\partial_jA_i~x^i\p_{\tau}x^j\right|_{\s=\pi}\right]
$$
where in the second line we integrated the boundary term by parts
and dropped a total $\tau$-derivative.
For a constant magnetic field threading the plane,
$F_{ij}=B\e_{ij}$.
Varying $x^i $ to get the equations of motion and integrating by parts we obtain
\be
\delta S=-{1\over 2\pi \a'}\int d\tau~d\s~\delta
x^i\left(\p_{\tau}^2+\p_{\s}^2\right)x^i+
\ee
$$+{1\over 2\pi \a'}
\int
d\tau~\left[\left.-\delta x^i\p_{\s}x^i\right|_{\s=0}+\left.\delta
x^i\left(\p_{\s}x^i+B\e_{ij}\p_{\tau}x^j\right)\right|_{\s=\pi}\right]
$$
we thus obtain the bulk free equations
$\left(\p_{\tau}^2+\p_{\s}^2\right)x^i=0$ along with the modified
boundary conditions

\be
\left.\p_{\s}x^1\right|_{\s=0}=0\sp
\left.\p_{\s}x^2\right|_{\s=0}=0
\ee
\be
\left.\p_{\s}x^1+B~\p_{\tau}x^2\right|_{\s=\pi}=0\sp \left.\p_{\s}x^2-B~\p_{\tau}x^1\right|_{\s=\pi}=0
\ee

We will now perform a T-duality in the coordinate $x^2$ which
exchanges $\p_{\s}x^2\leftrightarrow \p_{\tau}x^2$.
The T-dual boundary conditions now read
\be
\left.\p_{\s}x^1\right|_{\s=0}=0\sp
\left.\p_{\tau}x^2\right|_{\s=0}=0
\ee
\be
\left.\p_{\s}x^1+B~\p_{\s}x^2\right|_{\s=\pi}=0\sp \left.\p_{\tau}x^2-B~\p_{\tau}x^1\right|_{\s=\pi}=0
\label{bou}\ee
The boundary conditions at the $\s=0$ end-point describe a string
attached to a D-brane localized along the line $x^2=0$.
On the other hand, the boundary conditions at the $\s=\pi$ end-point describe a string
attached to a D-brane rotated by an angle $\theta$ counter-clockwise, where $B=-\cot\theta$.
Otherwise stated the second D-brane is localized along the line
$\sin\theta~x^1+\cos \theta~x^2=0$.

The net result is that an open string coupled to a constant
magnetic field is T-dual to a string stretched between two
D-branes intersecting at an angle $\theta=-{\rm arccot} (B)$.

The effect of the boundary conditions (\ref{bou}) on the
oscillator expansion of the string is to eliminate momentum and
winding terms, and to shift the oscillators $a_{n}\to
a_{n+\theta}$.
Thus, the overall effect is  the same as that of an orbifold  of the plane by a angle of rotation
$\theta$.
Similar remarks apply to the world-sheet fermions.

\subsection{Intersecting $D_6$-branes\label{id6}}
\setcounter{equation}{0}

To obtain chiral four-dimensional spinors, no internal two-torus
should be left intact. Thus, the full picture involves $D_6$
branes that intersect at each of the three internal
tori\footnote{We consider a $T^6$ with moduli such that it
factorizes into a product, for simplicity.}
\cite{b1}-\cite{poly}.

$D_6$-branes can be obtained by performing a T-duality along 3
directions of the standard type-I string. A T-duality along a
direction changes the orientifold projection from $\Omega$ to
$\Omega~I$ where $I$ flips the sign of the T-dualized coordinate
\cite{carlo}.

We are thus led to consider IIB on $T^6$ moded by $\Omega~I_3$.
Introducing coordinates $x^1_i,x^2_i$, $i=1,2,3$ for the three two-tori,
$I_3$ flips the sign of the $x_i^2$ coordinates.
Turning-on magnetic fields $B_i$ in the type-I theory corresponds
to this T-dual version to $D_6$-branes intersections at angles
$\theta_i$ on each of the three internal two-tori.

The $D_6$-branes stretch in the 3+1 Minkowski directions and wrap a
one-cycle in each of the three internal tori.
Introducing the standard $a_i,b_i$ cycles on each torus, a given $D_6$
brane wraps the homology 3-cycle
\be
[~\Pi_A]=\prod_{i=1}^3(m_i[a_i]+n_i[b_i])
\ee
 where the product is a wedge product.

Since $\Omega~I_3$ reflects the D-branes along the $x^2_i$
coordinates, it maps a given D-brane {A} to its mirror image {A'} wrapped
on the cycle

\be
[~\Pi_{A'}]=\prod_{i=1}^3(m_i[a_i]-n_i[b_i])
\ee

At each (oriented) intersection there is a chiral fermion
(localized in four dimensions) which is left-moving (right-moving) if the intersection numbers is 1(-1).
Thus the net number of chiral fermions coming from the string
between a brane A and a brane B is equal to the total intersection
number
\be
I_{AB}=\int_{T^6}~[~\Pi_A]\wedge
[~\Pi_B]=\prod_{i=1}^3~(m^i_An^i_B-n^i_Am^i_B)
\label{intersection}\ee
where we have used
\be
\int_{T^2}~[a]\wedge [b] =1
\ee
In the general case, a $D_p$ brane wrapping a p-dimensional
sub-torus of $T^N$, and a $D_{N-p}$ brane wrapping a $(N-p)$-dimensional
sub-torus of $T^N$ intersect at points in $T^N$.
Choosing coordinates $X^I$, $I=1,2,\cdots, N$ for $T^N$, $x^a$, $a=1,2,\cdots,p$ for $D_p$ and
$\xi^i$, $i=1,2,\cdots, N-p$ for $D_{N-p}$, all of them periodic up to integers, the wrapping numbers
are defined as
\be
X^I=N^I_{a}~x^a\sp X^I=M^I_i~\xi^i\sp N^I_a,M^I_i\in \Zint
\ee
Then, the intersection number is
\be
I_{p,N-p}=\epsilon_{I_1I_2\cdots I_N}~\e^{a_1a_2\cdots a_p}~\e^{i_1i_2\cdots i_{N-p}}~N^{I_1}_{a_1}\cdots N^{I_p}_{a_p}
~M^{I_{p+1}}_{i_1}\cdots N^{I_N}_{i_{N-p}}
\ee

Considering now stacks of $N_A$ $D_6$-branes, from the AA strings
we obtain gauge bosons in the adjoint of $U(N_A)$ as well as
scalars that are generically massive (and maybe tachyonic)
\cite{b1,b3}.

From the AB+BA strings we obtain $I_{AB}$ chiral fermions in the $(N_A,{\bar N_B})$ of $U(N_A)\times U(N_B)$.
From the AB'+B'A strings we obtain $I_{AB'}$ chiral fermions in the $(N_A,N_B)$ of $U(N_A)\times U(N_B)$.
Finally, from the AA'+A'A strings we obtain chiral fermions in the symmetric or antisymmetric representation
of $U(N_A)$. Their number is model-dependent.

The tadpole cancellation conditions specify that the D-branes
should be put on top of the orientifold planes (wrapping the cycle $[~\Pi_O]$) namely
\be
\sum_{\rm all ~~branes}~N_A~[~\Pi_A]+[~\Pi_O]=0
\ee

By arranging the multiplicities $N_A$ and the wrapping
numbers
$m_A^i, n_A^i$, models can be obtained with the standard model
gauge group, and the correct spectrum \cite{b1,b2,b3,madrid0}-\cite{madrid3}.

It should be mentioned that such intersecting configurations of
D-branes break supersymmetry completely for arbitrary angles. If
however $\sum_{i=1}^3\theta_i=0$ in a given intersection, then N=1
supersymmetry is preserved locally.

\subsection{Supersymmetry breaking}

Supersymmetry in string theory is broken by compactification (ie.
background fields).
For the flat (orbifold or orientifold) compactifications
supersymmetry is typically broken by the orbifold/orientifold
projection.

We distinguish two types of orbifold actions on toroidal compactifications.

$\bullet$ Non-free susy-breaking orbifolds. In this case the orbifold action has fixed points and the compactification
manifold is singular. At the singularities (orbifold planes) the
twisted fields are localized. The breaking of supersymmetry here
is ``explicit" in the sense that there are no massive gravitini
with masses below the string scale. Another way of saying this is
that the supersymmetry breaking scale is the string scale, $M_{\rm susy}\sim M_s$.

$\bullet$ Freely-acting susy-breaking orbifolds. Here the orbifold action has no
fixed points, and the compactification manifold is flat and
smooth. The lightest gravitino has a mass proportional to an
inverse internal radius and thus the supersymmetry breaking scale
is $M_{\rm susy}\sim 1/R$.
Such compactifications are essentially stringy \cite{sss1}-\cite{sss4} Scherk-Schwarz
compactifications \cite{ss}, (as  shown in  \cite{sss4}).
Moreover, the nature of the supersymmetry breaking is spontaneous,
since (broken) supersymmetric Ward identities are satisfied
\cite{porrati}.
In the limit $R\to \infty$ supersymmetry is restored in
one higher dimension.

\renewcommand{\theequation}{\arabic{section}.\arabic{subsection}.\arabic{subsubsection}.\arabic{equation}}
\subsubsection{Supersymmetry breaking in low-string scale
orientifolds}
\setcounter{equation}{0}

As it was mentioned earlier, supersymmetry must break in a
realistic string ground-state. When the string scale is low (in
the TeV region) the supersymmetry must break around that scale.
In the context of orientifold ground states this may be
accomplished at the string level by a combination of the following
mechanisms \cite{openss1}-\cite{openss7,lau1,non-madr1,non-madr2}

$\bullet$ Non-freely acting supersymmetry-breaking orbifolds in
the bulk, that generically induce breaking in the open sector.

$\bullet$ Freely-acting (SS) supersymmetry breaking orbifolds in the
bulk that may also induce breaking in the open sector. The
breaking could affect all modes in the open sector, or only the
massive ones.

$\bullet$ Intersecting branes induce supersymmetry breaking in the
open sector. This also is related as we have seen to non-trivial
internal magnetic fields on the branes \cite{bachas,bdl}.

$\bullet$ The presence of anti-branes induces supersymmetry
breaking in the open sector.

It is obvious that at tree level, supersymmetry may break in the open sector and/or
in the bulk. We may imagine more involved situations in which a
tree level breaking in the closed sector may be communicated by
loop effects to the open sector and vice versa. Although this is
possible, calculability is reduced in such cases.

It is also possible that different sectors of open strings may be
invariant under different supersymmetries, and finally
supersymmetry will be broken in a given sector by radiative
corrections \cite{iba}. Calculability is generically impaired  also in such
cases. It should be remembered that in the open string case,
unlike the closed string case, essentially nothing is known on
two-
and higher-loop diagrams.

Many different mechanisms and combinations of supersymmetry
breaking patterns have been investigated in toy models \cite{stoy1}-\cite{stoy3}.

Thus, it seems that at the simplest level, supersymmetry must
break at tree level in the open sector, namely in the standard
model part of the open sector (The open sector can in principle
contain hidden matter, interacting to the SM fields via particles
coming from stretched/massive
strings).

A priori we can imagine that  supersymmetry may be intact at tree level
in the closed sector. We will however argue  that for (NS) tadpole
cancellation, supersymmetry should also be broken in the closed
sector.

Imagine an N=1 bulk compactification (orientifold) in the closed
sector with an open sector that breaks supersymmetry completely.
The (open) supersymmetry breaking can be interpreted as coming
from  D and F auxiliary field expectation values that always give
a positive contribution to the effective potential.
The only term that could cancel such contributions and make the
vacuum energy (dilaton/graviton tadpole) vanish is the
supergravity auxiliary field \cite{super}.
Thus, supersymmetry should be also broken in the bulk.

It is interesting that this effective supergravity picture fits
nicely with the function of orientifold planes in orientifolds.
Orientifold planes do not only cancel RR charges but also energy
(since they carry negative tension).
They can be interpreted as localized expectation values of the (higher-dimensional)
supergravity auxiliary fields.

This picture is  vague in the context of N$>$1 four-dimensional
supersymmetry since the full off-shell structure of the
supergravity and the associated auxiliary fields are not well
understood.
However, we expect that it can be made precise in the N=1 case.
It has been also argued \cite{lau1,lau2} that the breaking of bulk
supersymmetry cannot be hard ($M_{SUSY}=M_s$) since this induces
quadratic instabilities ($\sim M_P^2$) in the theory when large
dimensions are present. This can be seen as follows:
The ten-dimensional one-loop vacuum energy in this case is $\sim
M_s^{10}$ and the four-dimensional one
\be
\Lambda_4\sim M_s^4~ V_6\sim M_s^2 M_P^2
\ee
where we have used (\ref{hetp}). This indicates a hierarchy
problem when $M_P>>M_s$.

The cancellation of RR tadpoles in a open string ground state is
necessary for the consistency of the theory \cite{cai}.
The cancellation of the NS tadpoles, on the other hand is
equivalent to the equilibrium condition of the brane
configuration, or in general of the ground-state.
When equilibrium is not present, one-loop amplitudes are infinite
and calculability is lost.

Since, in constructing a stringy SM, one-loop corrections
are
crucial for comparisons with data, such a situation seems
doomed.
Although we may consider shifting the background data to cancel
the tadpoles perturbatively, this does not seem to be a
calculationally promising avenue.
It is thus imperative to cancel NS tadpoles at tree-level.

A final issue is that of stability of the ground-state alias the
absence of tachyons. We should consider separately the open and
closed tachyons.

$\bullet$. Closed string tachyons signal an instability in the
closed sector and must be canceled.

$\bullet$ Open string tachyons signal an instability of the brane
configuration but that is what the SM wants!
The standard Higgs (as well as other possible Higgses) are
tachyons in the symmetric $SU(3)\times SU(2)\times U(1)$ vacuum.
One may thus allow the appropriate tachyons at tree level, and
this is the case in some semi-realistic models \cite{non-madr1,non-madr2,b1,b2,b3,madrid0}-\cite{madrid3}.
Then the rolling of the Higgs to the minimum of its potential can
be interpreted as a rearrangement of the brane configuration
(brane recombination \cite{iba1}). The flip side of this is that in
the final ground state the branes do not seem flat anymore and
further progress is needed in order to be able to calculate.

The other possibility \cite{akt,akt1}, which may be easily realized in orientifolds
is that the candidate Higgs field is massless at tree level (and
typically has a quartic term in its potential.
At one-loop radiative corrections to its mass can be negative
overall. In particular this is the case in the non-supersymmetric
SM since fermions give negative contributions, and massive
fermions dominate. Thus,  one-loop correction to the Higgs mass is
dominated by the top and is negative.
The flip side of this situation is that two-loop calculations are
beyond reach.

Finally we should mention that one may construct models with a
string scale close to the Planck scale and unbroken supersymmetry
\cite{cvetic1}-\cite{cvetic6}.

\section{Anomalous U(1)s in six and four dimensions and their role in string model building}

In string theory, we often encounter U(1) gauge symmetries with an
anomalous spectrum, namely $Tr[Q]\not= 0$, $Tr[Q^3]\not= 0$ etc.
The theory is however free of the associated anomalies due to the
presence of the generalized Green-Schwarz mechanism.
The anomalous variation of the action is canceled by
non-invariant, classical contributions of antisymmetric tensor
fields.
This mechanism works in various dimensions, with a prototypical
example, the cancellation of mixed and reducible anomalies in the
ten-dimensional heterotic and type-I string theories.

In four dimensions, anomalous U(1)'s have their anomalies canceled
via their Stuckelberg coupling to a pseudoscalar axion. As a result, the gauge symmetry is broken
and the anomalous U(1) gauge boson obtains a mass.

There are two reasons to be interested in this issue: First, it seems
that (several) anomalous U(1) symmetries are
unavoidable \cite{akt} in brane constructions of the SM. Second, the
fate of the associated global symmetries (that may include baryon
and lepton number) is of paramount phenomenological importance in
such realizations of the SM.
Finally, anomalous U(1) gauge bosons mix at tree level with
the massive standard model gauge bosons and thus there are
constraints coming from the SM $\rho$-parameter \cite{Z}.
Also, the anomalous U(1) gauge bosons mediate at loop level
several SM processes, like $g-2$ and data provide constraints on
their masses \cite{g-2}.

A recent review on anomalies and anomalous U(1)s in field theory
with extra dimensions can be found in \cite{scrucca1}

\renewcommand{\theequation}{\arabic{section}.\arabic{subsection}.\arabic{equation}}
\subsection{Anomalies and symmetries in four dimensions}
\setcounter{equation}{0}

We will now consider an anomalous U(1) gauge field $A_{\m}$ in a four-dimensional
theory.
Due to non-vanishing triangle diagrams

\bea
\unitlength=0.6mm
\begin{fmffile}{anomalyQRR}
\begin{fmfgraph*}(40,30)
\fmfpen{thick} \fmfleft{i1} \fmfright{o1,o2}
\fmftop{v1} \fmfbottom{v2}
\fmf{plain}{i1,v1,v2,i1}
\fmf{zigzag}{v1,o2}
\fmf{zigzag}{v2,o1}
\fmffreeze
\fmfv{decor.shape=circle, decor.filled=empty, decor.size=.20w}{i1}
\fmffreeze \fmfdraw \fmfv{d.sh=cross,d.size=.20w}{i1}
\fmffreeze
\fmflabel{$U(1)_i~$}{i1} \fmflabel{$g_{\m\n}$}{o1}
\fmflabel{$g_{\m\n}$}{o2}
\end{fmfgraph*}
\end{fmffile}
&~~~~~~~~~~~~
\unitlength=0.6mm
\begin{fmffile}{anomalyQGG}
\begin{fmfgraph*}(40,30)
\fmfpen{thick} \fmfleft{i1} \fmfright{o1,o2}
\fmftop{v1} \fmfbottom{v2}
\fmf{plain}{i1,v1,v2,i1}
\fmf{gluon}{v1,o2}
\fmf{gluon}{v2,o1}
\fmffreeze
\fmfv{decor.shape=circle, decor.filled=empty, decor.size=.20w}{i1}
\fmffreeze \fmfdraw \fmfv{d.sh=cross,d.size=.20w}{i1}
\fmffreeze
\fmflabel{$U(1)_i~$}{i1} \fmflabel{$G^\a$}{o1}
\fmflabel{$G^\a$}{o2}
\end{fmfgraph*}
\end{fmffile}
&~~~~~~~~~~~~
\unitlength=0.6mm
\begin{fmffile}{anomalyQQQ}
\begin{fmfgraph*}(40,30)
\fmfpen{thick} \fmfleft{i1} \fmfright{o1,o2}
\fmftop{v1} \fmfbottom{v2}
\fmf{plain}{i1,v1,v2,i1}
\fmf{photon}{v1,o2}
\fmf{photon}{v2,o1}
\fmffreeze
\fmfv{decor.shape=circle, decor.filled=empty, decor.size=.20w}{i1}
\fmffreeze \fmfdraw \fmfv{d.sh=cross,d.size=.20w}{i1}
\fmffreeze
\fmflabel{$U(1)_i~$}{i1} \fmflabel{$U(1)_k$}{o1}
\fmflabel{$U(1)_j$}{o2}
\end{fmfgraph*}
\end{fmffile}
\nonumber\eea

the effective action has
an
anomalous variation under a gauge transformation $A_{\m}\to
A_{\m}+\p_{\m}\e$
\be
\delta S_{\rm triangle}=C \int d^4 x~\e~\left(Tr[Q]~R\wedge
R+{1\over 3}Tr[Q^3]~F\wedge F+Tr[Q T^aT^a]Tr[G\wedge G]\right)
\ee
where $R$ is the gravitational two-form, $G$ the
field strength of a non-abelian gauge boson and $C$ is a universal coefficient.

This one-loop anomalous variation is cancelled by an axion field
$a$ that couples appropriately to the action:
\be
S_{eff}=\int d^4 x~\left[-{1\over 4g^2}F_{\m\n}F^{\m\n}-{M^2\over
2}(\p_{\m}a+A_{\m})^2+\right.
\label{anom}\ee$$\left.
+C~a\left(Tr[Q]~R\wedge
R+{1\over 3}Tr[Q^3]~F\wedge F+Tr[Q T^aT^a]Tr[G\wedge
G]\right)\right]
$$
Under the U(1) gauge transformation the axion transforms as $a\to
a-\e$, and since $\delta S_{eff}=-\delta S_{\rm triangle}$ the
full effective action, including the one-loop determinants is
invariant.
By choosing the physical gauge $a=0$ we can see that the U(1)
gauge boson has acquired an (unormalized) mass $M$. We will
indicate how to calculate this mass further in these lectures.

In heterotic N=1 vacua there is a unique anomalous U(1) and its
anomalies are cancelled by the universal two-index anti-symmetric
tensor.
In the type-I vacua, we may have several anomalous U(1)s and in
four-dimensional compactifications their anomalies are cancelled
by a combination of RR untwisted and twisted axions \cite{anom6,anom4}.

There are however, further effects associated with the anomalous
U(1)'s and they are best visible when N=1 supersymmetry is present.
In this case, the axion belongs to a chiral multiplet, and has a
CP-even partner $s$. When the axion comes from the twisted
RR-sector, its partner is a scalar from the twisted NS sector.
Its non-zero expectation value blows-up the orbifold singularity.

First, supersymmetry implies that there is a CP-even partner to
the anomaly-cancelling CP-odd coupling of the axion.
Thus, the D-brane gauge couplings have a non-trivial tree level
correction of the form
\be
{1\over g^2}={V_{||}\over g_s}+\sum_{i}\lambda_i~s_i
\ee where the sum is on the various twisted scalar partners of the
axions that couple to the given gauge field and $\lambda_i$
are computable coefficients \cite{ab}.

As it was shown in \cite{pop}, at the orbifold fixed point
$<s_i>=0$. Once the orbifold is blown up, the tree-level gauge
coupling changes.

Second, there is a D-term potential of the form
\be
V\sim {1\over g^2}~D^2={1\over g^2}(s+\sum_i~Q_i~|\phi_i|^2)^2
\ee
where
$Q_i$ are the U(1) charges of the scalars $\phi_i$.
The minima of this potential are qualitatively distinct depending
on the expectation value of $s$.

$\bullet$ $<s>\not=0$. Then, at least one charged scalar acquires a
non-zero expectation value in order to minimize the D-term and the
global U(1) symmetry is broken.
This is the case in the heterotic string where
$<s>=<e^{\phi}>=g_s$, and for non-trivial (interacting) vacua,
$<s>\not= 0$.

$\bullet$ $<s>=0$. Here, the D-term is generically minimized by vanishing
expectation values of the charged scalars and thus, although the
gauge U(1) symmetry is broken (and the gauge boson massive), the
associated global symmetry remains intact in perturbation theory.

This possibility is of utmost importance in D-brane model
building, when the string scale is low. Potential baryon and
lepton number violating operators are suppressed only by $M_s\sim
$TeV and would be catastrophic unless baryon number is a good
symmetry.
It can be arranged that baryon number is an anomalous U(1) gauge
symmetry, and its associated scalar stays at the orbifold
point so that the symmetry remains as a global symmetry, to
protect against fast proton decay \cite{proton1,proton2}.

Non-perturbative effects (instantons) break the global symmetry.
For anomalous U(1)s there are generically mixed anomalies
with non-abelian groups. This indicates that the associated
non-abelian instantons violate the global symmetry.

We should distinguish two cases here.

$\bullet$. The non-abelian group is unbroken and the interaction
is
strong. This implies a strong breaking of the global symmetry.
This is the case of the axial U(1) flavor symmetry in QCD.

$\bullet$ The non-abelian group symmetry is broken and the
interaction necessarily weak. In this case the breaking is small.
An example of this is the breaking of baryon number by weak SU(2)
instantons, a process estimated by 't Hooft to be highly
suppressed \cite{hoo}.

An important issue for phenomenology is the value for the
physical mass of the anomalous U(1) gauge bosons.
The physical mass depends on the following effective theory data.

(i) The UV mass $M$ in (\ref{anom}). This can be calculated only
by a string calculation and we will describe it in a following
section.

(ii) The mass has an additive contribution from the Higgs effect
if the SM Higgs is charged under the anomalous U(1) in question
(and it always is charged under at least one anomalous U(1)).

(iii) The mass depends multiplicatively on the gauge coupling of
the anomalous U(1).

An important question is, how much of the above discussion applies
in the absence of unbroken supersymmetry?

The qualitative structure of anomaly cancellation remains the
same.
In orientifolds at tree level the D-term potential is as before,
but it is corrected (typically mildly) at one-loop, and this is welcome
 in order to generate a negative mass-square for the Higgs.
The UV mass remains the
same (one-loop) since it is tied to anomalies. However, the
contributions from the Higgs effect and the gauge coupling change
in general at one loop.

\subsection{Anomalies in six dimensions}
\setcounter{equation}{0}

An analysis of four-dimensional mixed U(1) anomalies, which can be
done at the level of the massless (effective field theory)
spectrum, is not enough to determine which of them obtain masses, \cite{scrucca,akr}.

The reason is analogous to the fact that in the standard model
massive quarks and leptons contribute to anomalies. As shown by 't
Hooft, any fermion that can potentially become massless at some
ground state of the theory, contributes to anomalies. For example,
although SM fermions are massive in the broken vacuum, they are
massless in the symmetric vacuum, and thus contribute to
anomalies.

In our case, possible vacua of the theory correspond also to
expectation values of the moduli scalars becoming infinite and the
theory decompactifies to a higher-dimensional string theory.
Typically there is a part  of the original spectrum which decompactifies uniformly
(some states remain localized in four dimensions).
This higher-dimensional theory may have anomalies not present in
the four-dimensional theory. They are induced by the
KK-descendants of the fermions that although they are massive in
four dimensions, they become massless in higher dimensions.
It is thus possible, that a non-anomalous U(1) of the
four-dimensional theory, becomes an anomalous higher dimensional
U(1). Thus,  it is important to investigate the effect of the
presence of higher-dimensional anomalies.

The first non-trivial higher-dimensional case arises in six
dimensions.
Here the leading anomalous diagrams are quadrangles with one
insertion of the divergence of the U(1) current and the three
further
insertions of gauge fields (gravitons,abelian and non-abelian gauge
fields) in various combinations.

Moreover, the available form gauge fields to cancel the anomalies
are (up to duality) pseudo-scalars (axions) and two-index
antisymmetric tensors (of which typically there are plenty in
six-dimensional vacua) \cite{anom6,klein}.

It is enough to consider mixed anomalies of a single anomalous U(1) with a single non-abelian
group
for our purposes.

\bea
\unitlength=0.6mm
\begin{fmffile}{anomalyQGGG}
\begin{fmfgraph*}(40,30)
\fmfpen{thick} \fmfleft{i1} \fmfright{o1}
\fmftop{v1,v2,v7,v8,v9} \fmfbottom{v4,v5,v10,v11,v12}
\fmf{plain}{i1,v2,v3,v5,i1}
\fmf{gluon}{v3,o1}
\fmffreeze \fmfdraw
\fmf{gluon}{v2,v8}
\fmf{gluon}{v5,v11}
\fmfv{decor.shape=circle, decor.filled=empty, decor.size=.20w}{i1}
\fmffreeze \fmfdraw \fmfv{d.sh=cross,d.size=.20w}{i1}
\fmffreeze
\fmflabel{$U(1)_i~$}{i1} \fmflabel{$G^\alpha$}{v8}
\fmflabel{$G^\gamma$}{v11} \fmflabel{$G^\beta$}{o1}
\end{fmfgraph*}
\end{fmffile}
~~~~~~~~~~~~&~~~~~~~~~~~~
\unitlength=0.6mm
\begin{fmffile}{anomalyQQGG}
\begin{fmfgraph*}(40,30)
\fmfpen{thick} \fmfleft{i1} \fmfright{o1}
\fmftop{v1,v2,v7,v8,v9} \fmfbottom{v4,v5,v10,v11,v12}
\fmf{plain}{i1,v2,v3,v5,i1}
\fmf{gluon}{v3,o1}
\fmffreeze \fmfdraw
\fmf{photon}{v2,v8}
\fmf{gluon}{v5,v11}
\fmfv{decor.shape=circle, decor.filled=empty, decor.size=.20w}{i1}
\fmffreeze \fmfdraw \fmfv{d.sh=cross,d.size=.20w}{i1}
\fmffreeze
\fmflabel{$U(1)_i~$}{i1} \fmflabel{$U(1)_j$}{v8}
\fmflabel{$G^\alpha$}{v11} \fmflabel{$G^\alpha$}{o1}
\end{fmfgraph*}
\end{fmffile}
\nonumber\eea

 Gravitational and U(1) anomalies have a similar
structure \cite{anom6}.

In six dimensions there are two distinct U(1)/non-abelian mixed
anomalies.

$\bullet$ Those corresponding to $Tr[Q~T^a\{T^b,T^c\}]$, where Q
is the U(1) charge generator and
the diagrams involved have three insertions of  gluons.

Such an anomaly induces a non-invariance of the effective action,
under U(1) gauge transformations $A_{\m}\to
A_{\m}+\p_{\m}\e$,
\be
\delta S_{\rm quadrangle}=C \int d^6 x~\e~Tr[Q ~T^a\{T^b,T^c\}]~G^a\wedge G^b\wedge G^c]
\ee
where $G^a$ is the
field strength of the non-abelian gauge boson.

This one-loop anomalous variation is cancelled by a pseudoscalar axion
$a$ that couples as
\be
S_{eff}=\int d^6 x~\left[-{1\over 4g_6^2}F_{\m\n}F^{\m\n}-{M_6^2\over
2}(\p_{\m}a+A_{\m})^2
+C~a ~Tr[Q ~T^a\{T^b,T^c\}]~G^a\wedge G^b\wedge G^c\right]
\label{anom63}\ee
Under the U(1) gauge transformation the axion transforms as $a\to
a-\e$, and again since $\delta S_{eff}=-\delta S_{\rm quadrangle}$ the
full effective action, including the one-loop determinants is
invariant.

In the presence of this type of six-dimensional anomaly, the
anomalous U(1) gauge boson picks a six-dimensional unormalized
mass $M_6$. With a gauge field scaling as mass and the axion being dimensionless, $g_6$  scales as length and
$M_6$ as (mass)$^2$. Upon compactification to four-dimensions on a
two-dimensional compact manifold of volume $V_2$ the unormalized mass
and the gauge coupling become
\be
{1\over g^2_4}={V_2\over g_6^2}\sp M_4^2=V_2 ~M_6^2
\ee
and the normalized physical four-dimensional mass is
\be
M_{\rm phys}=g_4~M_4=g_6~M_6
\ee

$\bullet$ Anomalies corresponding to $Tr[Q^2~T^aT^a]$,
the diagrams involved having  two insertions of  gluons and one anomalous U(1) gauge boson.
The one-loop non-invariance of the effective action, here is

\be
\delta S_{\rm quadrangle}=C \int d^6 x~\e~Tr[Q^2 ~T^aT^a]~F\wedge G^a\wedge G^a
\ee
and the anomaly is cancelled by a two-index antisymmetric tensor
$B_{\m\n}$
\be
S_{eff}=\int d^6 x~\left[-{1\over 4g_6^2}F_{\m\n}F^{\m\n}-{1\over
3!}\hat H_{\m\n\r}\hat H^{\m\n\r}+Tr[Q^2 ~T^aT^a]~B\wedge G^a\wedge G^a\right]
\label{anom61}\ee
where
\be
\hat H_{\m\n\r}=\p_{\m}B_{\n\r}+2~A_{\m}\p_{\n}A_{\r}+{\rm
cyclic~~permutations}\sim dB+CS_3
\ee
where $CS_3$ is the Chern-Simons 3-form of the U(1) gauge field.
Under the U(1) gauge transformation the
antisymmetric tensor transforms as
\be
B_{\m\n}\to B_{\m\n}-\e~F_{\m\n}
\ee
so that $\hat H_{\m\n\r}$ is gauge invariant.
The gauge variation is non-zero for the last term
and since $\delta S_{eff}=-\delta S_{\rm quadrangle}$ the
full effective action, including the one-loop determinants is
invariant.

Note that in this case, there is no mass generated for the
anomalous U(1) gauge boson, and the associated U(1) gauge symmetry
is not broken by the anomaly.

The general case involves a combination of the two types of mixed
anomalies. A six-dimensional mass is generated only in the
presence of the first type of anomaly.

Although there are no known eight-dimensional ground states with
anomalous U(1)s we will briefly describe the effects of anomalies in
this case.

$\bullet$ A $Tr[Q^3~T^aT^a]$ anomaly is canceled by a four-form gauge
field $C_4$ with a field strength $dC_4+F\wedge CS_3(A)$ and
anomaly-canceling term $C_4\wedge G^a \wedge G^a$.
The anomalous gauge boson remains massless.

$\bullet$ A $Tr[Q^2~T^a\{T^b,T^c\}]]$ anomaly is canceled by a two-form gauge
field $C_2$ with a field strength $dC_2+CS_3(A)$ and
anomaly-canceling term $C_2\wedge G^a \wedge G^b\wedge G^c$.
The anomalous gauge boson remains massless.

$\bullet$ A $Tr[Q~T^4]$ anomaly is canceled by an axion $a$
with kinetic term $(\p a+A)^2$ and anomaly canceling term $a~G^a\wedge G^b \wedge G^c\wedge G^d$.
In this case the anomalous U(1) gauge boson acquires a mass.

\subsection{String calculation of anomalous U(1) masses}
\setcounter{equation}{0}

In view of the fact that anomalous U(1) symmetries are generic in
D-brane realizations of the SM, their detailed study is of
phenomenological importance for two reasons.

$\bullet$ Anomalous massive gauge bosons are a generic prediction
and their masses can be at most as large as the string scale (or
lower) which for low-scale vacua is in the TeV range. Moreover, it is possible
that their UV masses be much smaller than the string scale \cite{akr} unlike the heterotic case.
Thus, such
gauge bosons can be produced in colliders, and they also affect,
via their virtual effects, SM processes \cite{g-2}.
Moreover, they generically mix with weak gauge bosons affecting
their tree level couplings and may have measurable
consequences \cite{Z}.
If their masses turn out to be much lower than the
string scale, they may make such ground-states incompatible with
existing experimental data.

$\bullet$ Four-dimensional U(1)'s free of four-dimensional
anomalies, namely the hypercharge of the SM, may suffer in a given
ground state of higher-dimensional anomalies that induce a mass
term, rendering the model incompatible with experimental data.

It is thus obvious that a string calculation of their masses is
important. Although we have argued that the presence or absence of
mass can be detected by a careful study of anomalies in various
decompactification limits, the precise value of the mass
depends on the UV structure of the open (matter) sector and via
UV/IR (open/closed) correspondence to IR data of the gravitational
(closed) sector.

The mass can be indirectly calculated by a disk calculation of the
mixing between the relevant axion and the anomalous U(1) gauge
boson, or  by a one-loop calculation of the two-gauge boson
amplitude.
Although the first method seems simpler, normalizing the amplitude is
very difficult, except for the case of axions descending from the
ten-dimensional RR forms (untwisted axions).
Thus, generally the second calculation is preferable and we will
sketch it below. The anomalous U(1) masses were calculated in
\cite{akr} for supersymmetric vacua and in \cite{panasta2} for
non-supersymmetric vacua.

The two possible surfaces that can contribute to terms quadratic in the
gauge boson at the one-loop level are the annulus and the M\" obius strip.
Of those, only the annulus with the gauge field vertex operators inserted
at the two opposite ends has the appropriate structure to contribute to
the mass-term. Indeed, vertex operators inserted at the same boundary will
be proportional to $Tr[\gamma_{k}\lambda^a\lambda^b]$, where $\gamma_k$ is
the representation of the orbifold group element in the $k$-th orbifold
sector acting on the Chan--Paton (CP) matrices $\lambda ^a$.
On the other hand, for gauge fields inserted on opposite boundaries, the
amplitude will be proportional
to $Tr[\gamma_{k}\lambda^a]Tr[\gamma_{k}\lambda^b]$ and it is this form of
traces that determines the anomalous $U(1)$s \cite{anom4}.
As we show below, the mass comes from an UV contact term.
The potential  UV divergences that come from vertex operators
inserted on the same boundary (both in cylinder and M\"obius strip) cancel
because of tadpole cancellation \cite{abd}.

We must concentrate on the {\em CP}-even part of the amplitude
which receives contributions only from even spin structures. This implies
that we need the gauge boson vertex operators in the zero-ghost picture:
\be
V^a=\lambda^a\epsilon_{\mu}(\partial X^{\mu}+
i(p\cdot \psi)\psi^{\mu})e^{ip\cdot X}\, ,
\ee
where $\lambda$ is the Chan--Paton matrix and $\epsilon^{\mu}$ is the
polarization vector.
Due to the structure of the vertex operators, the annulus amplitude
for the two anomalous gauge bosons is ${\cal O}(p^2)$ before
integration over the annulus modulus $t$.
A contribution to the mass will be generated if a $1/p^2$ pole
appears after the $t$ integration. This can only come from the UV
region $t\to \infty$. The IR region, only produces the standard
$\log(p^2)$ running behavior of the effective gauge couplings.
Thus, the UV tadpole responsible for the mass can be calculated
directly in the closed string channel ($t\to 1/t$).
For N=0,1 sectors we obtain for the unormalized mass\cite{akr,panasta2}
\be
M^2_{ab}|_{N=1}={1\over \pi^3}\sum_{N=1~~\rm  sectors}Tr[\gamma_k\lambda^a]
Tr[\gamma_k\lambda^b]Str_{k}\left[{1\over 12}-s^2\right]_{\rm massless~closed~~
channel}
\ee
where $s$ is the four-dimensional helicity.
When the gauge fields come from different branes, then, the
associated $\gamma$ matrices appear in the formula above.

The contribution of N=2 sectors is volume-dependent.
If the boundary conditions along the untwisted torus are NN and its volume (in string units) $V_2$, then
\be
\left.
M^2_{ab}\right|_{N=2}=-{2V_2\over \pi^3}\sum_{N=2~\rm  sectors}Tr[\gamma_k\lambda^a]
Tr[\gamma_k\lambda^b]Str_{k}\left[
{1\over 12}-s^2\right]_{\rm massless~~open~~ channel}
\ee
where here the helicity supertrace is the same in the open and
closed channels.
This in particular implies that volume dependent contributions to
the anomalous U(1) masses can be calculated in the effective
(open) field theory.

In the DD case, relevant for mass matrix elements coming from
$D_{p<9}$ branes, the mass is similar as above with ${V_2\over
\alpha'}\to {\alpha'\over V_2}$ ($V_2\to 1/(4V_2)$ for the $\alpha'=1/2$ choice used here.

We can intuitively understand the volume-depended contributions of
the anomalous U(1) masses from the following geometrical picture.
The contribution to a given quadratic term $A^i_{\m}A^j_{\m}$ is
mediated by all axions that mix at the disk level with both
$A^i_{\m}$ and $A^j_{\m}$

Consider first the case of the two anomalous U(1) gauge bosons
coming from the same stack of branes that stretch in the four-dimensional Minkowski
space, $M_4$
and wrap an internal manifold $G_A$.
The axion that couples to the U(1)'s lies on an orbifold plane
stretching along $M_4$, and wrapping an internal manifold $G_a$.
Let $G_c$  be the common submanifold of $G_A$ and $G_a$ with volume  $V_c$,
$V_A$ the volume of $G^t_A$ (the part of $G_A$ transverse to $G_c$) and $V_a$ the volume
of $G^t_a$ (the part of $G_a$ transverse to $G_c$).
Upon T-duality in the directions spanned by $G^t_a$ we obtain a
D-brane that wraps the directions of $G_a$ but with volume
$V_c/V_a$. The axion and the gauge fields now overlap over all
directions along $G_a$.

Standard dimensional reduction now implies that
\be
M_{ij}^2\sim {V_c\over V_a}
\label{mass}\ee
and this captures the general behavior.

When the two gauge fields come from different stacks of branes, then
(\ref{mass}) is still valid if in the previous discussion we substitute the D-brane world-volume with the
common part of the two D-brane world-volumes.

The N=2 contributions, linear in the internal volumes are
effectively higher-dimensional effects and thus sensitive to
higher dimensional anomalies as advocated earlier.

\subsection{An example: anomalous U(1) masses in the $Z_6'$
orientifold}
\setcounter{equation}{0}

The orbifold rotation vector is $(v_1,v_2,v_3)=(1,-3,2)/6$. There
is an order two twist ($k=3$) and we must have one set of
D5-branes. Tadpole cancellation then implies the existence of 32
D9-branes and 32 D5-branes that we put together at one of the
fixed points of the $Z_2$ action (say the origin).
More details on this orientifold can be found in \cite{4ori}.
 The gauge group has a factor of $U(4)\times U(4)\times U(8)$
coming from the D9-branes and an isomorphic factor coming from the
D5-branes. The $N=1$ sectors correspond to $k=1,5$, while for
$k=2,3,4$ we have $N=2$ sectors.

The potentially anomalous $U(1)$s are the abelian factors of the gauge
group. The four-dimensional
anomalies of these $U(1)$s (and their cancellation mechanism) were
computed in \cite{anom4}. The mixed anomalies with the six
non-abelian groups are given by the matrix\footnote{Here
we use a different normalization for the $U(1)$ generators than in
\cite{anom4}.}
\be
\left(\begin{matrix} 2&2&4\sqrt{2}&-2&0&-2\sqrt{2}\\
-2&-2&-4\sqrt{2}&0&2&2\sqrt{2}\\ 0&0&0&2&-2&0\\
-2&0&-2\sqrt{2}&2&2&4\sqrt{2}\\ 0&2&2\sqrt{2}&-2&-2&-4\sqrt{2}\\
2&-2&0&0&0&0\end{matrix}\right)\, ,
\ee
where the columns label the $U(1)$s while the
rows label the non-abelian factors $SU(4)_9^2\times SU(8)_9\times
SU(4)^2_5\times SU(8)_5$. The upper 3$\times$3 part corresponds
to the 99 sector and the lower one to the 55 sector. As can be seen
by this matrix, the two linear combinations $\sqrt{2}(A_1+A_2)-A_3$
and $\sqrt{2}(\tilde A_1+\tilde A_2)-\tilde A_3$ are free of mixed
four-dimensional non-abelian anomalies. It can also be shown that they are also
free of mixed $U(1)$ anomalies.

The un-normalized mass matrix has eigenvalues and eigenvectors \cite{akr}:
\be
m_1^2=6V_2\sp -A_1+A_2\, ;
\label{bubu}
\ee
\be
m_2^2={3\over 2V_2}\sp -\tilde A_1+\tilde A_2\, ;
\label{brbu}
\ee
\be
m_{3,4}^2={5\sqrt{3}+48V_3\pm\sqrt{3(25-128
\sqrt{3}V_3+768V_3^2)}\over 12}\, ,
\ee
with respective eigenvectors
\be
\pm a_{\pm}(A_1+A_2-\tilde A_1-\tilde A_2)-A_3+\tilde A_3
\ee
where
\be a_{\pm}={\mp 3+\sqrt{25-128\sqrt{3}V_3+768V_3^2}\over 4\sqrt{2}
(4\sqrt{3}V_3-1)}\, ;
\ee
\be
m_{5,6}^2={15\sqrt{3}+80V_3\pm\sqrt{5(135-384\sqrt{3}V_3+1280V_3^2)}
\over 12}\, ,
\ee
with respective eigenvectors
\be
\pm
b_{\pm}(A_1+A_2+\tilde A_1+\tilde A_2)+A_3+\tilde A_3
\ee
where
\be
b_{\pm}={\pm 9\sqrt{3}-\sqrt{5(135-384\sqrt{3}V_3+1280V_3^2)}
\over 4\sqrt{2}(20V_3-3\sqrt{3})}\, .
\ee
$V_{1,2,3}$ are the volumes of the three two-dimensional internal
tori.

Note that the eigenvalues are always positive. They are also
invariant under the T-duality symmetry of the theory $V_2\to 1/4V_2$.
Thus, all $U(1)$s become massive, including the two anomaly-free
combinations. The reason is that these combinations are anomalous in
six dimensions \cite{panasta}. Observe however, that in the limit $V_3\to 0$, the two
linear combinations that are free of four-dimensional anomalies
become massless. This is consistent with the fact that the
six-dimensional anomalies responsible for their mass cancel locally
in this limit \cite{scrucca,akr}.

More examples can be found in \cite{akr,panasta2}.
The relationship between four-dimensional masses and six dimensional
anomalies has been analyzed in detail in \cite{panasta}

\subsection{Physical axions}
\setcounter{equation}{0}

When the Higgs effect is also at work in the presence of
anomalous U(1)'s then there may remain physical massless axions in the
spectrum. They are linear combinations of the closed string axions
responsible for anomaly cancellation and the Higgs phases. Such
axions will obtain small masses from instanton effects. However,
their couplings to matter are strongly constrained by experimental
data, and it is thus important to understand to what extent they
can be problematic.
In D-brane realizations of the standard model we must have at least two  Higgs
fields in order to be able to give masses to all quarks and leptons \cite{akt}.
Moreover, each Higgs is  charged under hypercharge and an  anomalous
U(1).

We will analyze here the case of a single anomalous U(1) coupled
to a Higgs field in order to discuss the relevant effects.

Consider an anomalous U(1) gauge boson $A_{\m}$ coupled to a complex Higgs $H$,
an anomaly-canceling
axion $a$, other (non)-abelian gauge fields $G$ and matter fermions $\psi$.
\be
S=-{1\over 4g^2}F_{\m\n}^2-{1\over 2}(k\p_{\m} a+M~A_{\m})^2+{\cal
A}_i~aTr[G_i\wedge G_i]-
\ee
$$-
{1\over 2}\left|\p_{\m}
H+ie~A_{\m}~H\right|^2+V(|H|)+\gamma~H~\psi\bar \psi~
$$
where ${\cal A}_i$ is the mixed anomaly ${\cal A}_i\sim Tr[Q~T^a_iT^a_i]$.
Parameterize $H=r~e^{i\theta}$.

The relevant U(1) gauge transformations are
\be
A_{\m}\to A_{\m}+\p_{\m} \e\sp a\to a-{M\over k}\e\sp \t\to \t-e~\e
\ee

The minimum of the Higgs potential imposes $r=v$ and expanding around it (ignoring the radial fluctuations)
 the relevant action becomes
\be
S=-{1\over 4g^2}F_{\m\n}^2-{1\over 2}(k\p_{\m} a+M~A_{\m})^2+{\cal
A}_i~aTr[G_i\wedge G_i]-{v^2\over 2}\left|\p_{\m}\t+e~A_{\m}\right|^2+\gamma v~e^{i\t}~\psi\bar \psi~
\ee

We now redefine the scalars
\be
\phi=kM~a+ev^2~\t\sp \c=-ek^2~a+kM~\t
\ee
\be
a={kM~\phi-ev^2~\c\over k^2(M^2+e^2v^2)}\sp \t={ek~\phi+M~\c\over k(M^2+e^2v^2)}
\ee
The scalar $\c$ is gauge invariant while $\phi$ transforms under
gauge transformations.

The action now becomes
\be
S=-{1\over 4g^2}F_{\m\n}^2-{1\over 2}{(\p_{\m} \phi+(M^2+e^2 v^2)~A_{\m})^2\over M^2+e^2v^2}-{v^2\over 2k^2(M^2+e^2v^2)}
(\p_{\m}\c)^2+\ee
$$
+{{\cal A}_i(kM\phi-ev^2\c)\over k^2(M^2+e^2v^2)}~Tr[G_i\wedge G_i]+\gamma v~e^{{i(ek\phi+M\c)\over k(M^2+e^2v^2)}}~\psi\bar \psi~
$$

We will gauge fix $\phi=0$ (physical gauge) and rescale $\c$
\be
\ch={v\over k\sqrt{M^2+e^2v^2}}~\c
\ee
to finally obtain

\be
S=-{1\over 4g^2}F_{\m\n}^2-{M^2+e^2
v^2\over 2}~A_{\m}^2-{1\over 2}(\p_{\m}\ch)^2-\ee
$$
-{{\cal A}_i ev\over k\sqrt{M^2+e^2v^2}}~\ch~Tr[G_i\wedge G_i]+
\gamma v~e^{{iM~\ch\over v\sqrt{M^2+e^2v^2}}}~\psi\bar \psi~
$$
The Yukawa coupling between the physical axion and the fermions is
\be
\gamma_{eff}=\gamma{M\over \sqrt{M^2+e^2v^2}}
\ee
In order to suppress this interaction we must have $M<<ev$.
Then

\be
S=-{1\over 4g^2}F^2-{e^2
v^2\over 2}~A^2-{1\over 2}(\p\ch)^2
-{{\cal A}_i\over k}~\ch~Tr[G_i\wedge G_i]+\gamma v~e^{{iM~\ch\over ev^2}}~\psi\bar \psi~
\ee

Consider the anomalous gauge boson to come from a D-brane, and the
axion from an orbifold plane. Call the internal volume of the
common intersection $V_c$, while the totally transverse volumes
$V_A$ and $V_a$.

Then
\be
{1\over g^2}={V_cV_A\over g_s}\sp M^2={V_c\over V_a}M_s^2\sp
k^2={V_cV_a\over g_s^2}\sp {\cal A}_i\sim {a_i\over M_s}
\ee
and
\be
M_A=\sqrt{g_s}{e~v\over \sqrt{V_cV_A}}\sp \gamma_{eff}=\gamma{M_s\over
ev}\sqrt{V_c\over V_a}\sp \mu_i={g_s\over \sqrt{V_a V_c}}{{a}_i\over M_s}
\ee
where $\mu_i$ is the axionic trilinear coupling to other gauge
bosons. Doing a chiral transformation we can transfer the axionic
couplings to the Yukawas:
\be
\hat\gamma_{eff}={m_{\psi}~M_s\over
ev^2}\sqrt{V_c\over V_a}+{m_{\psi}\over M_s}{g_s\over \sqrt{V_a~V_c}}
\ee
where we have used $m_{\psi}=\gamma ~v$.
Thus, for suppression, we must have large $V_a$ and we can set $V_c\sim {\cal O}(1)$ since, as we will argue later,
we want a minimum of
large dimensions.

If $V_A\sim 1$, then the axion coupling is ${\cal O}(1)$ and the anomalous gauge-boson mass is
of the
order of the $Z^0$ mass. This is experimentally excluded.
Thus $V_A$ should be large. Then, we end up with a gauge boson stretching in 4 large dimensions
and we expect problems from supernovae energy-loss.
To estimate it, let us recall the amplitude for KK-graviton energy
loss from supernovae, \cite{sav2}
\be
P_{g}\sim {1\over M_P^2}~(RT)^n\sim g_s^2~{T^n\over M_s^{n+2}}
\ee
where we assumed n large dimensions of common size R, T is the temperature at the supernova core and the
four-dimensional Planck scale is $M_P^2=R^nM_{s}^{n+2}/g_s^2$.
The factor $RT$ counts the number of KK states that can be
radiated away (per large dimension).

The similar rate for the emission of KK-states of a (massive)
vector is
\be
P_A\sim g^2~(RT-M_A)^{n_{A}}~{1\over T^2}
\ee
where $g$ is the four-dimensional gauge coupling $g^2=g_s/(M_sR)^{n_A}$ and $n_A$
is the number of large dimensions felt by the
gauge field.
When $M_A\geq RT$ then the emission rate is suppressed
kinematically.
In the opposite case $M_A<<RT$ we have
\be
P_A\sim g_s~{T^{n_A-2}\over M_s^{n_A}}
\ee
and
\be
{P_A\over P_g}\sim {1\over g_s}\left({M_s\over T}\right)^{n+2-n_A}
\ee
Since $n-n_A\geq 0$, for low scale string models ${P_A\over
P_g}\geq 10^8-10^{10}$ and gauge boson emission provides more stringent
constraints than KK-graviton emission.

These constraints can be avoided if $M_A\sim 30$ MeV. This requires $V_A\sim
10^7$ with ${g^2\over 4\pi}\sim 10^{-8}$.
This may be allowed, however in this case there are again 4 large
dimensions and there are only two string-size dimensions in which standard model branes can stretch.

A way out of this impasse is to have additional U(1) symmetry
breaking effects  in the potential, by moving, for example, away from the
orbifold point.

\renewcommand{\theequation}{\arabic{section}.\arabic{equation}}

\section{D-brane Standard Model building}

\setcounter{equation}{0}

Our purpose here is to investigate closely the configuration of
D-branes that can lead to the SM spectrum and gauge groups.
In general, an orientifold ground state consists of a
six-dimensional compactification manifold (with potential orbifold
singularities). Embedded are $D_{p\geq 3}$ branes who stretch along Minkowski space and wrap
the extra p-3 dimensions on appropriate cycles of the compact
manifold. Also included are Orientifold planes that cancel the
tadpoles of the theory.

Since masses of open strings are proportional to their lengths, it
is obvious that the branes that give rise to the SM fields must be
very close together in the internal space. Thus, we can talk about
the local group of SM D-branes and we may focus our discussion on
this. The presence of other branes further away may affect global
rather than local properties of the model (but can be important
for the overall stability of the configuration).

For simplicity we will assume that the compact space is an
orbifolded torus, but our discussion applies also to curved
compactifications.

We will also focus on ground states with a string scale in the TeV
range, because such ground states may have easily reachable
experimental consequences for the future collider experiments.

The standard relation between the string scale and the Planck
scale, namely $M_P^2={V_6\over g_s^2}M_s^2$  implies that the
internal volume must be very large in string units. The
 hierarchy problem in this context is the question of what
 stabilises the value of $V_6>>1$. No compelling answer exists to
 this question so we will bypass it and move on.

 Since $M_s\sim $TeV, supersymmetry can break at the string scale.

An important constraint comes from the fact that a SM D-brane
wrapping a large dimension automatically implies a multitude of KK
states for the appropriate gauge bosons with small masses. These
are experimentally unacceptable. Thus, D-branes generating the SM
gauge group should wrap string-sized dimensions.

The interesting question is how many large dimensions the internal
manifold can have. We have two competing effects. If there are
several large dimensions, the space where we can stretch the SM
branes is very much reduced, and it is difficult to generate the
small differences of the SM gauge couplings (that in the simplest case are proportional
to the internal volumes).
Thus, it would seem that the optimum would correspond to one large
dimension. However as shown in \cite{anba}, in this case UV/IR
duality implies power corrections to the effective field theory
that destroy decoupling. The next best is two large dimensions,
since in this case quantum corrections are soft (logarithmic) and
this is what we will assume in the sequel.

There is another important ingredient in the SM and this is
neutrinos with naturally light (e.g. $10^{-3}-1$ eV) masses. The
only known mechanism in open string theory that can achieve this,
stipulates that right-handed neutrinos come from a brane wrapping
one or more large dimensions \cite{neutrino1}-\cite{neutrino7}.

The relevant effective action is
\be
S_{\rm neutrino}\sim \int d^4 x~~( V~\bar
\nu_{R}~\sla{\partial}~\nu_R+m~\bar \nu_R~\nu_L)
\ee
where $V$ is large and $m\sim{\cal O}(1)$ since the brane carrying $\nu_L$
 is intersecting the one of $\nu_R$ only along the four-dimensional Minkowski space.
 Normalizing the kinetic term to one, we obtain a mass
 $\sim ~m/\sqrt{V}$ which is naturally small. For this to be
 protected, large ($\sim ~M_s$) Majorana masses for the neutrinos
 must be forbidden, and this is taken care of by good lepton
 number conservation.

Thus, in order to have naturally light neutrinos, at least one of
the SM branes should wrap the large dimensions.
There are several further questions that we will subsequently address.
Concerning the structure of generations, as advocated earlier they
may be obtained either from multiplicities in the case of branes
at singularities or multiple intersections of branes.

\renewcommand{\theequation}{\arabic{section}.\arabic{subsection}.\arabic{equation}}
\subsection{Symmetries and Charges }
\setcounter{equation}{0}

Extensions of the SM with a low string scale may have a priori
severe problems with proton decay unless baryon number is a good
global symmetry. It is similar with lepton number since it is an
approximate symmetry of the low energy physics and is needed to
protect small neutrino masses.

As we have argued earlier, in orientifolds, anomalous U(1)
symmetries may remain as unbroken global symmetries in
perturbation theory, broken only by non-perturbative effects.
It is thus a requirement that the brane configuration guarantees
that baryon and lepton number are such anomalous U(1) gauge
symmetries.

We may now move on to discuss the embedding of the SM charges in the
D-brane configuration.

The minimal $D$-brane configuration
that can successfully accommodate the SM gauge group
consists of three sets of branes with gauge
symmetry $\u3_c\times\u2_L\times\u1_1$. The first set contains three
coincident branes (``color" branes). An open string with one end
attached to this set, transforms as an $\su3_c$ triplet (or anti-triplet),
but also carries an additional $\u1_c$ quantum number which
can be identified with the (gauged) baryon number.
Similarly, $\u2_L$ is realized by a set of two coincident branes
(``weak" branes) and open strings attached to them from the one end are $SU(2)_L$
doublets characterized by an additional
$\u1_L$ quantum number, the (gauged) weak ``doublet" number. Moreover,
consistency of the SM embedding requires
the presence of an additional $\u1_1$ factor, generated by a single
brane. This is needed for several reasons, the most important being mass generation
 for all quarks and leptons of the
heaviest generation.

In all the above brane configurations there exist states (e.g. the $SU(2)_L$
singlet anti-quarks) which correspond to open strings with only one of their ends
attached to one of the three sets of D-branes. The other end must go to another
brane.
This requires the existence of at least one more U(1) brane.
Moreover, this brane can wrap a large dimension and can provide
right-handed neutrinos with small masses.

Thus, we consider an additional D-brane (in the bulk)
giving rise to an extra abelian gauge factor $\u1_b$.

\begin{table}[!ht]
\center
\begin{tabular}{|c|c|c|c|c|c|c|}
\hline
particle&${U(1)}_c$&${U(1)}_L$&${U(1)}_1$&${U(1)}_b$\\
\hline
$Q(\b3,\b2,\hphantom{+}\frac{1}{6})$&$+1$&$w$&$0$&$0$\\
$u^c(\bb3,\b1,-\frac{2}{3})$&$-1$&$0$&$a_1$&$a_2$\\
$d^c(\bb3,\b1,+\frac{1}{3})$&$-1$&$0$&$b_1$&$b_2$\\
$L\;(\b1,\b2,-\frac{1}{2})$&$0$&$+1$&$c_1$&$c_2$\\
$e^c(\b1,\b1,+1)$&$0$&$d_L$&$d_1$&$d_2$\\
\hline
\end{tabular}
\caption{\label{tta} SM particles with their generic charges under
the abelian part of
the gauge group $U(3)_c\times{U(2)}_L\times{U(1)}_1\times{U(1)}_b$.}
\end{table}

The total gauge group is
\ba
G &=& U(3)_c\times{U(2)}_L\times{U(1)}_1\times{U(1)}_b\nonumber\\ &=&
SU(3)_c\times{U(1)_c}\times{SU(2)}_L\times{{U(1)}_L}\times{U(1)}_1\times{U(1)}_b
\ea
and contains four abelian factors. The assignment of the SM particles is
partially fixed from its non-abelian structure. The quark doublet $Q$ corresponds
to an open string with one end on the color and the other on the weak set of
branes. The anti-quarks $u^c, d^c$ must have one of their ends attached to the
color branes. The lepton doublet and possible Higgs doublets must have one end on
the weak branes. However, there is a freedom related to the abelian structure,
since the hypercharge can arise as a linear combination of all four abelian
factors. In a generic model, the abelian charges can be expressed without loss of
generality in terms of ten parameters displayed in Table~\ref{tta}.

We should stress here that we assume the minimal SM spectrum without
superpartners and other exotics.

In a convenient parameterization, normalizing the $U(N)\sim SU(N)\times U(1)$
generators as ${\rm Tr} T^a T^b=\delta^{ab}/2$, and
measuring the corresponding $U(1)$ charges with respect to the coupling
$g/\sqrt{2N}$, the ten parameters are integers:
$a_{1,2},b_{1,2},c_{1,2},d_2=0,\pm1$, $d_{1}=0,\pm1,\pm2$, $d_L=0,\pm2,w=\pm1$
satisfying
\ba
\sum_{i=1,2}|a_i|=\sum_{i=1,2}|b_i|=\sum_{i=1,2}|c_i|=1,\
\sum_{i=1,2,L}|d_i|=2\, .  \label{range}\
\ea
The first three constraints in (\ref{range}) correspond to the
requirement that the $u^c$ and $d^c$ anti-quarks, as well as the lepton doublet,
must come from open strings with one end attached to one of the abelian
D-brane sets. The fourth constraint forces the positron $e^c$ open string
to be stretched either between the two abelian branes, or to have both ends
attached to the abelian $\u1_1$ brane, or to the weak set of branes. In the
latter case, it has $U(1)_L$ charge $\pm 2$ and is an $SU(2)_L$ singlet
arising from the antisymmetric product of two doublets.
The parameter $w$ in Table~\ref{tta}
refers to the $\u1_L$ charges of the quark-doublets, that we can choose to
be $\pm1$, since doublets are equivalent with anti-doublets.
Note that a priori one might also consider the case in which one of the $u^c$
and $d^c$ anti-quarks arises as a string with both ends on the
color branes $({\bf 3}\times{\bf 3}={\bf \bar 3}+{\bf 6})$, so that its
$\u1_c$ charges would be $\pm2$. This, however, would invalidate the identification of
$\u1_c$ with the baryon number and forbid the presence of quark mass terms, since
one of the combinations $Q u^c$ and $Q d^c$ would not be neutral under
$\u1_c$.

The hypercharge can in general be a linear combination
of all four abelian group factors. However, we must restrict ourselves to
models in which the bulk ${U(1)}_b$ does not contribute to the
hypercharge.
Since the ${U(1)}_b$ gauge coupling is tiny (because the brane
wraps a large volume) if it participates in the hypercharge, it
will force the hypercharge gauge coupling to be tiny, contrary to
the experimental results.

Hence,
\ba
Y= k_3\,Q_c+k_2\,Q_L+k_1\,Q_1\, .\label{ourh}
\ea
The correct assignments for SM particles are reproduced, provided
\ba
k_3+k_2\,w=\frac{1}{6}\sp
-k_3+a_1 k_1=-\frac{2}{3}\sp
-k_3+b_1\,k_1=\frac{1}{3}\label{hyp}\\
k_2+c_1\,k_1=-\frac{1}{2}\sp
k_2\,d_L+d_1\,k_1=1.\nonumber
\ea
Notice that the  above equations imply that $k_1\ne0$.

The next step, after assigning the correct hypercharge to the SM particles,
is to check for the existence of  candidate fermion mass terms.
Here, we discuss only the question of masses for one generation (the heaviest) and
we do not address the general problem of flavor. To lowest order, the mass terms
are of the form $Q d^c H^\dagger_d$, $Q u^c H_u$ and
$L e^c H_e^\dagger$ where $H_d, H_u, H_e$ are scalar Higgs doublets
with appropriate charges.
For the generic charge assignments of Table~\ref{tta}, the required Higgs
charges are
\ba
H_u&=&\left(\b1,\b2,0,-w,-a_1,-a_2\right)\nonumber\\
H_d&=&\left(\b1,\b2,0,+w,+b_1,+b_2\right)\label{allh}\\
H_e&=&\left(\b1,\b2,0,1+d_L,c_1+d_1,c_2+d_2\right)\, .
\nonumber
\ea

Provided the constraints (\ref{range}) are satisfied, both $H_u$
and $H_d$ have the right charges  and correspond to strings
stretched between the weak and one of the abelian branes. Thus, (\ref{range})
guarantees the existence of tree-level quark masses. On the other hand, the
existence of $H_e$ depends on the particular choice of parameters, e.g. for
$c_1+d_1=2$, $H_e$ does not exist
and a tree-level lepton mass term ($L e^c H^\dagger$)
is forbidden. The generic constraint that guarantees tree-level
lepton masses is
\ba
\sum_{i=1,2}\left|c_i+d_i\right|=
\left|1+d_L\right|=1\ .\label{masset}
\ea

The hypercharge constraints  (\ref{hyp}) can be easily solved.
They require $a_1\ne b_1$ and
\ba
k_3&=&\frac{{a_1} + 2\,{b_1}}
  {3\,\left( {b_1} -{a_1} \right) }\sp
k_2=-\frac{\left( {a_1} + {b_1} \right)
      }{2\,\left(  {b_1}-{a_1}
      \right)\, w }\sp
k_1=\frac{1}{b_1-a_1}\\
c_1&=&-\frac{{b_1} - {a_1}}{2} +
  \frac{\left( {a_1} + {b_1}
       \right) }{2w}\sp
d_1= {b_1}-{a_1}  +
  \frac{\left( {a_1} + {b_1}
       \right) \,{d_L}}{2w}\, .
\ea
 Using charge conjugation, it is sufficient to search for solutions with
$(a_1,b_1) \in \left\{(-1,0),(-1,1),(0,1)\right\}$.
Solving for these choices, we get three allowed hypercharge embeddings:
\ba
(i) \ \ \ a_1=-1, b_1=1 &:& Y=\frac{1}{6} Q_c +\frac{1}{2} Q_1 \label{emba}\\
(ii) \ \ a_1=-1, b_1=0 &:& Y=-\frac{1}{3} Q_c +\frac{w}{2} Q_L+  Q_1 \label{embb}\\
(iii) \ \ \ a_1=0, b_1=1 &:& Y=\frac{2}{3} Q_c -\frac{w}{2} Q_L+  Q_1\, .
\label{embc}
\ea
Case (i)  leads to
$c_1=-1, c_2=0, d_1=2, d_1=d_L=0$. This is a special solution where the $\u1_b$ brane
decouples from the model since no SM particles are attached to it.  It satisfies
(\ref{masset}) and thus leads to tree level lepton masses. The solution exists
for both $w=\pm1$, as the value of
$w$ does not play an important role when $k_2=0$.
In case (ii), we have $c_1=-(1+w)/2$,
$d_L=0, d_1=1$ or $c_1=(1+w)/2, d_L=2w, d_1=d_2=0$, while
case (iii) leads to
$c_1=(w-1)/2, d_L=0, d_1=1$ or $c_1=(1+w)/2, d_L=2w, d_1=d_2=0$.

Combining the above three cases with the constraints (\ref{range}) and
(\ref{masset}), we get nine distinct configurations with tree-level quark and lepton
masses. Out of these only four have a lepton number symmetry.
We display these models in  table \ref{ptab} \cite{akt,akt1}.

\begin{table}[!ht]
\center
\begin{tabular}{|r|c|c|c|c|c|c|c|c|c|c|c|c|}
\hline
&$a_1$&$a_2$&$b_1$&$b_2$&$c_1$&$c_2$&$d_1$&$d_2$&$d_L$&$w$&$Y$&$n_h$\\
\hline
$1$&$-1$&$0$&$0$&$-1$&$0$&$-1$&$1$&$1$&$0$&$-1$&
$-  \frac{1}{3}\,Q_c -  \frac{1}{2}\,Q_L+Q_1$
&$2$\\
$2$&$0$&$1$&$1$&$0$&$0$&$-1$&$1$&$1$&$0$&$1$&
$\hphantom{+}\frac{2}{3}\,Q_c-  \frac{1}{2}\,Q_L+Q_1$
&$2$\\
$3$&$-1$&$0$&$0$&$-1$&$0$&$-1$&$0$&$0$&$-2$&$-1$&
$-  \frac{1}{3}\,Q_c-  \frac{1}{2}\,Q_L+Q_1$&$2$\\
$4$&$0$&$1$&$1$&$0$&$0$&$-1$&$0$&$0$&$-2$&$\hphantom{+}1$&
$\hphantom{+}\frac{2}{3}\,Q_c-  \frac{1}{2}\,Q_L + Q_1$&$2$\\
\hline
\end{tabular}
\caption{\label{ptab} The four brane configurations consistent with baryon and lepton number conservation.}
\end{table}

We should note that in order for lepton number to appear as a
gauged symmetry, the presence of $U(1)_b$ is crucial.

\subsection{Gauge couplings}
\setcounter{equation}{0}

We will now match the gauge couplings of the configurations
described in the previous section with the measured ones
in order to determine the string scale.

We will have to be a bit more explicit about the brane
configuration. The branes we can use are either $D_3/D_7$, or $D_5/D_9$ branes.
The two sets can be interchanged by duality. We will focus on the
second set. Then the Standard model particles must be associated
with $D_5$ branes except for the $U(1)_b$ brane that can be either $D_5$ or $D_9$.
We will take it to be a $D_5$ without loss of generality.
We split, for simplicity, the six internal dimensions $4,5,\cdots,9$ into three
two-tori. The third two-torus along $8,9$ is the one that has large
volume.

Without loss of generality we take the world-volume of the U(3) branes to stretch along
$0,1,2,3,4,5$, and that of the U(2) branes along $0,1,2,3,6,7$. We
need the internal volumes wrapped to be distinct, in order to
create the difference between the weak and the strong coupling
constant.
The $U(1)_b$ stretches then along $0,1,2,3,8,9$ and wraps the two
large dimensions.
the final U(1) brane must wrap the string size dimensions. We have
two distinct options:

$\bullet$ It is parallel to the U(3) branes. Then (\ref{ym})
indicates that it must have the same gauge coupling at the string scale: $g_1=g_3$.

$\bullet$ It is parallel to the U(2) branes. Then  $g_1=g_2$.

Thus the strategy is to relate the two independent parameters
$g_2,g_3$ and the string scale $M_s$ to the measured values of the
three SM couplings.
This will determine $g_{2,3}$ and the string scale.

In our normalizations, the hypercharge coupling
$g_Y$ at the string scale
is expressed as
\ba
\frac{1}{g_Y^2}=\frac{6\,k_3^2}{g_3^2}+\frac{4\,k_2^2}{g_2^2}+\frac{2\,k_1^2}{g_1^2}\, .
\label{yy}
\ea
The one loop coupling evolution is given by ($\alpha_i={g_i^2}/{4\pi}$),
\ba
\frac{1}{\alpha_i(M_s)}=\frac{1}{\alpha_i(M_Z)}+\frac{b_i}{4 \pi}
\ln\frac{\Delta^I M_s}{M_Z}\, ,
\ea
where $b_3=-7, b_2=-10/3+n_h/6,
b_Y=20/3+n_h/6$ and  $n_h$ is the number of scalar Higgs doublets.

 The results are
presented in Table~\ref{rt1}. In the above calculations we have assumed that the number of
doublets $n_h$ is the minimum $n_h=2$ required by the model.

\begin{table}
\centering
\begin{tabular}{|r|c|c|c|c|c|c|c|}
\hline
&$|k_3|$&$|k_2|$&$|k_1|$&$M_U (TeV)$&$g_2(M_U)/g_3(M_U)$&$g_2(M_U)g_3(M_U)$\\
\hline
&$\frac{1}{6}$&$0$&$\frac{1}{2}$&$4.6\times 10^{20}$&$1.1$&$0.21$\\
\cline{2-7}
$g_1=g_3$&$\frac{1}{3}$&$\frac{1}{2}$&$1$&$2.4\times 10^3$&$0.76$&$0.48$\\
\cline{2-7}
&$\frac{2}{3}$&$\frac{1}{2}$&$1$&$7.2$&$0.65$&0.61\\
\hline
&$\frac{1}{6}$&$0$&$\frac{1}{2}$&$1.5\times 10^{22}$&$1.1$&$0.26$\\
\cline{2-7}
$g_1=g_2$&$\frac{1}{3}$&$\frac{1}{2}$&$1$&$0.32$&$0.57$&$0.73$\\
\cline{2-7}
&$\frac{2}{3}$&$\frac{1}{2}$&$1$&$-$&$-$&$-$\\
\hline
\end{tabular}
\caption{\it \label{rt1}
The string  scale $M_s=2.5~M_U$ and the two
independent gauge couplings for the two possible brane configurations and the
various hypercharge embeddings.}
\end{table}

It is obvious that the models that have consistent (small) values
of the string scale are the third and the fifth in Table \ref{rt1}.
Such values for the string scale are indicative since they may be
modified by non-trivial threshold corrections.

On the other hand the coupling of $U(1)_b$ can be calculated from
(\ref{hetp},\ref{ym}) to be
\be
g_b^2={2g_s\over \alpha_2~\alpha_3}{M_s^2\over M_P^2}
\ee
giving $g_b\sim 10^{-16}-10^{-14}$ when $M_s\sim 1-10$ TeV.

There is finally  another possibility that is relevant
{\em only} for D-brane models with a string scale close to the
four-dimensional Planck scale \cite{Lust}.
For such models the U(1)$_b$ brane does not wrap a large
dimension.
It can consequently participate in the hypercharge. Taking this
into account there is one more acceptable hypercharge embedding
(up to signs) given by
\be
Q_Y={1\over 6}Q_3-{1\over 2}(Q_1+Q_b)\label{newhyper}
\ee

The spectrum of a model in this class is
\ba
&~&Q\left(\b3,\b2,+1,-1,0,0\right)\sp
u^c(\bb3,\b1,-1,0,1,0)\sp
d^c(\bb3,\b1,-1,0,-1,0)\nn\\
&~&~L(~\b1,\b2,0,+1,0,1)\sp
e^c(\b1,\b1,0,0,-1,-1)\sp \nu_R(\b1,\b1,0,0,1,-1)\nn
\ea
An economic solution  of solving the tadpole conditions  is putting the
$U(1)_1$ brane on top of the color branes indicating again a ``petite
unification": $g_3=g_1$. This also produces the appropriate intersection
numbers needed for family replication. Asking for the hypercharge
gauge boson to remain massless indicates a symmetric configuration
with $g_b=g_2$. We are thus left with two independent gauge
couplings at the string scale. Fitting them to gauge coupling one
obtains a ``consistent" string scale of $M_s\sim 2\times 10^{16}$
GeV \cite{Lust}.

\subsection{Viable D-brane configurations}
\setcounter{equation}{0}

So far, we have classified all possible
$\u3_c\times\u2_L\times\u1_1\times\u1_b$ brane models that can successfully
accommodate the SM spectrum. The quantum numbers of each model as well as the
hypercharge embedding are summarized in Table~\ref{ptab}. Furthermore,
compatibility with type I string theory with string scale in the TeV region,
requires the bulk to be two-dimensional of (sub)millimeter size, and leads to
two possible configurations: Place the $\u1_1$ brane on top of the weak
$\u2_L$ stack of branes or on top of the color $\u3_c$ branes. These impose
two different brane coupling relations at the string (unification) scale:
$g_1=g_2$ or $g_1=g_3$, respectively. For every model, using the hypercharge
embedding of Table~\ref{ptab}, the one loop gauge coupling evolution and one of
the above brane coupling conditions, we can determine the unification (string)
scale that reproduces the weak angle at low energies. The results are
summarized in Table~\ref{rt1}.

We will now describe in more detail the four viable brane
configurations that we label A,A',B,B'.

\medskip
\noindent{\it Models $A$ and  $A'$}

\begin{figure}
\center
\epsfxsize=6cm
\epsfbox{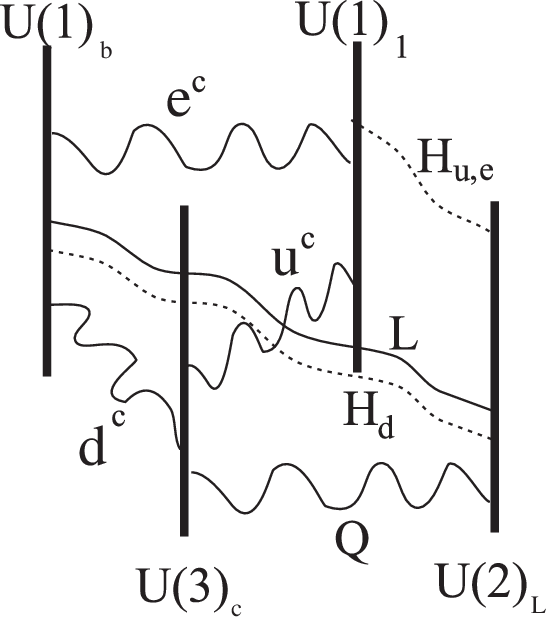}
\epsfxsize=6cm
\epsfbox{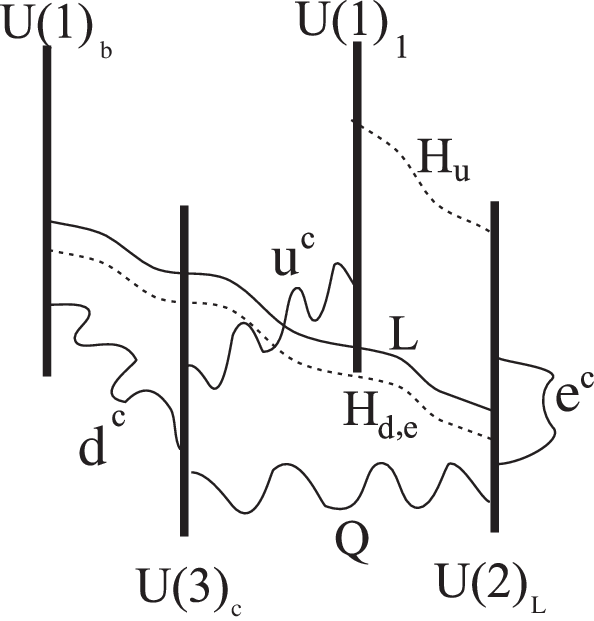}
\caption{\label{figa}{\it Pictorial representation of models} $A, A'$.}
\end{figure}

We consider here  the models 1 and 3 of Table~\ref{ptab},
hereafter referred as models $A$ and $A'$ respectively.
They are characterized by the common  hypercharge embedding
\ba
Y= -\frac{1}{3}\,Q_c-\frac{1}{2}\,Q_L+Q_1\label{hypa}
\ea
but they differ slightly in their spectra.
The spectrum of model $A$ is
\ba
&~&Q\left(\b3,\b2,+1,-1,0,0\right)\sp
u^c(\bb3,\b1,-1,0,-1,0)\sp
d^c(\bb3,\b1,-1,0,0,-1)\nn\\
&~&~L(~\b1,\b2,0,+1,0,-1)\sp
e^c(\b1,\b1,0,0,+1,+1)\sp \nu_R(\b1,\b1,0,0,0,\pm 2)\nn\\
&~&H_u(\b1,\b2,0,+1,+1,0)\sp
H_d(\b1,\b2,0,-1,0,-1)\nn
\ea
while in model $A'$ the right-handed electron $e^c$ is replaced by an open string
with both ends on the weak brane stack, and thus $e^c=(\b1,\b1,0,-2,0,0)$. The two
models are presented pictorially in Figure~\ref{figa}.

Apart from the hypercharge combination (\ref{hypa}) all remaining abelian
factors are anomalous. Indeed, for every abelian generator $Q_I, I=(c,L,1,b)$, we can
calculate the mixed gauge anomaly $K_{IJ}\equiv {\rm Tr} Q_I T^2_J$ with
$J=SU(3),SU(2),Y$, and gravitational anomaly $K_{I4}\equiv{\rm Tr} Q_I$
for both models $A$ and $A'$:
\ba
K^{(A)}=
\begin{pmatrix}
  0 & -1 & -\frac{1}{2} & -\frac{1}{2} \\
  \frac{3}{2} & -1 & 0 & -\frac{1}{2} \\
  -\frac{3}{2} & \frac{1}{3} & -\frac{1}{3} & \frac{1}{6} \\
  0 & -4 & -2 & -4
\end{pmatrix}\ ,\
K^{(A')}=
\begin{pmatrix}
  0 & -1 & -\frac{1}{2} & -\frac{1}{2} \\
  \frac{3}{2} & -1 & 0 & -\frac{1}{2} \\
  -\frac{3}{2} & -\frac{5}{3} & -\frac{4}{3} & -\frac{5}{6} \\
  0 & -6 & -3 & -5
\end{pmatrix}\label{anoma}
\ea It is easy to check that the matrices $K K^T$ for both models
have only one zero eigenvalue corresponding to the hypercharge
combination (\ref{hypa}) and three non vanishing ones
corresponding to the orthogonal $U(1)$ anomalous combinations. As discussed earlier,
 the three extra gauge bosons become massive, leaving behind the corresponding
global symmetries unbroken in perturbation theory. The three extra $U(1)$'s can be expressed
in terms of known SM symmetries: \ba
\mbox{Baryon number}\,\ \ \ B&=&\frac{1}{3} Q_c\nn\\
\mbox{Lepton number}\,\,\ \ \ L&=&\frac{1}{2}\left(Q_c+Q_L-Q_1-Q_b\right)\label{glob}\\
\mbox{Peccei--Quinn}\ \ \ Q_{PQ}&=&-\frac{1}{2}\left(Q_c-Q_L-3\,Q_1-3\,Q_b\right)\nn
\ea
Thus, our effective SM inherits baryon and lepton number
as well as Peccei--Quinn (PQ) global symmetries from the anomaly cancellation
mechanism.

Accordind to our previous discussion of anomalous U(1) masses:

\begin{enumerate}
\item
The two $U(1)$ combinations, orthogonal to the hypercharge and localized on the
strong and weak D-brane sets, acquire in general masses of the order of the
string scale from contributions of $N=1$ sectors, in agreement with effective
field theory expectations based on 4d anomalies.
\item
Such contributions are not sufficient though to make heavy the third $U(1)$
propagating in the bulk, since the resulting mass terms are localized and
suppressed by the volume of the bulk. In order to give string scale mass, one
needs instead $N=2$ contributions associated to 6d anomalies along the two large
bulk directions.

\item
Special care is needed to guarantee that the hypercharge remains massless despite
the fact that it is anomaly free.
\end{enumerate}

The presence of massive gauge bosons associated to anomalous abelian gauge symmetries
is generic. Their mass is given by $M_A^2\sim g_s M_s^2$,
up to a numerical model dependent factor and is somewhat smaller that the string scale.
When the latter is low, they can affect low energy measurable data, such as $g-2$
for leptons \cite{g-2} and the $\rho$-parameter \cite{Z}, leading to additional bounds
on the string scale.

An extension of the model is the introduction of a right-handed neutrino.
A natural candidate state would be an open string ending on the
${U(1)}_b$ brane. Its charge is then fixed to $+2$ by the requirement of
existence of the single possible neutrino mass term $L\,H_d\,\nu_R$. The
suppression of the brane-bulk couplings due to the wave function of $\nu_R$ would
thus provide a natural explanation for the smallness of neutrino masses.

Coming to the issue of gauge couplings, in this configuration  we must take the
$\u1_1$ brane on top of the weak branes, leading to $g_1=g_2$. The required
string scale is  low $M_s\sim{\cal O}(500)$ GeV (300-800 GeV, depending on the
threshold corrections), and could account for the stability of the hierarchy.

\medskip
\noindent{\it Models $B$ and $B'$}

Another  phenomenologically promising pair of configurations
consists of solutions 4 and 9  of Table~\ref{ptab}, named hereafter $B$ and
$B'$, which corresponds to the hypercharge embedding
\ba
Y= \frac{2}{3}\,Q_c-\frac{1}{2}\,Q_L+Q_1\, .
\label{hypb}
\ea
The spectrum is
\ba
&~&Q(\b3,\b2,+1,+1,0,0)\sp
u^c(\bb3,\b1,-1,0,0,1)\sp
d^c(\bb3,\b1,-1,0,1,0)\nn\\
&~&L(\b1,\b2,0,+1,0,-1)\sp
e^c(\b1,\b1,0,0,+1,+1)\sp \nu_R(\b1,\b1,0,0,0,\pm 2)\nn\\
&~&H_u(\b1,\b2,0,-1,0,-1)\sp
H_d(\b1,\b2,0,+1,+1,0)\nn
\ea
for model $B$, while in $B'$ $e^c$ is replaced by $e^c(\b1,\b1,0,-2,0,0)$.
\begin{figure}
\center
\epsfxsize=6cm
\epsfbox{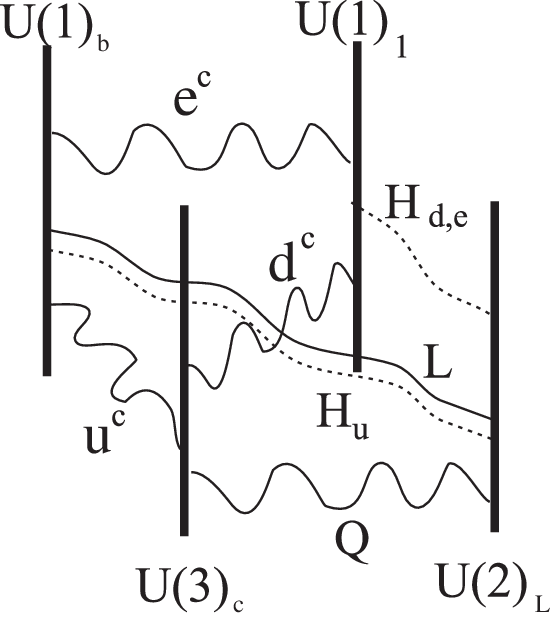}
\epsfxsize=6cm
\epsfbox{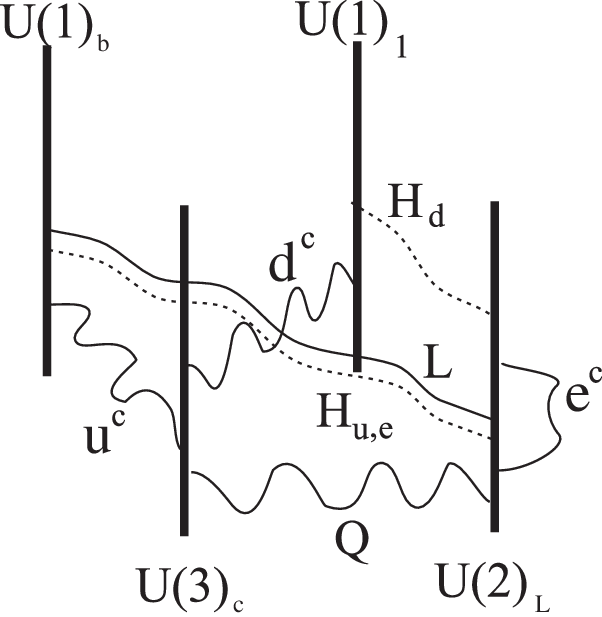}
\caption{\label{figb}\it Pictorial representation of models $B$ and $B'$.}
\end{figure}
The two models are represented pictorially  in Figure~\ref{figb}.
The four abelian gauge factors are anomalous. Proceeding as in the analysis
(\ref{anoma}) of models $A$ and $A'$, the mixed gauge and
gravitational anomalies are
\ba
K^{(B)}=
\begin{pmatrix}
  0 & 1 & \frac{1}{2} & \frac{1}{2} \\
  \frac{3}{2} & 2 & 0 & -\frac{1}{2} \\
  -\frac{3}{2} & \frac{2}{3} & \frac{4}{3} & \frac{11}{6} \\
  0 & 8 & 4 & 2
\end{pmatrix}\ ,\
K^{(B')}=
\begin{pmatrix}
  0 & 1 & \frac{1}{2} & \frac{1}{2} \\
  \frac{3}{2} & 2 & 0 & -\frac{1}{2} \\
  -\frac{3}{2} & -\frac{4}{3} & \frac{1}{3} & \frac{5}{6} \\
  0 & 6 & 3 & 1
\end{pmatrix}\label{anomb}
\ea
It is easy to see that the only anomaly free combination is the hypercharge (\ref{hypb})
which survives at low energies.
All other abelian gauge factors are anomalous and will be broken, leaving behind global symmetries. They can be
expressed in terms of the usual SM global symmetries as the following $\u1$
combinations:
\ba
\mbox{Baryon number}\,\ \ \ B&=&\frac{1}{3}\,Q_c\\
\mbox{Lepton number}\ \ \ \ L&=&-\frac{1}{2}\left(Q_c-Q_L+Q_1+Q_b\right)\\
\mbox{Peccei-Quinn}\ \ \
Q_{PQ}&=&\frac{1}{2}\left(-Q_c+3\,Q_L+Q_1+Q_b\right)
\ea

The right handed neutrino can also be accommodated as an open string
with both ends on the bulk abelian brane:
\ba
&~&{\nu_R}(\b1,\b1,0,0,0,+2)+{\nu^c_R}(\b1,\b1,0,0,0,-2)\nn
\ea

According to the RGE running results of Table~\ref{rt1}, there is
only one brane configuration, for the models under discussion,
that reproduces the weak mixing angle at low energies. This consists
of placing the $\u1_1$ brane on the top of the color branes, so that
$g_1=g_3$, which leads to $M_s\sim{\cal O}(10)$ TeV (7-17 TeV, depending on the
threshold corrections).

Finally we should stress that the triplication of families, can be obtained by the mechanisms of
branes at singularities \cite{bsing} or intersecting branes
\cite{b1}-\cite{b3},\cite{madrid0}-\cite{madrid3}.

\subsection{The quark and lepton mass structure}
\setcounter{equation}{0}

An important ingredient in the realization of the SM is the structure of the
quark and lepton masses and the associated Yukawa
couplings. Although this problem remains open in the context of D-brane
 models, several features are evident at this
level. The essential feature is that  the Yukawa couplings relevant to
fermion masses  are constrained by the
various U(1) symmetries and can present interesting patterns.

We will describe below the structure of the masses (and associated
Yukawas) for the heaviest generation in the four low-scale string
models of the previous section.

\bigskip
$\bullet$ {\bf Model A}. The relevant Yukawa couplings compatible with all the gauge symmetries are
\ba
M_A=\lambda_u\,Q\, u^c\, H_u +\lambda_d\,Q\,d^c\,H_d^\dagger+
\lambda_e\,L\,e^c\,H_u^\dagger+\lambda_\nu\,L\,H_d\,\nu_R
\ea
Here, charged
leptons and up quarks (of the heaviest generation) obtain masses from the same
Higgs ($H_u$).

When all Yukawa couplings arise at the lowest (disk) order, it is easy to check
that in the simplest case (absence of discrete selection rules, etc), they
satisfy the following relations:
\ba
\lambda_u =\lambda_e =\sqrt{2}g_2\ ,\quad \lambda_d =\sqrt{2g_s}\ ,\quad
\lambda_\nu =\sqrt{2}g_b\, .
\label{A}\ea

The central idea behind such relations is that couplings between
two or three different branes scale with the common
(intersection) volume.

Consequently the top and bottom quark masses are given by:
\ba
m_t=g_2 v \sin\beta\qquad ;\qquad m_b={\sqrt g_s}v\cos\beta\, ,
\label{beta}\ea
where $\tan\beta=v_u/v_d$, with $v_u$ and $v_d$ the vacuum expectation
values (VEVs) of the two higgses $H_u$ and $H_d$, respectively, and
$v=\sqrt{v_u^2+v_d^2}=246$ GeV.
Note that in the case where the color branes are identified
with D3 branes, one has $\sqrt{g_s}=g_3$, and in any case $g_s\ge g_3^2$.
Note also that since the string scale in this model is relatively low, $M_s\simlt
1$ TeV, there is no much evolution of the low energy couplings from the
electroweak to the string scale. Thus, using the known value of the bottom mass
$m_b\simeq 4$ GeV, one obtains for the top quark mass $m_t\simeq 162$ GeV
which is less than 5\% below its experimental value
$m_t^{\rm exp}=174.3\pm 5.1$ GeV. In addition, the Higgs VEV ratio turns
out to be large, $\tan\beta\simeq 100$. Note that such a large value is
not in principle problematic as in the supersymmetric case, but it can
lead to important higher order corrections.

On the other hand the $\tau$-mass is of the same order as the top
mass, which is unrealistic.
However, there is still the possibility that
the lepton Yukawa coupling $\lambda_e$ vanishes to
lowest order due to additional string discrete selection rules, and is generated by a higher dimensional operator of
the form $Le^c(H^\dagger_u H^\dagger H)$ providing the appropriate
suppression.\footnote{Models with similar properties have been considered in the past
in the perturbative heterotic string framework.}

\bigskip
$\bullet$ {\bf Model A'}. The Yukawa couplings here are
\ba
M_{A'}=\lambda_u\,Q\, u^c\, H_u + \lambda_d\,Q\,d^c\,H_d^\dagger+
\lambda_e\,L\,e^c\,H_d^\dagger+\lambda_\nu\,L\,H_d\,\nu_R
\ea
with the same relation for the tree-level couplings as in
(\ref{A}).
Using the parametrization in (\ref{beta}) we see that the
relation of $m_{t}$ to $m_{b}$ is the same as in model A and the
same remarks apply.
Since here the lepton and down quark acquire their
masses from the same Higgs, one obtains the phenomenologically interesting
relation: $m_b/m_\tau =\sqrt{g_s}/g_2=g_3/g_2$, when strong interactions are on
D3 branes. Thus, from Table~\ref{rt1}, $m_b/m_\tau\simeq 1.75$ at the (string)
unification scale, which is in the upper edge of the experimentally
allowed region at the $Z$-mass, $1.46\simlt m_b/m_\tau|_{\rm exp} \simlt
1.75$. This relation could replace the successful GUT prediction
$m_b=m_\tau$ of the conventional unification framework, in low scale string
models.
In conclusion model A' seems to be able to generate the required
hierarchy of masses for the third generation.

\bigskip
$\bullet$ {\bf Model B}.
The relevant trilinear Yukawa couplings are,
\ba
M_B=\lambda_u\,Q\, u^c\, H_u + \lambda_d\,Q\,d^c\,H_d^\dagger+
\lambda_e\,L\,e^c\,H_d^\dagger+\lambda_\nu\,L\,H_u\,\nu_R
\ea
The tree-level Yukawa couplings satisfy
\ba
\lambda_e=\lambda_u =\sqrt{2g_s}\ ,\quad \lambda_d =\sqrt{2}g_3\ ,\quad
\lambda_\nu =\sqrt{2}g_b\
\ea
and we have
\ba
m_t={\sqrt g_s}v \sin\beta\qquad ;\qquad m_b=g_3 v\cos\beta\, .
\label{rgebd}
\ea

The first relation implies again a heavy top, while the bottom to tau mass
ratio is now predicted, with a value
$m_b/m_\tau=g_3/\sqrt{g_s}\simlt 1$ which is apparently far from its
experimental value. However, in this case, the string scale is
relatively high and therefore one should take into account the
renormalization group evolution above the weak scale.
Solving the associated RGEs with the boundary conditions (\ref{rgebd})
and assuming $g_3=\sqrt{g_s}$, we obtain acceptable $m_b$ and $m_\tau$
masses for $M_s\sim 3 \times 10^3$ TeV and $\tan\beta\sim 80$.
Note that the successful prediction of $m_b$ and $m_\tau$ is related to
the condition $m_b=m_\tau$ at the (string) unification scale,
which in the case of non-supersymmetric Standard Model is obtained at
relatively low energies~\cite{ara1,ara2}.
%======changed Jan 03
\begin{figure}[t]
\center \epsfxsize=12cm \epsfbox{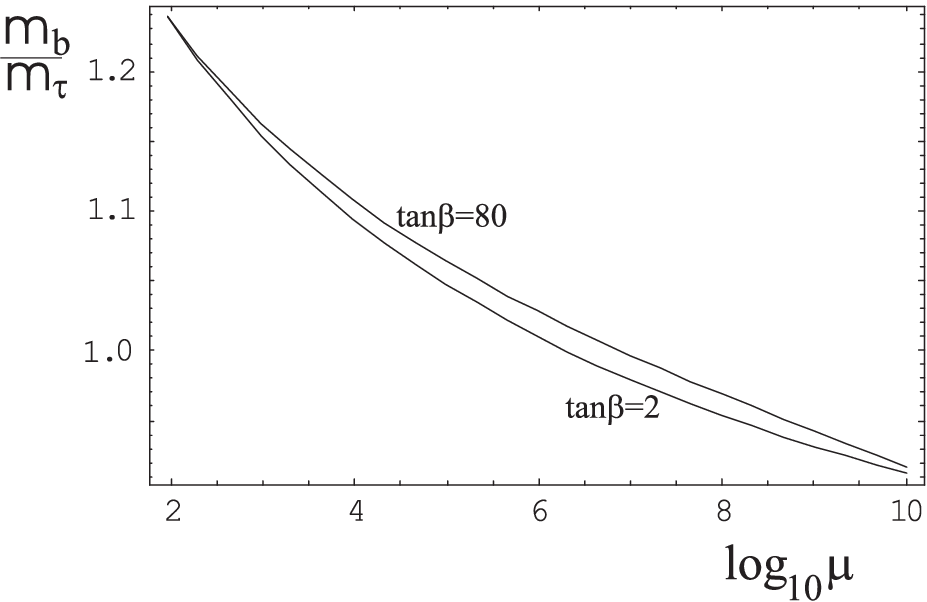}
\caption{\label{mbomtau}{\it Evolution of the ratio $m_b/m_\tau$
as a function of the energy $\mu$ for $\tan \beta=2$ and $\tan
\beta=80$. We have used as low energy parameters $m_b=4$ {\rm
GeV}, $m_{top}=174$ {\rm GeV} $a_3(M_z)=0.12$,
$\sin^2\theta_W=0.23113$.}}
\end{figure}
Indeed, in Figure~\ref{mbomtau}, we plot the mass ratio $m_b/m_\tau$ as a
function of the energy, within the non supersymmetric Standard Model with
two Higgs doublets.
Nevertheless, the resulting value of $M_s$ is still significantly
higher than the unification scale required from the analysis of gauge couplings
in section 3. Moreover, the top quark mass turns out to be rather high,
$m_t\sim 220$ GeV. It is an open question wether this discrepancy can be
attributed to threshold corrections that can be important in the case
of two dimensional bulk \cite{anba}.

\bigskip
$\bullet$ {\bf Model B'}.
The relevant Higgs couplings are given by
\ba
M_{B'}=\lambda_u\,Q\, u^c\, H_u +
\lambda_d\,Q\,d^c\,H_d^\dagger+\lambda_e\,L\,e^c\,H_u^\dagger+
\lambda_\nu\,L\,H_u\,\nu_R
\ea
while the tree-level Yukawa couplings by
\ba
\lambda_u =\sqrt{2g_s}\ ,\quad \lambda_d =\sqrt{2}g_3\ ,\quad
\lambda_\nu =\sqrt{2}g_b\ {\rm and}\ \lambda_e =\sqrt{2}g_2
\ea

Here, as in model A, the $\tau$ and top mass are of the same
order and thus in conflict with experiment. As in model A,
vanishing leading order coupling could be a way out.

In the discussion above we have paid attention to the heaviest
generation. However, an important feature of the SM spectrum is
the large hierarchy between the masses of the different
generations. Can this be generated in the models above?

A concrete idea in this direction si to generate such a hierarchy
by exploiting the presence of world-sheet (disk) instanton
corrections in models where branes intersect in the internal
space \cite{yuk}. It relies on the fact that the BCFT three-point
function responsible for the Yukawa couplings may have a prefactor
coming from world-sheet instantons, if the relevant fields come
from three branes that interest by-two in the internal space in
such way as to create a triangle. Then the open string world-sheet
must wrap this triangle and there is an exponential suppression
due to the area of this triangle in space-time. Adjusting such
areas it seem possible that an appropriate hierarchy of Yukawa
couplings can be generated \cite{yuk}.

A further comment is due concerning neutrinos. By now there are
stringent experimental limits in the mass matrix of neutrinos. A
natural question is: how neutrinos with the right properties can
be accommodated in D-brane models?

Right-handed neutrinos appear naturally in D-brane models. To
generate (small) masses compatible with experiment, either one
will have to wrap the associated brane around a large cycle, or
produce the suppression from world-sheet instantons.
Moreover, there are several possibilities in the neutrino sector.

A minimal possibility is to have a single right-handed neutrino
which provides the mixing along with it KK-descendants.
This possibility is rather constrained and was analyzed  in
\cite{akt1}. Since in this case the KK-neutrinos are acting almost as
sterile, there are stringent experimental constraints that
indicate that one needs fine  tuning in order that such a possibility
to be compatible with data.
On the other hand, if one has one right-handed neutrino per family
then a priori there are are no such strong constraints. It remains
however to be seen if there are brane models with a neutrino mass
-matrix compatible with current data.

\begin{table}[htb] \footnotesize
\renewcommand{\arraystretch}{2.5}
\begin{center}
\begin{tabular}{|c||c|c|c|}
\hline
 $N_i$    &  $(n_i^1,m_i^1)$  &  $(n_i^2,m_i^2)$   & $(n_i^3,m_i^3)$ \\
\hline\hline $N_3=3$ & $(1,0)$  &  $(2,1)$ &
 $(1 ,  1/2)$  \\
\hline $N_2=2$ &   $(0,-1)$    &  $ (1 ,0)$  &
$(1,3 /2)$   \\
\hline $N_1=1$ & $(1,3)$  &
 $(1,0)$  & $(0,1)$  \\
\hline $N_{1'}=1$ &   $(1,0)$    &  $(0,-1 )$  &
$(1, 3/2)$   \\
\hline \end{tabular}
\end{center} \caption{Example of  D6-brane wrapping numbers giving rise to
the right number of SM families (from (\cite{madrid2,madrid3}).
\label{sol} }
\end{table}

\subsection{The Standard Model structure from intersecting
D-branes}
\setcounter{equation}{0}

We will describe in this section, as an example a non-supersymmetric D-brane model utilizing
intersecting D6-branes that gives the correct spectrum \cite{madrid2,madrid3}.

We will be using the results and notation of section \ref{id6}. The important
ingredients are four stacks of branes that generate the minimal
brane gauge group $U(3)\times U(2)\times U(1)\times U(1)'$. We
will label the associated stack of branes by the indices
$3,2,1,1'$.
The hypercharge embedding is given by the possibility
(\ref{newhyper})
\be
Y={1\over 6}Q_3-{1\over 2}(Q_1-Q_{1'})
\ee

To generate the three families of the SM we will need some
specific intersection numbers (defined in (\ref{intersection}))
among the four stacks of branes
\bea
I_{3,2}\ & = &  \ \phantom{-} 1 \ ;\ I_{3,2*}\ =\  \phantom{-}2  \nonumber \\
I_{3,1}\ & = &  \ -3 \ ;\ I_{3,1*}\ =\ -3  \nonumber \\
I_{2,1'}\ & = &  \ \phantom{-}0  \ ;\ I_{2,1'*}\ =\ -3  \nonumber \\
I_{1,1'}\ & = &  \ -3 \ ;\ I_{1,1'*}\ =\ \phantom{-}3
\label{intersection2}
\eea
These can be achieved by choosing the wrapping numbers on the
three  two-tori as in table \ref{sol}.
A negative number denotes that the
corresponding fermions should have opposite chirality to those
with positive intersection number.

\begin{table}[htb] \footnotesize
\renewcommand{\arraystretch}{1.25}
\begin{center}
\begin{tabular}{|c|c|c|c|c|c|c|c|}
\hline Intersection &
 Matter fields  &   &  $Q_3$  & $Q_2 $ & $Q_1 $ & $Q'_1$  & Y \\
\hline\hline (3,2) & $Q_L$ &  $(3,2)$ & 1  & -1 & 0 & 0 & 1/6 \\
\hline (3,2*) & $q_L$   &  $2( 3,2)$ &  1  & 1  & 0  & 0  & 1/6 \\
\hline (3,1) & $u_R$   &  $3( {\bar 3},1)$ &  -1  & 0  & 1  & 0 & -2/3 \\
\hline (3,1*) & $d_R$   &  $3( {\bar 3},1)$ &  -1  & 0  & -1  & 0 & 1/3 \\
\hline (2,1'*) & $ L$    &  $3(1,2)$ &  0   & -1   & 0  & -1 & -1/2  \\
\hline (1,1') & $e_R$   &  $3(1,1)$ &  0  & 0  & -1  & 1  & 1   \\
\hline (1,1'*) & $\nu_R$   &  $3(1,1)$ &  0  & 0  & 1  & 1  & 0 \\
\hline \end{tabular}
\end{center} \caption{ Standard model spectrum and $U(1)$ charges
(from \cite{madrid2,madrid3})
\label{spec} }
\end{table}

The massless fermions living at the intersections  are
shown in  table \ref{spec}.
As we have argued in section \ref{id6}, due to the orientifold operation  one has to include
the D6-branes which are ``mirror" under that operation and have the same
wrapping numbers except for a flip in sign for the $m_a^i$'s.
We have denoted the
mirror D6-branes with a star.

All U(1)'s of the model obtain masses except for the hypercharge.
Thus, the low energy gauge group is that of the standard model
($SU(3)\times SU(2)\times Y$).

Generic intersecting D6-brane models are non-supersymmetric due
to non-trivial intersections. To each
of the intersections there are  associated massive scalar fields
which may be considered  as ``supersymmetric-partners", squarks and sleptons,
of the massless chiral fermions. They have the same multiplicity $|I_{ij}|$
 and carry the same gauge quantum numbers.
The lightest of those states have  masses
{\small \bea
\begin{array}{cc}
t_1:  & \alpha' {\rm (Mass)}^2 =
\frac 12(-\vartheta_1+\vartheta_2+\vartheta_3) \\
t_2:  & \alpha' {\rm (Mass)}^2 =
\frac 12(\vartheta_1-\vartheta_2+\vartheta_3) \\
t_3:   & \alpha' {\rm (Mass)}^2 =
\frac 12(\vartheta_1+\vartheta_2-\vartheta_3)
\label{tachyon}
\end{array}
\eea}
Here $\vartheta_i $ are the intersection angles (in units of $\pi $)
at each of the three sub-tori.
Thus, the masses depend on the moduli of the internal torus.
Although  in  principle some of the scalars could  be tachyonic,
  in general it is possible
   to vary the compact radii in order  to get rid of all tachyons
of a given model \cite{madrid2,madrid3}.

On the D6-branes there is at the tree level, an $N=4$
SUSY gauge multiplet for each of the groups of the SM. It is expected that
loop effects involving the fields at the intersections render the
charged SUSY partners of the gauge bosons massive, with masses of the order
of the
string scale.

We also have possible Higgs fields
coupling to quarks coming in  four varieties
with charges under $Q_2,Q_1$ and hypercharge given in  table \ref{hig}.

\begin{table}[htb] \footnotesize
\renewcommand{\arraystretch}{1.25}
\begin{center}
\begin{tabular}{|c|c|c|c|}
\hline
 Higgs   &  $Q_2$  &  $Q_1$   & Y \\
\hline\hline $h_1$ & 1  &  -1 & 1/2  \\
\hline $h_2$ &   -1    &  1  &  -1/2   \\
\hline\hline $H_1$ & -1  &  -1 & 1/2  \\
\hline $H_2$ &   1    &  1  &  -1/2   \\
\hline \end{tabular}
\end{center} \caption{ Electroweak Higgs fields
\label{hig} }
\end{table}
Such Higgs fields can trigger the electroweak symmetry breaking.

\bigskip

There has been a lot of progress  in constructing D-brane
models that come quite close to the SM model.
They involve non-supersymmetric intersecting brane models in a flat space
 \cite{b1}-\cite{b3}, \cite{madrid0}-\cite{madrid3},\cite{non-madr1}-\cite{non2} or with a nontrivial internal space \cite{com1}-\cite{com2} as well as supersymmetric models,
\cite{cvetic1}-\cite{cvetic6}. The effective low energy couplings of such models were also investigated \cite{c1}-\cite{c3}.
Although none so far can be claimed to be fully successful, progress in model building gives promise that fully realistic string
models may be available soon.
\newpage

\renewcommand{\theequation}{\arabic{section}.\arabic{equation}}
\section{Novel realizations of four-dimensional gravity\label{novel}}
\setcounter{equation}{0}

We will now turn to another ingredient of fundamental interactions
namely gravity.
In D-brane constructions, gravity is obtained from the closed string
sector and propagates in ten dimensions before compactification.
Four-dimensional gravity emerges at large distance after six
dimensions become compact. For future comparison we will study
here the gravitational potential of a 4+n dimensional gravity theory
compactified on an n-dimensional torus with (equal) radii $R$.
\be
V(r,\theta_i)=\sum_{n_i\in Z}{M^{-2-n}\over
\left[r^2+R^2\sum_{i=1}^n(\theta_i+n_i)^2\right]^{n+1\over 2}}
\label{compa}\ee
Here $V$ is the static gravitational potential of a point source
located at $r=\theta_i=0$, M is the (n+4)-dimensional Planck
scale, r is the four-dimensional radial distance and $\theta_i$,
$i=1,2\cdots,n$
are angular coordinates on the internal torus with uniform period 1.

$\bullet$ At four-dimensional distances much smaller than the size
of the internal torus gravity is (n+4)-dimensional. The infinite
sum in (\ref{compa}) is dominated by the $n_i=0$ term and thus
\be
V\sim {M^{-2-n}\over
\left[r^2+R^2\sum_{i=1}^n(\theta_i)^2\right]^{n+1\over 2}}\sp r<<R
\ee

$\bullet$ On the other hand, for distances much longer than the size of the
internal torus gravity becomes four-dimensional.
Upon a Poisson resummation of (\ref{compa}) and then keeping the $\tilde n_i=0$ term we obtain
\be
V\sim {1\over (MR)^n}{M^{-2}\over r}\sp r>>R
\ee
which gives also the effective four-dimensional Planck scale as
$M_P^2=M^2~(MR)^n$ as expected.

Thus, compactification changes the behavior of gravity in the IR.
The characteristic turn-over scale is the compactification scale.

\renewcommand{\theequation}{\arabic{section}.\arabic{subsection}.\arabic{equation}}
\subsection{Randall-Sundrum localization\label{RS}}
\setcounter{equation}{0}

We  describe next an alternative realization of four-dimensional gravity
that comes under the name of RS localization \cite{rs2}. Recent  lectures describing in detail
large dimensions and
the RS universe can be found in  \cite{review2,csaki,brax}.

We will consider the case of a five-dimensional bulk space
with coordinates $x^M=(y,x^{\mu})$, $\mu=0,1,2,3$.
We also consider a three-brane located at $y=0$.
Apart from the five-dimensional Einstein term we also have a
constant energy density on the brane, and a constant energy
density in the bulk.
We can summarize the effective action as

\be
S=\int dy~d^4x~\sqrt{g}~[M^3~R-\Lambda]-\delta(y)~\int d^4\xi \sqrt{\hat
g}~V_b
\label{rs}\ee
where $\hat g_{ab}$ is the induced metric on the brane $\hat
g_{ab}=g_{MN}{\p x^{M}\over \p{\xi^a}}{\p x^{N}\over \p{\xi^b}}$. We will pick a
static gauge for the brane coordinates $\xi^a=x^a$.
To simplify matters we will also consider an
orbifold structure under $y\to -y$. Thus, the two parts of
space-time separated by the brane are mirror symmetric around the position of the brane.

We would like to solve the equations of motion stemming from the
action (\ref{rs}).
Let us first seek solutions invariant under the orbifold action
which are flat along the brane, and depend only on the
fifth-coordinate $y$.

The ansatz for the five-dimensional metric is
\be
ds^2=e^{2A(y)}dx^{\m}dx_{\m}+dy^2
\label{m1}\ee
In order for the flat-brane ansatz to provide a solution, the two vacuum
energies must be related
\be
V_b^2=-12~\Lambda~M^3
\ee
This implies that the bulk vacuum energy $\Lambda$ must be negative.
We will also define the RS (AdS) energy scale
\be
K=-{\Lambda\over V_b}
\ee
The gravitational equations have the solution
\be
A(y)=-K|y|
\label{m2}\ee
The space on the one-side of the brane is a slice of $AdS_5$
patched-up with its mirror image at $y=0$. Indeed defining
$r=e^{Ky}$ for $y>0$ and scaling $x^{\mu}\to x^{\m}/k$ we obtain
\be
ds^2={1\over K^2r^2}[dr^2+dx^{\m}dx_{\m}]\sp r\geq 1
\label{ads}\ee
which is the $r\geq 1$ slice of $AdS_5$ in Poincar\'e coordinates with AdS energy scale $K$.
Note that the orbifold has removed the boundary of $AdS_5$ at
$r=0$.

An interesting further question concerns the effective
interactions mediated by gravity in this background. To find them we must study the
small fluctuations around this solution \cite{rs2,gravf1,gravf2,gravf3}.
Direct variation of the equations along the brane longitudinal directions
and gauge fixing gives a scalar  equation for the static graviton propagator
\be
M^3\left[(-e^{-2A}\bbox_x-\p_y^2-4A'\p_y\right]~G(x;y)=\delta^{(3)}(x)~\delta(y)
\ee where we placed the source on the brane.
This can be Fourier transformed along the $x^{i}$ coordinates
obtaining
\be
M^3(e^{-2A}\vec p^2-\p_y^2-4A'\p_y]~G(\vec p;y)=\delta
(y) \label{5}\ee
Changing variables to $w=e^{K|y|}$ we obtain away from $y=0$ the
solution (imposing the symmetry $G(\vec p;y)=G(\vec p;-y)$
\be
G(\vec p;y)=B~w^2~K_2\left({wp\over K}\right)
\ee
where $p\equiv |\vec p|$ and $K_2$ is the standard Bessel function.

The multiplicative constant $B$ is fixed by the requirement that the
solution  satisfies the discontinuity condition
\be
{{\partial G({\vec p},y)} \over{\partial y}}\Big|_{y=0+} -
{{\partial G({\vec p},y)} \over{\partial y}}\Big|_{y=0-} = - {1\over{M^3}}
\label{disc}
\ee
obtained from (\ref{5}) by integrating both sides in the interval
$(-\e, +\e)$ and taking the limit $\e \to 0$.
We obtain
\be
B=-{1\over 2M^3}{1\over 2K~K_2(p/K)+pK_2'(p/K)}
\ee
We may use the identity $K'_2(z)+2K_2(z)/z=-K_1(z)$ to rewrite
\be
B={1\over 2M^3~p~K_1\left({p\over K}\right)}
\ee

We can now investigate the force mediated by the graviton
fluctuations on the brane by evaluating
\be
G(\vec p,0)={1\over 2M^3 ~p}{K_2\left({p\over K}\right)\over K_1\left({p\over
K}\right)}
\ee
The static gravitational potential between two unit sources on the
brane (upon transforming back to configuration space)  becomes
\be
V(r)={1\over 2\pi^2~r}\int_0^{\infty}pdp~\sin (pr)~G(\vec p;0)
\ee
$$
={1\over 4\pi}{K\over M^3}{1\over r}+\delta V(r)
$$
with
\be
\delta V(r)={1\over 4\pi}{K\over M^3~r}\int_0^{\infty}dq~{K_0\left(q\right)\over
K_1\left(q\right)}~\sin(qr)
\label{dV}\ee
where here $r^2=\vec x^2 $ the spatial distance on the brane, not
to be confused with radial coordinate of $AdS_5$ space in
(\ref{ads}). To obtain the second equality above, we used the identity
$K_2(z)=2K_1(z)/z+K_0(z)$.

For $q\to\infty$ the ratio $K_0(q)/K_1(q)\to 1$
and the integral in (\ref{dV})
reduces to the ill-defined integral of $\sin~ q$ over the positive real axis.
In order to evaluate (\ref{dV}) we multiply the integrand by
$e^{-\zeta q}$, perform the integration and then take the limit $\zeta\to
0$.
The integral depends on the  class of regulators (see \cite{irs} for a discussion).

For $Kr\gg 1$, the strongly oscillatory behavior of $\sin(q Kr)$ results
in a negligible contribution to the integral from large $q$.
This means that we can employ the expansion of the Bessel functions for
small $q$: $K_0(\tilde p)/K_1(\tilde p)=-\tilde p \log\tilde p+\cdots$.
We obtain in this  regime \cite{rs2}
\be
\delta V(r)\simeq\frac{1}{8\pi}\,\frac{1}{M^3K}\,\frac{1}{r^3}.
\label{dv1} \ee
We thus  reproduce the leading and sub-leading behavior of the potential
at long distances in the RS setup.
\be
V(r)={1\over 4\pi}{K\over
M^3~r}\left[1+\frac{1}{2}\,\frac{1}{K^2r^2}+{\cal
O}\left({1\over K^4r^4}\right)\right]\sp Kr>>1
\ee
Thus, at long distances gravity is four-dimensional (with
sub-leading corrections). The effective four dimensional Planck
scale reads $M_P^2=M^3/K$.

For $Kr\ll 1$ the main contribution to the integral comes from large $q$, for which
$K_0(q)/K_1(q)=1$. We find
\be
\delta V(r)\simeq \frac{1}{4\pi^2}\,\frac{1}{M^3}\,\frac{1}{r^2}.
\label{dv2} \ee
so that the total potential is
\be
V(r)={1\over 4\pi^2~M^3~r^2}\left[1+\pi~Kr+{\cal O}(K^2r^2\log
r)\right]\sp Kr<<1
\ee

Thus, at short distances gravity is five-dimensional  \cite{Kakushadze:2001rz,irs}.
This is completely analogous to compactification with radius
$1/K$. The RS setup is thus an alternative mechanism to
compactification for transforming five-dimensional gravity into
four-dimensional at large distances.

We have neglected the scalar part of the fluctuations. This will
give an extra interaction. For such an interaction to be unobservable, the extra scalar mode (the radion)
should acquire a mass of the order of an eV or more.

\subsection{Brane Induced Gravity}
\setcounter{equation}{0}

We now describe  an alternative realization of four-dimensional gravity
that comes under the name of Brane Induced Gravity (BIG)\cite{big1}-\cite{big5}.
The idea of induced gravity has a long history \cite{ig}. The
context here is however different since gravity is induced on a brane (in general a submanifold of space-time) rather
than the full space-time.

We  consider first  the simplest case of a five-dimensional bulk
with coordinates $y,x^{\mu}$, $\mu=0,1,2,3$.
We also consider a three-brane located at $y=0$.
Apart from the five-dimensional Einstein term we would like to study
the effects of a four-dimensional Einstein term localized on the
brane. The presence of such an (induced) term will be motivated
below.

The relevant action is
\be
S=M^3\int dy~d^4x~\sqrt{g}~R+\delta(y)M^3 \rc~\int d^4\xi \sqrt{\hat g}~\hat
R
\label{big}\ee
where $\hat g_{\mu\nu}$ is the induced metric on the brane and
$\hat R$ the induced Ricci scalar.
We also parameterized the coefficient of the four-dimensional term
in terms of a new length scale $\rc$.

We are interested in the gravitational interaction, generated by
the action (\ref{big}), as perceived on the brane.
We will evaluate first the static propagator of (\ref{big}).
Although there is interesting physics in the tensor structure, we
will neglect it for the moment and consider instead the scalar
propagator. Placing the source on the brane (at the origin) we
must solve
\be
M^3(\bbox_3+\p_{y}^2+\rc ~\delta(y)\bbox_3)G(\vec
x;y)=-\delta(y)\delta^{(3)}(x)
\ee
Fourier transforming in the 3-spatial coordinates $\vec x$ we
obtain
\be
M^3(\vec p^2-\p_{y}^2+\rc~\delta(y)~\vec p^2)G(\vec p;y)=\delta(z)
\ee
The solution can be found by first solving the equations away from
the position of the brane, and then matching along $y=0$. The
result is
\be
G(\vec p;y)={e^{-|\vec p|~|y|}\over M^3(2|\vec p|+\rc~\vec p^2)}
\ee
For the source and the probe being on the brane $y=0$ the
static propagator becomes

\be
G(\vec p;0)={1\over M^3(2|\vec p|+\rc~\vec p^2)}
\ee

By Fourier transforming back we obtain the static gravitational potential
\be
V(r)={1\over{2\pi^2 r}}\int_0^\infty dp\, p \sin{pr}\, G(\vec p;0)
\label{V(r)}.
\ee
where we have set $p\equiv |\vec p|$ for simplicity.

We are now ready to study the behavior of the gravitational force
in various regimes.

$\bullet$  Long distances: $p<<1/\rc$.
Then the propagator and the potential can be approximated as
\be
G\sim {1\over 2M^3p}\sp V(r)\sim {1\over M^3~r^2}
\ee
This is the behavior of five-dimensional gravity with Planck scale $M$.

$\bullet$ Short distances: $p>>1/\rc$.
We obtain in this case
\be
G\sim {1\over M^3~\rc~p}\sp V(r)\sim {1\over M^3~\rc~r}
\ee
This is the behavior of four-dimensional gravity with effective Planck scale $M_P^2=M^3~\rc$.

We have thus a situation which is inverted with respect to normal
compactification: four-dimensional gravity at short distance and
five dimensional gravity at long distance.

If this mechanism is to play a role in realistic situations, we
must tackle an immediate problem: we know from table-top to solar
system to cosmological observations that gravity is four-dimensional
on length scales ranging from 100 $\mu$m to $10^{26}$
m.
In view of this there are two possibilities:

(a) The length scale $\rc$ is cosmologically large. We will
comment below on whether this possibility can be realized in a
concrete theory.

(b) The theory must also include another mechanism that forces
four-dimensional gravity at long distances. For example, the extra
coordinate $y$ can be compact with radius $R$.

We will now examine these two possibilities in turn.

(a) Branes appearing in superstring theories are the best
controlled candidates for the (solitonic)  branes we are
considering. However, such branes have no tree level
induced term\footnote{Such a term can appear in D-branes
of bosonic string theory \cite{corley} but its effects are masked by
stringy effects. They are not visible macroscopically.}.
We do expect though that branes with localized degrees of freedom
on them will develop a localized  induced gravity term due to
quantum corrections. In fact in any theory, one-loop diagrams of
matter fields generate a quadratically divergent correction to the
Einstein term (in four dimensions).
Although this calculation is tricky \footnote{The only gauge
invariant regularization of gravity at one-loop is dimensional
regularization, which is not sensitive to quadratic divergences.}
we expect that the coefficient of the induced four-dimensional
Einstein term due to a given particle going around in the loop is
given by $\zeta~\Lambda^2$ where $\Lambda$ is an ultraviolet
cutoff and $\zeta$ depends on the particle. If we regularize using the heat kernel method, a
 scalar contributes
${1\over 6}-\xi$ where $\xi$ is the conformal coupling,
 a Weyl fermion contributes
$-{1\over 6}$, a massless vector $-{2\over 3}$ \cite{ig}.
However, the individual contributions as calculated in string
theory, (where the generic cutoff is $M_s$) are different
\cite{kkk,narain,kohlprath}.
Such a discrepancy in gravitational contributions has been noticed
before \cite{instanton}. We are bound to conclude that the only
reliable method to calculate such corrections is given by string
theory, as other regularizations are partial, and do not imply
global consistency of the gravitational theory.

In string theory, UV finiteness indicates that typically (in the
absence of large values for the moduli) the one-loop correction,
when non-zero, is proportional to $M_s^2$.
However, KK states can transform this contribution so that it
becomes moduli dependent.

In heterotic string theory and asymmetric type II vacua , such corrections are zero for $N\geq
1$ supersymmetry \cite{anton,kkk,book}.
In symmetric type II vacua corrections are non-zero for $N\leq 2$
supersymmetry \cite{kkk}. In particular, for CY compactifications,
the correction is non-zero at one-loop only and is proportional to
the Euler number \cite{narain,kohlprath,amv}.
In type-I/orientifold vacua there are non-zero one-loop induced
terms on D-branes when $N\leq 2$ \cite{abf,fat}.

It is in principle conceivable that the induced term can be large (the
transition scale $\rc$ is large). We can imagine for example a
D-brane wrapping a small cycle \cite{fat} or a non-compact CY with an
astronomical Euler number (that does not seem to be forbidden by
geometric considerations) or a large N $Z_N$ non-compact orbifold \cite{Kohl2}.
 It is not obvious that this will be
stable in the absence of supersymmetry. Thus, at this point a
parametrically large $\rc$ which is stable against radiative
corrections in the absence of supersymmetry
is at the speculative level.

(b) Here we have two possible contexts depending on the
relation between the induced gravity scale $\rc$ and the
compactification scale $R$.

$\bullet$ $\rc>>R$. In this case gravity is four-dimensional at
all distance scales. The four-dimensional Planck constant is
$M_P^2=M^3~\rc$, independent of the compactification radius.

$\bullet$ $\rc<<R$ Here gravity is four-dimensional at
length scales much larger than R or much smaller than $\rc$. For
length scales in between, $\rc<<l<<R$ gravity is five dimensional.
The effective four-dimensional Planck mass for $l>>R$ is
$M_P^2=M^3~R$ while for $l<<\rc$, $M_P^2=M^3~\rc$.

A relevant question in case (b) is: if we need compactification
why worry about BIG.
The answer rests in the special properties of the coupling of
graviton KK states to the brane.
To understand this, consider a standard compactification with
radius R.
An important process is the emission of the KK states of the
graviton from the brane fields. At energies small compared to the
KK masses, $\sim 1/R$, such emission is suppressed for kinematical
reasons. On the other hand at energies $>>1/R$ this emission is
important, and energy is flowing from the brane to the bulk. Such
processing place stringent constraints on brane-models. In order for example
that the energy loss from supernovae is compatible with data an upper
bound can be put on $R$.

This behavior can be understood intuitively as follows:
In the regime of four-dimensional gravity $l>>R$ (or $E<<1/R$)
gravitons cannot easily leave the brane (otherwise gravity would
have been higher-dimensional), and their emission is suppressed.
In the opposite regime, $l<<R$ (or $E>>1/R$) gravity is
higher-dimensional and graviton emission in the bulk is
unsuppressed. This intuition is applicable in more general
circumstances as we will see below.

Let us now return to case (b), namely the combination of
compactification with BIG. Since BIG is dominating in the UV, it
suppresses the emission of KK gravitons in that regime, that would
have been otherwise unsuppressed.
A calculation of the value of the wave-function of graviton KK
states of mass m, in this case gives $|\phi_m(0)|^2\sim
(4+m^2\rc^2)^{-1}$ \cite{dvali}.
This value is the classical coupling constant, controlling the
emission of the KK states.
It is obvious that it is suppressed (compared to the toroidal
case $\rc=0$), for masses (energies) $>>1/\rc$.
This eases quite a bit the
phenomenological constraints of such models, and greatly affects the early cosmology.

To indicate the effect  of BIG, we quote that a five-to-four
dimensional compactification with $R=10^{16}$ m and $R/\rc\sim
10^{-4}$ would not be in conflict with today's experimental data,
\cite{dvali}. Thus, a new dimension the size of the solar system would
have been invisible!

We will now investigate BIG when there are $n>1$ transverse dimensions to the
3-brane since this case is qualitatively different.

We start again from the simplified action
\be
S=M^{n+2}\int d^ny~d^4x~\sqrt{g}~R+\delta^{(n)}(y)M^{n+2} \rc^n~\int d^4\xi \sqrt{\hat g}~\hat
R\label{bign}\ee
The equation for the scalar propagator in this case is

\be
M^{n+2}(\bbox_x+\bbox_y+\rc^n ~\delta^{(n)}(y)\bbox_x)G(\vec
x;\vec y)=-\delta^{(n)}(y)\delta^{(3)}(x)
\ee
Fourier transforming with respect to $\vec x$ we obtain
\be
M^{n+2}(\vec p^2-\bbox_y+\rc^n ~\delta^{(n)}(y)\vec p^2)G(\vec
p;\vec y)=\delta^{(n)}(y)
\ee
Going through the same steps as before we can evaluate the
propagator on the brane to be
\be
G(\vec
p;\vec y=\vec 0)={D_n(\vec p;\vec 0)\over M^{n+2}\left[1+\rc^n~\vec
p^2~D_n(\vec p;\vec 0)\right]}
\ee
with
\be
D_{n}(\vec p;\vec y)=\int d^n q~{e^{i\vec q\cdot \vec y}\over \vec
p^2+\vec q^2}
\ee

In this case, since $n\geq 2$, $D_n$ is UV divergent and thus
infinite. We then obtain that $G(\vec
p;\vec y=\vec 0)=1/(M^{n+2}\rc^n~p^2)$ which indicates
four-dimensional behavior at all distances. The bulk term is
completely diluted when $n\geq 2$.
This however, is due to the zero thickness of the brane. We can
introduce an analog of finite thickness by cutting off $D_n$  in
the UV,
\be
D_{n}(\vec p;\vec y;\Lambda)=\int_0^{\Lambda}q^{n-1}dq~ d\Omega_{n-1}~{e^{i\vec q\cdot \vec y}\over \vec
p^2+\vec q^2}
\ee

To see the effects of the brane thickness we will analyse the case
$n=4$ for which
\be
D_4(\vec p;\vec 0;\Lambda)=\Lambda^2-p^2\log{\Lambda^2+p^2\over
p^2}+{\cal O}\left({1\over \Lambda^2}\right)
\ee
so that the propagator becomes
\be
G(\vec p;\vec 0)\simeq {\Lambda^2\over
M^{6}(1+\rc^4\Lambda^2~p^2)}\simeq {1\over
M^6\rc^4\left[p^2+{1\over \Lambda^2\rc^4}\right]}
\ee

We thus see that at distances $\Lambda^{-1}<<l<<\Lambda \rc^2$,
gravity is four-dimensional. On the other hand at large distances
$l>>\Lambda\rc^2$ gravity is screened and the graviton has an
effective mass \cite{fat}.

Even at codimension one, a brane-thickness $w$ affects the
gravitational interaction.
It can be shown that for distances shorter than $\sqrt{w\rc}$ the
equivalence principle breaks down unless the theory is fine tuned \cite{fat,rubakov}.

So far we have neglected the tensor structure of gravity. This can
be taken into account and problems may arise from the scalar
component of the higher-dimensional graviton. It can be shown that there is
no vDVZ discontinuity in this case  \cite{por1}-\cite{por3}. However,
a new threshold scale appears where the linearized theory breaks down since
the scalar graviton interactions become non-perturbative, \cite{luty,rub2},
although there is no consensus yet on the dependence of this scale
on the four- and five-dimensional Planck scale. Further work in
this direction can be found in \cite{dpg1}-\cite{dpg2}.

\subsection{Randall-Sundrum  meets Brane-Induced Gravity\label{rsbig}}
\setcounter{equation}{0}

In this subsection we will investigate what happens when both mechanisms (RS+BIG)  are at work
simultaneously.

The relevant effective action now is (\ref{rs}) with the addition
of the four-dimensional Einstein term localized on the brane.
\be
S=\int dy~d^4x~\sqrt{g}~[M^3~R-\Lambda]+\delta(y)~\int d^4\xi \sqrt{\hat
g}~\left[M^3~\rc~\hat R-V_b\right]
\label{rsbig}\ee

We will again solve the equations of motion stemming from the
action (\ref{rsbig}).
The crucial observation here is that since the RS solution is flat
on the brane, it is not affected by the presence of the localized
Einstein term.
Thus with
\be
V_b^2=-12~\Lambda~M^3
\ee
the solution (\ref{m1},\ref{m2}) is still valid.

Now the equation for the static (scalar) graviton propagator is
modified to
\be
M^3\left[(-e^{-2A}\bbox_x-\p_y^2-4A'\p_y-\rc\delta(y)~\bbox_x\right]~G(x;y)=\delta^{(3)}(x)~\delta(y)
\ee
This can be Fourier-transformed along the $x^{i}$ coordinates
obtaining
\be
M^3(e^{-2A} p^2-\p_y^2-4A'\p_y+\rc\delta(y)p^2]~G(\vec p;y)=\delta
(y) \label{5big}\ee
Changing again variables to $w=e^{K|y|}$ we obtain away from $y=0$ the
solution (imposing the symmetry $G(\vec p;y)=G(\vec p;-y)$
\be
G(\vec p;y)=B~w^2~K_2\left({wp\over K}\right)
\ee

The multiplicative constant $B$ is fixed by the requirement on the
solution to satisfy the discontinuity condition (\ref{disc})
from which we obtain
\be
B={1\over M^3~p~\left[2K_1\left({p\over K}\right)+\rc~p~K_2\left({p\over K}\right)\right]}
\ee

We investigate the force mediated by the graviton
fluctuations on the brane by evaluating
\be
G(\vec p,0)={1\over M^3p}~{K_2\left({p\over K}\right)
\over 2K_1\left({p\over K}\right)+\rc~p~K_2\left({p\over K}\right)}
\ee

The nature of the force can be directly discerned in momentum
space.
We have the following limits
\be
{\rm for}~~~p \ll K, ~~~~~~~~~~~~~~~~
G({\bf p},z=0)\simeq \frac{1}{M^3\left(r_c+\frac{1}{K}\right)p^2},
\label{g1}
\ee
\be
{\rm for}~~~p \gg K, ~~~~~~~~~~~~~~~~
G({\bf p},z=0)\simeq \frac{1}{M^3\left(r_c p^2+2p \right)}.
\label{g2}
\ee

We shall distinguish two separate cases:
\bigskip

(a){\bf Strong BIG},  $Kr_c \gg 1$: Both for
$p\ll K$, as well as for $p\gg K$ we have $G^{-1} \simeq M^3r_cp^2$.
Thus we expect four-dimensional behavior $\sim 1/r$ for the potential at
all distances on the brane, with an effective
Planck constant $M^2_{Pl} \simeq M^3 r_c$.
The leading corrections to $V(r)$ can also be evaluated
by employing the full propagator \cite{irs}.

\bigskip

(b){\bf Weak BIG}  $Kr_c \ll 1$: For $p\ll K$, we have $G^{-1} \simeq M^3p^2/K$. We find that
at large distances ($r \gg 1/K$) the potential displays four-dimensional behavior
with ${\tilde M_{\rm P}^2}\simeq M^3/K$, as in the standard RS scenario.
For $k \ll p \ll 1/r_c$, we have $G^{-1} \simeq 2 M^3 p$.
Thus, for distances
$r_c \ll r \ll 1/K$ we find a
five-dimensional behavior $\sim 1/r^2$ for the potential.
This is in agreement with
the direct evaluation of the potential for $r_c=0$. Finally, for $p \gg 1/r_c$,
$G^{-1} \simeq M^3 r_c p^2$. At short distances
$r \ll r_c$ the behavior becomes again four-dimensional $\sim 1/r$, with
$M^2_{P} \simeq M^3 r_c$.

To summarize, the four-dimensional Einstein term
induced quantum mechanically on the 3-brane affects considerably
the gravitational interactions. Specifically, the gravitational
potential on the brane exhibits the four-dimensional behavior $V(r)\sim 1/r$ at all scales except in the
intermediate region $r_c\ll r\ll 1/K$, in which it is effectively
five-dimensional and given by $V(r)\sim 1/r^2$. Furthermore, for $Kr_c\ll 1$ the strength
of the gravitational interaction, i.e. the value of the effective $M_{Pl}$,
depends on the distance between the interacting masses. It is stronger for short distances,
the ratio of its value for $r\ll r_c$ to the one for large $r\gg 1/K$ being equal to $Kr_c$.

\subsection{Graviton emission in the bulk and brane energy-loss\label{energyloss}}
\setcounter{equation}{0}

The KK spectrum of gravitons in the case of four-dimensional
gravity descending from five dimensions via a combination of the
RS and BIG mechanisms plays an important role in the physics.
We naively expect that in energy regions where gravity is
four-dimensional on the brane, KK emission will be suppressed. In
the opposite case of regions of five-dimensional gravity, we expect
that KK emission will be significant.
We will  show that this picture is correct.

We will ignore as before the tensor structure of the metric and denote by $\Phi(x^\alpha,z)$
its small fluctuation field around the background
(\ref{m1},\ref{m2}). More details can be found in \cite{irs}.
The equation of motion at the linearized level is
\be
M^3\Bigl[{1\over\sqrt{-g}}\partial_\mu(\sqrt{-g}g^{\mu\nu}\partial_\nu)
+{{r_c}\over\sqrt{-{\hat g}}}\partial_\alpha(\sqrt{-{\hat g}}{\hat g}^{\alpha\beta}
\partial_\beta)\Big]\Phi(x^\alpha,z)=0.
\label{Phieqn}
\ee
As suggested by the symmetries of the background, we look for solutions in the form
$\Phi(x^\alpha,z)=\sum_n \phi_n(z)\sigma_n(x^\alpha)$, where
the $\sigma_n(x^\alpha)$ satisfy the four-dimensional
Klein-Gordon equation $(\partial^\alpha
\partial_\alpha+m_n^2)\sigma_n=0$. Using this in (\ref{Phieqn}), one is led to the field
equation
\begin{equation}
\left( \partial_z^2+e^{-2A}m_n^2+4A'\partial_z
+r_c \delta(z)m_n^2 \right) \phi_n(z)=0
\label{31}
\end{equation}
for the mode function $\phi_n(z)$.

The zero mode, (the solution corresponding to $m^2=0$), is not affected by
the presence of the term proportional to $r_c$ and consequently is identical
to the one in reference \cite{rs2}.

The KK modes,
analogous to those of \cite{rs2},
are defined as $\psi_n=\exp(3A/2)\,\phi_n$.
For $A(z)=-K|z|$ eq. (\ref{31}) gives
(for simplicity we omit the index $n$ from
$m_n$ and $\psi_n$)
\begin{equation}
\psi(z)=N(\mt) w^{1/2} \Bigl[Y_2(\mt w) + F(\mt) \, J_2 (\mt w)\Bigr],
\label{solkk}
\end{equation}
with $w\equiv\exp(K|z|)$, $\mt=m/K$. The constant $F(\mt)$
is fixed by the discontinuity
in $\partial_z \phi(0)$ due to the presence of the $\delta$-function
\be
F(\mt)=-\frac{2Y_1(\mt)+\rct \mt Y_2(\mt)}{2J_1(\mt)+\rct \mt J_2(\mt)},
\label{ccc}
\ee
with $\rct=r_cK$.

For $w \to \infty$ the KK modes become approximate plane waves
\begin{eqnarray}
\lefteqn{
\psi(w)\simeq N(\mt) \sqrt{\frac{2}{\pi \mt}}
\left[
\sin \left( \mt w - \frac{5}{4}\pi \right)
+ F(\mt) \cos \left( \mt w - \frac{5}{4}\pi \right)
\right] }
\nonumber \\
&\simeq&
N(\mt) \sqrt{\frac{2\left(1+F^2(\mt)\right)}{\pi \mt}}
\sin \left( \mt w - \frac{5}{4}\pi  + \beta(\mt) \right),
\label{plane}
\end{eqnarray}
with $\beta = \arctan F$.
As a result,
for a non-compact fifth dimension, the KK modes have a continuous spectrum and
their normalization is
approximately that of plane waves
\be
N(\mt) \sim \sqrt{\frac{\mt}{1+F^2(\mt)}},
\label{norm}
\ee
where we have neglected factors of order 1.
The strength of the interaction of the KK graviton modes with the other fields
on the brane is determined by
the square of their wave-function at the position $z=0$ of the brane. We find
\be
\psi(z=0) \sim \sqrt{\frac{\mt}{1+F^2(\mt)}}
\left[Y_2(\mt)+F(\mt) \, J_2(\mt)\right].
\label{supp}
\ee

A careful examination of the low energy effective action reveals the
presence of an additional suppression factor. The second term in
the action (\ref{rsbig}) results in a non-canonical kinetic term for
the fields $\sigma_n(x^\alpha)$. In order to render this term canonical
we must absorb a factor $(1+\rct |\psi(0)|^2)^{1/2}$ into the redefinition
of the fields.

After the KK kinetic terms have been properly normalized,
the coupling of the KK modes to matter on the brane is
given by $\sqrt{k/M^3}$. This coupling is squared in the
calculation of quantities such as KK mode production rates etc.
It is also accompanied by the integration over all KK states with a
plane-wave measure $dm/k$. As a result, the combination $dm/M^3$
appears in all the estimates of KK mode production in the following.

For $\mt \ll 1$ eq. (\ref{supp}) gives
\be
\psi(z=0) \sim \frac{\sqrt{\mt}}{1+\rct}.
\label{rang1} \ee
We thus recover the suppression $\sim \sqrt{m/K}$ of the
standard Randall-Sundrum model, which is further enhanced for
large $Kr_c$.

For $\mt \gg 1$, on the other hand, we find
\be
\psi(z=0) \sim \sin \left[
\arctan \left( \frac{\rct \mt}{2} \right) -\frac{\pi}{2}
\right].
\label{rang2}
\ee
For $m\gg 1/r_c$ there is a significant suppression factor
$\sim 1/(mr_c)$, while for $m \ll 1/r_c$
the wave-function on the brane is unsuppressed of order 1.

As we will show next, these results are consistent with our naive expectation and,
in addition, crucial to clarify the origin
of the behavior of the effective gravitational interaction on the brane.
We will again separate the two different cases:
(a) Strong BIG,  $Kr_c\gg 1$, where gravity was found to be four-dimensional at all distances;
(b) Weak BIG, $Kr_c\ll 1$, where gravity again behaves as in four dimensions,
except in the intermediate range $(K,1/r_c)$ where it is five-dimensional.

\bigskip

(a) {\bf Strong BIG} $Kr_c\gg 1$:

The wave function scales as

\begin{equation}
\psi(0)\sim\left\{ \begin{array}{llll}
\displaystyle  \phantom{llllll} &\sqrt{m\over r_c^2 K^3}={1\over Kr_c}\sqrt{m\over K} &\phantom{llllll} & m\lta K  \\
\\
\displaystyle \phantom{llllll} &{1\over mr_c}={1\over Kr_c}{K\over m},&\phantom{llllll} &m\gta K.\\
\end{array}\right.
\end{equation}

Thus, the gravitational
potential is dominated by the exchange of the zero mode and falls off
$\sim 1/r$ for all $r$.
The effective squared Planck constant is $M^3(r_c + 1/K)\simeq M^3r_c$.

\bigskip
(b) {\bf Weak BIG} $Kr_c\ll 1$:
The wave-function here scales as
\begin{equation}
\psi(0)\sim\left\{ \begin{array}{llll}
\displaystyle  \phantom{llllll} &\sqrt{m\over K} &\phantom{llllll} & m\lta K  \\
\\
\displaystyle \phantom{llllll} &1,&\phantom{llllll} &K \lta m \lta 1/r_c,\\
\\
\displaystyle \phantom{llllll} &{1\over mr_c}, &\phantom{llllll} & m \gta 1/r_c.
\end{array}\right.
\end{equation}

\bigskip

{\bf i)} For $r\gta 1/K$
the corrections to the four-dimensional
potential are dominated by
modes with $m\lta K$ because the contribution of modes with $m\gta K$
is exponentially suppressed.
The contribution of massive modes
is negligible relative to that of the
zero mode. Thus we expect a
fall-off $\sim 1/r$
with a squared Planck constant $M^3(r_c + 1/K)\simeq M^3/K$.\\

{\bf ii)} For $r_c \lta r \lta 1/K$ only the modes with $m \lta 1/r_c$ contribute
significantly.
Those with
$K \lta m \lta 1/r_c$
generate a term in the potential
\be
\delta V(r) \sim \frac{1}{M^3} \int_K^{1/r_c} dm \frac{e^{-mr}}{r} \psi^2(0)
\sim \frac{1}{M^3} \int_K^{1/r_c} dm \frac{e^{-mr}}{r}
\simeq \frac{1}{M^3r^2}.
\label{corr1} \ee
This contribution is much larger than those from
the modes with $m \lta K$ and the zero mode.
For example the modes with $m \lta K$ give
\be
\delta V(r) \sim \frac{1}{M^3} \int_0^{K} dm \frac{e^{-mr}}{r} \frac{m}{K}
\simeq \frac{K}{M^3r}.
\label{corr2} \ee
Thus, for distances
$r_c \lta r \lta 1/K$ we expect
five-dimensional behavior $\sim 1/r^2$ for the potential. \\

{\bf iii)} Finally, at short distances
$r \lta r_c$ the modes with $m \gta 1/r_c$ give a contribution
\be
\delta V_1(r) \sim \frac{1}{M^3} \int_{1/r_c}^\infty dm \frac{e^{-mr}}{r}
\frac{1}{m^2 r^2_c}
\simeq \frac{1}{M^3r_cr}.
\label{corr3} \ee
Those with $1/r_c \gta m \gta K$ give
\be
\delta V_2(r) \sim \frac{1}{M^3} \int^{1/r_c}_K dm \frac{e^{-mr}}{r}
\simeq \frac{1}{M^3r_cr}.
\label{corr4} \ee
These dominate over the contribution of the zero mode, as well as
of the modes with $m \lta K$.
Thus, the potential obtains again the four-dimensional form $\sim 1/r$, with
a squared Planck constant $M^3 r_c$. It is remarkable that this behavior is not
due to the zero mode, as one might have guessed, but instead it is attributed to
the exchange of massive modes with masses $m\gta K$.
Similar behavior was also observed in \cite{big1}-\cite{big5},\cite{Karch:2001ct}.

\section{The brane-universe and its cosmology\label{rsbigcosm}}
\setcounter{equation}{0}

There several ways of studying the cosmology of a brane-universe
that depend largely on the formalism that describes the brane
dynamics and its interaction with the bulk.
There is the RS description \cite{rs2,rs1}, appropriate from
gravitating branes, that has the advantage of simplicity since
many of the intricacies of the UV description are truncated.
There is the probe description (mirage cosmology, \cite{mirage})
which uses the DBI action to study the brane-geodesics in bulk
geometries. Finally there is the BCFT description of D-branes,
which for cosmological backgrounds is in its infancy and will not
be treated here.

\subsection{Brane cosmological evolution in the Randall-Sundrum
 paradigm.\label{crs}}
\setcounter{equation}{0}

The RS context was described (in the static case) in section
\ref{RS}.
Here we will allow the brane metric to depend also on time, in
order to derive effective equations for the associated cosmology.

Thus, the  model is described by the five-dimensional action
\be
S=\int d^5x~ \sqrt{-g} \left( M^3 R -\Lambda +{\cal L}_B^{mat}\right)
+\int d^4 x\sqrt{-\hat g} \,\left( -V+{\cal L}_b^{mat} \right),
\label{001}
\ee
where $R$ is the curvature scalar of the five-dimensional metric
$g_{AB}, A,B=0,1,2,3,5$,
$\Lambda$ is the bulk cosmological constant, and
${\hat g}_{\alpha \beta}$, with $\alpha,\beta=0,1,2,3$,
is the induced metric on the 3-brane.
We identify
$(x,z)$ with $(x,-z)$, where $z\equiv x_5$. However, following the conventions
of \cite{rs2} we extend the bulk integration over the entire interval
$(-\infty,\infty)$.
The quantity $V$ includes the brane tension as well as
quantum contributions to the
four-dimensional cosmological constant.
We have added an arbitrary matter action on the brane, as well as
in the bulk for future convenience.

We consider the following cosmological  ansatz for the metric
\begin{equation}
ds^{2}=-n^{2}(t,z) dt^{2}+a^{2}(t,z)\gamma_{ij}dx^{i}dx^{j}
+b^{2}(t,z)dz^{2},
\label{metric}
\end{equation}
where $\gamma_{ij}$ is a maximally symmetric 3-dimensional metric.
We use $\tilde k$ to parameterize the spatial curvature.

The non-zero components of the five-dimensional Einstein tensor are
\begin{eqnarray}
{G}_{00} &=& 3\left\{ \fda \left( \fda+ \fdb \right) - \frac{n^2}{b^2}
\left(\fppa + \fpa \left( \fpa - \fpb \right) \right) + \tilde k \frac{n^2}{a^2} \right\},
\label{ein00} \\
 {G}_{\ii\jj} &=&
\frac{a^2}{b^2} \gamma_{ij}\left\{\fpa
\left(\fpa+2\fpn\right)-\fpb\left(\fpn+2\fpa\right)
+2\fppa+\fppn\right\}
\nonumber \\
& &+\frac{a^2}{n^2} \gamma_{ij} \left\{ \fda \left(-\fda+2\fdn\right)-2\fdda
+ \fdb \left(-2\fda + \fdn \right) - \fddb \right\} -\tilde k \gamma_{ij},
\label{einij} \\
{G}_{05} &=&  3\left(\fpn \fda + \fpa \fdb - \frac{\dot{a}^{\prime}}{a}
 \right),
\label{ein05} \\
{G}_{55} &=& 3\left\{ \fpa \left(\fpa+\fpn \right) - \frac{b^2}{n^2}
\left(\fda \left(\fda-\fdn \right) + \fdda\right) - \tilde k \frac{b^2}{a^2}\right\}.
\label{ein55}
\end{eqnarray}
Primes indicate derivatives with respect to
$z$, while dots derivatives with respect to $t$.

The five-dimensional Einstein equations take the usual form
\be
G_{AC}
= \frac{1}{2 M^3} T_{AC} \;,
\label{einstein}
\ee
where $T_{AC}$ denotes the total energy-momentum tensor.

Assuming a perfect fluid on the brane and, possibly an additional energy-momentum
$T^A_C|_{m,B}$ in the bulk, we write
\begin{eqnarray}
T^A_{~C}&=&
\left. T^A_{~C}\right|_{{\rm v},b}
+\left. T^A_{~C}\right|_{m,b}
+\left. T^A_{~C}\right|_{{\rm v},B}
+\left. T^A_{~C}\right|_{m,B}
\label{tmn1} \\
\left. T^A_{~C}\right|_{{\rm v},b}&=&
\frac{\delta(z)}{b}{\rm diag}(-V,-V,-V,-V,0)
\label{tmn2} \\
\left. T^A_{~C}\right|_{{\rm v},B}&=&
{\rm diag}(-\Lambda,-\Lambda,-\Lambda,-\Lambda,-\Lambda)
\label{tmn3} \\
\left. T^A_{~C}\right|_{{\rm m},b}&=&
\frac{\delta(z)}{b}{\rm diag}(-\tilde\rho,\tilde p,\tilde p,\tilde p,0),
\label{tmn4}
\end{eqnarray}
where $\tilde \rho$ and $\tilde p$ are the energy density and pressure on the brane, respectively.
The
behavior of $T^A_C|_{m,B}$ is in general complicated in the presence
of flows, but we will not  specify it further at this point.

We wish to solve the Einstein equations at the location
of the brane following \cite{betal1}-\cite{betal3}. We indicate by the subscript $o$ the value of
various quantities on the brane.
Integrating equations (\ref{ein00}), (\ref{einij})
with respect to $z$ around $z=0$, enforcing also the $Z_2$ symmetry,  gives the
jump conditions
\begin{eqnarray}
a_{o^+}'=-a_{o^-}'  &=& -\frac{1}{12M^3} b_o a_o \left( V +\tilde \rho \right)
\label{ap0} \\
n'_{o^+}=-n_{o^-}' &=&  \frac{1}{12M^3} b_o n_o \left(- V +2\tilde \rho +3 \tilde p\right).
\label{np0}
\end{eqnarray}

The other two Einstein equations (\ref{ein05}), (\ref{ein55})
give
\begin{equation}
\frac{n'_o}{n_o}\frac{\dot a_o}{a_o}
+\frac{a'_o}{a_o}\frac{\dot b_o}{b_o}
-\frac{\dot a'_o}{a_o} =
\frac{1}{6M^3}T_{05}
\label{la1}
\end{equation}
\begin{equation}
\frac{a'_o}{a_o}\left(
\frac{a'_o}{a_o}+\frac{n'_o}{n_o}\right)
-\frac{b^2_o}{n^2_o}\left(
\frac{\dot a_o}{a_o} \left( \frac{\dot a_o}{a_o}-\frac{\dot n_o}{n_o}\right)
+\frac{\ddot a_o}{a_o}\right)
-\tilde k\frac{b^2_o}{a^2_o} =-\frac{1}{6M^3}\Lambda b^2_o
+ \frac{1}{6M^3}T_{55},
\label{la2}
\end{equation}
where $T_{05}, T_{55}$ are the $05$ and $55$ components of $T_{AC}|_{m,B}$
evaluated on the brane.
Substituting (\ref{ap0}), (\ref{np0})
in equations (\ref{la1}), (\ref{la2}) one obtains
\begin{equation}
\dot {\tilde\rho} + 3 \frac{\dot a_o}{a_o} (\tilde\rho +\tilde p)
= -\frac{2n^2_o}{b_o}
T^0_{~5}
\label{la3}
\end{equation}
\begin{eqnarray}
\frac{1}{n^2_o} \Biggl(
\frac{\ddot a_o}{a_o}
+\left( \frac{\dot a_o}{a_o} \right)^2
&-&\frac{\dot a_o}{a_o}\frac{\dot n_o}{n_o}\Biggr)
+\frac{\tilde k}{a^2_o}
=\frac{1}{6M^3} \Bigl(\Lambda + \frac{1}{12M^3} V^2
\Bigr)
\nonumber \\
&-&\frac{1}{144 M^6} \left(
V (3\tilde p-\tilde \rho ) +\tilde \rho (3\tilde p +\rho)
\right)
- \frac{1}{6M^3}T^5_{~5}.
\label{la4}
\end{eqnarray}

In the model that reduces to the Randall-Sundrum
vacuum \cite{rs2} in the absence of matter, the first
term on the right hand side of equation (\ref{la4}) vanishes. This is the effective cosmological constant on the brane.
A new mass scale $K$
may be  defined through the relations
$V=-\Lambda/K=12M^3 K$. This is the inverse characteristic length
scale of AdS. Sometimes, we may keep the effective cosmological
constant non-zero.

At this point we find it convenient to employ a coordinate frame in
which $b_o=n_o=1$ in the above equations. This can be achieved by using Gauss
normal coordinates with $b(t,z)=1$, and by going to the temporal gauge on the
brane with $n_o=1.$ The assumptions for the form of the energy-momentum
tensor are then specific to this frame.
Using $\beta\equiv M^{-6}/144$ and $\gamma\equiv V M^{-6}/144$,
and omitting the subscript o for convenience in the following, we rewrite
equations (\ref{la3}) and (\ref{la4}) in the equivalent first order form
\be
\dot{\tilde\rho}+3(1+w)\,{{\dot a}\over a} \,\tilde \rho = -\tilde
T\sp
{{{\dot a}^2}\over {a^2}}=\beta\tilde\rho^2+2\gamma (\tilde\rho+\tilde\chi) -
{\tilde k\over{a^2}}+\tilde \lambda
\label{aa}
\ee
\be
\dot{\tilde\chi}+4\,{{\dot a}\over
a}\,\tilde\chi=\left({\beta\over \gamma}\tilde \rho+1\right)\tilde T-{1\over 6\gamma~M^3}{{\dot a}\over
a}{T^5}_5,
\label{chi}
\ee
where $\tilde p=w\tilde \rho$,
$\tilde T=2T^0_{~5}$ is the discontinuity of the zero-five component of the bulk
energy-momentum tensor,
and $\tilde \lambda=(\Lambda+V^2/12M^3)/12M^3$ the effective
cosmological constant on the brane.

In the equations above, Eq. (\ref{aa}) is the $definition$ of the auxiliary density $\tilde\chi$.
With this definition, the other two equations are equivalent to the original system
(\ref{la3},\ref{la4}).

At this point we will specialize to the pure RS cosmology where
the only field that lives in the bulk is the metric, and thus
$\tilde T={T^5}_5=0$.
Then the cosmological equations on the brane (\ref{aa}), (\ref{chi}) become
\be
\dot{\tilde\rho}+3(1+w)\,{{\dot a}\over a} \,\tilde \rho =0\sp
\dot{\tilde\chi}+4\,{{\dot a}\over
a}\,\tilde\chi=0
\label{mat}
\ee
\be
{{{\dot a}^2}\over {a^2}}=\beta\tilde\rho^2+2\gamma (\tilde\rho+\tilde\chi) -
{\tilde k\over{a^2}}+\tilde \lambda
\label{frw}
\ee

We define the dimensionless quantities
\be \rho={\tilde\rho\over 2 M^3K}\sp \chi={\tilde\chi\over 2 M^3K}\sp \lambda
=144{\tilde\lambda \over K^2}={12\Lambda\over M^3K^2}+1 \label{rescale}\ee
as well as a rescaled cosmological time $\tau =K~t/6$. The combination $\tilde k/K^2$ is then dimensionless
and by scaling $a$ we can set it to
$ k=0,\pm 1$.
Then the cosmological equations
become
\be
\dot{\rho}+3(1+w)\,{{\dot a}\over a} \, \rho = 0\sp \dot\chi+4\,{{\dot a}\over
a}\,\chi=0
\label{rho1}
\ee
\be
{{{\dot a}^2}\over {a^2}}=\rho^2+\rho+\chi -
{k\over{a^2}}+\lambda
\label{a1}
\ee
where now dots stand for derivatives with respect to $\tau$.
\bigskip

There are two effective energy densities that drive the
cosmological evolution on the brane. The first, $\rho$, is the
localized energy density on the brane, and it is conserved.
The second, $\chi$ behaves like radiation, and although it affects
the cosmological evolution on the brane, it is rooted in the bulk
dynamics. It is the projection on the brane, of the bulk Weyl
tensor \cite{shiro}. It is an example of a "mirage" energy
density. It is also known as dark radiation.

An extra ingredient of the cosmological evolution is the $\rho^2$
term in the effective Friedman equation.

For $\rho<< 1$ which in dimensionfull units amounts to the localized energy density
being negligible compared to the brane tension,
 $\tilde
\rho<< V$, the $\rho^2$ term can be neglected and the cosmological
evolution is four-dimensional , with only extra ingredient the
mirage radiation term.

However, for $\rho>>1$ the $\rho^2$ term dominates and the
evolution is different $H\sim \rho$, and the universe slows down.
This "five-dimensional" region is analogous to the short-distance
region in the static RS case in section \ref{RS}.

There a complementary way of looking at this cosmological
evolution \cite{krauss} that is related to the one above by a
change of coordinates. In this case the bulk metric is that of a
Schwartzild-AdS black-hole in five dimensions, while now the brane
is moving in a geodesic. This is in fact very similar to the
mirage cosmology idea to be discussed in a subsequent section.

\subsection{How Brane-induced gravity affects the cosmological
evolution}
\setcounter{equation}{0}

The Friedman equations for the RS cosmology, \cite{betal1}-\cite{betal3}, as modified by
the induced Einstein term,
were derived in \cite{def,def2}.
The cosmology has been analyzed in different contexts in  \cite{dvg1,dvg2,irs}.

We add an induced Einstein term $M^3r_c\int d^4 \xi \sqrt{\hat g} ~ \hat
R$ to (\ref{001}).
The equation (\ref{einstein}) is modified by a localized
(four-dimensional) Einstein tensor on the right hand side. This
changes the Israel (jump) conditions (\ref{ap0},\ref{np0}) to
\begin{eqnarray}
a_{o^+}'=-a_{o^-}'  &=& -\frac{1}{12M^3} b_o a_o \left( V +\tilde \rho
\right)+{r_c\over 2}{a_ob_o\over n_o^2}\left({\dot a_o^2\over
a_o^2}+k{n_o^2\over a_o^2}\right)
\label{ap1} \\
n'_{o^+}=-n_{o^-}' &=&  \frac{b_o n_o}{12M^3}  \left(2\tilde \rho +3
 \tilde p-V\right)+{r_c b_o\over 2n_o}\left(2{\ddot a_o\over a_o}-{\dot a_o^2\over
a_o^2}-2{\dot a_o\dot n_o\over a_on_o}-k{n_o^2\over a_o^2}\right)
\label{np1}
\end{eqnarray}

Using (\ref{ap1},\ref{np1}) in the 05 and 55 equations, integrating once, and solving the quadratic equation for $H^2$ we obtain

\be
{r_c^2\over 2}\left(H^2+{k\over a^2}\right)
=1+{r_c\over 12M^3}(\tilde\rho+V)+\epsilon
\sqrt{1+{r_c\over 6M^3}(\tilde\rho+V)-{r_c^2\over 12 M^3}\Lambda-r_c^2\chi}.
\label{friedman1}
\ee
where $\chi$ is the mirage radiation density related to the value of
the five-dimensional Weyl tensor, and $\epsilon=\pm 1$ defining two possible branches
in the solution.

For simplicity, we shall assume for the time being, that there in no significant
flow of energy out of the brane through the decay of brane matter into KK modes of
the graviton \footnote{It will be shown later that this assumption is correct in periods of cosmological evolution where
the Friedman equation is approximately four dimensional.}.
Under this assumption, the energy density $\rho_b$ on the brane
satisfies the conservation equation
\be
\dot{\tilde\rho}=-3H(\tilde\rho+\tilde p).
\ee

We are studying here the effects induced by the presence
of brane matter and radiation on the
RS "vacuum" background (\ref{m1},\ref{m2}) with a given value of $K$,
i.e. with the parameters $\Lambda$ and $V$ of the theory satisfying
$V=-\Lambda/K=12M^3K$.
Using the above values for $\Lambda$ and $V$, (\ref{friedman1})
takes the form
\be
{r_c^2\over 2}\left(H^2+{k\over a^2}\right)
=1+Kr_c+{r_c\over 12M^3}\tilde\rho+\epsilon
\sqrt{(1+Kr_c)^2+{r_c\over 6M^3}\tilde\rho-r_c^2\chi}.
\label{friedman2}
\ee
The equation with $\epsilon=-1$ has a smooth limit as $r_c\to 0$
and gives the cosmological evolution
of a RS universe (\ref{frw})
\be
H^2={V\tilde\rho\over 72M^6}+{1\over 4}\left({\tilde\rho\over 6M^3}\right)^2
+\chi-{k\over a^2}.
\ee
On the contrary, the limit with vanishing bulk and brane vacuum energy
is given by $K\to 0$
\be
{r_c^2\over 2}\left(H^2+{k\over a^2}\right)=1+{r_c\over 12M^3}\tilde\rho
+\epsilon\sqrt{1+
{r_c\over 6M^3}\tilde\rho-r_c^2\chi}
\ee
and provides an effective vacuum energy \cite{dvg1,dvg2} when $\e=1$.
From now on, for simplicity,  we shall set the ``mirage" radiation density to zero: $\chi=0$.

In analyzing the physical content of (\ref{friedman2}) we shall distinguish
two cases.

\bigskip

{\bf (a)} {\bf Strong BIG ~:~ $Kr_c\gg 1$}. For the static
gravitational potential
this corresponds
to four-dimensional
behavior on the brane at all scales with $M_{\rm P}^2=M^3r_c$.
We define the dimensionless density $\hat\rho\equiv {r_c\tilde\rho/ M^3}$. We
obtain
\be
H^2\simeq {\tilde\rho\over 6\mpl^2}-{k\over a^2}\sp \hat\rho\gta
(Kr_c)^2
\label{kkrr1}\ee
\be
H^2\simeq {\tilde\rho\over 6\mpl^2}-{k\over a^2}+(\epsilon+1){2k
\over r_c}\sp \hat\rho<<
(Kr_c)^2
\label{kkrr2}\ee
Thus, for the $\epsilon=-1$ branch we obtain at all times the
standard Friedman equation.
For $\epsilon=1$ there is at late times
a "mirage" vacuum energy $\Lambda_4={24K\mpl^2/ r_c}$ \cite{dvg1,dvg2}.

\bigskip
{\bf (b)} {\bf Weak BIG ~:~ $Kr_c\ll 1$}.
The static gravitational potential on the brane is four-dimensional
at energies $E\ll K$ with $\tilde \mpl^2=M^3/K$ (RS regime),
and for $E\gg 1/r_c$ with $\mpl^2=M^3r_c$ (Induced Gravity (IG) regime).
At intermediate energies $K\ll E\ll 1/r_c$ gravity is five-dimensional
($5d$ regime).

\bigskip

The Friedman equation now behaves as
\be
H^2\simeq {\tilde\rho\over 6\mpl^2}-{k\over a^2}\sp \hat\rho\gg 1
\ee
corresponding to the IG regime.
and
\be
H^2\simeq (\epsilon+1+Kr_c){\tilde\rho\over 6M^3r_c}
-{\epsilon \over 4} \left( {\tilde\rho \over 6 M^3} \right)^2
-{k\over a^2}
+{2(\epsilon+1)\over r_c^2}\sp \hat\rho \ll 1
\ee

For $\epsilon=1$ this expression becomes
\be
H^2\simeq {\tilde\rho\over 3\mpl^2}-{k\over a^2}+{4(1+Kr_c)\over r_c^2},
\ee
indicating the late-time vacuum energy.
For $\epsilon = -1$, we recover the cosmology of an RS universe
\be
H^2\simeq {\tilde\rho\over 6\tilde \mpl^2}
+{1 \over 4} \left( {\tilde\rho \over 6 \tilde \mpl^2 K} \right)^2
-{k\over a^2}.
\label{5d}
\ee

We thus confirm that the rough cosmological evolution mimics the static
behavior of gravity on the brane.
There are further generalizations where the Gauss-Bonnet term is
added in the bulk, \cite{gb1}-\cite{gb8}which  will not be discussed
here.

\subsection{Brane-bulk energy exchange\label{exchange}}
\setcounter{equation}{0}

In our discussion so far we have neglected a potentially  important factor: the
energy that is radiated off the brane and into the bulk in the form of KK gravitons.
This can be studied as follows \cite{irs}: from the analysis of
the previous section we can compute the cross-section for the
emission of KK gravitons. This can be convoluted with the
appropriate matter densities to provide the rate of energy loss per unit
mass.
This rate can be integrated to provide the full rate of energy
loss as
a function of time. The rate of energy loss  can then be compared with the dilution of
the brane energy density due to the expansion.
We will describe  this calculation below.

To simplify the calculations we will have to
distinguish several cases.

{\bf (a)Strong BIG : $Kr_c \gg 1$}. This is the case discussed earlier in section
\ref{rsbig}. Here gravity is four dimensional at all distances
and
hence there are no stringent experimental constraints.
All KK modes are significantly
suppressed and do not affect standard processes. For example,
the rate of emission of KK modes from a star can be estimated
as \cite{sav2}
\be
\Gamma(T) \propto \frac{1}{M^3} \int_0^T dm\, \psi(0)^2
\sim \frac{1}{M^3} \int_0^T dm \frac{m}{K} \frac{1}{r_c^2K^2}
\sim \frac{1}{\mpl^2} \frac{T^2}{K^2} \frac{1}{r_cK}
\label{rate1}
\ee
for $T <K$. This is much smaller than the rate of production
of zero-mode gravitons $\Gamma_0(T) \propto 1/\mpl^2$, and thus it is negligible.
For $T > K$ the largest contribution to the rate is
\be
\Gamma(T) \propto
\frac{1}{M^3} \int_K^T dm \frac{1}{m^2r_c^2}
\sim \frac{1}{\mpl^2} \frac{1}{r_cK},
\label{rate2}
\ee
and is negligible again. Thus, we do not expect  severe constraints on such a case from current
astrophysical data, like supernovae.
The only  requirement is to reproduce the
value of the Planck constant $\mpl^2 =M^3 r_c$.

The cosmological evolution here turns out to be standard for
all densities. There is a small amount of energy loss to the
bulk but as we will show shortly it does not affect the Friedman evolution (\ref{kkrr1}), (\ref{kkrr2})
we found already in the previous section.

The change in energy density
per unit time is equal to the
rate of energy loss to KK gravitons per unit time and volume.
For a process $a+b \to c + KK$ it is given by
\be
\left( \frac{d\rho}{dt} \right)_{\rm lost}
= - \left\langle n_a \, n_b\, \sigma_{a+b \to c + KK}\, v \,E_{KK}
\right\rangle,
\label{energyloss1} \ee
where the brackets indicate thermal averaging.
For a radiation-dominated brane
we can take approximately $n_a,n_b\sim T^3$, $E_{KK}\sim T$,
and estimate
\be
\left\langle \sigma_{a+b \to c + KK}\, v \right\rangle
\sim \frac{1}{M^3} \int_K^T dm \frac{1}{m^2r_c^2}
\sim \frac{1}{\mpl^2} \frac{1}{r_cK},
\label{sigmavt} \ee
in agreement with eq. (\ref{rate2}).
This leads to
\be
\left( \frac{d\rho}{dt} \right)_{\rm lost}
\sim - \frac{T^7}{\mpl^2} \frac{1}{r_c K}
\sim - \frac{\rho^{7/4}}{\mpl^2} \frac{1}{r_c K}.
\label{enloss} \ee
We conclude that the energy loss is negligible because
$(d\rho/dt)_{\rm expansion}/(d\rho/dt)_{\rm lost} \sim
(\mpl/\rho^{1/4})r_cK\gg 1$. Thus,
we obtain a standard Friedman cosmological expansion with essentially
no energy loss.

\bigskip

{\bf (b) Weak BIG:  $Kr_c \ll 1$}. Here, the deviations from the standard RS physics appear at
energy scales much larger than $K$. For the gravitational potential,
we expect a transition from the four-dimensional form $\sim 1/r$
to the five-dimensional one $\sim 1/r^2$ at distances $r\lta  1/K$.
The experimental constraints require $k \gta (10~\mu{\rm m})^{-1} \simeq
10^{-11}$ GeV, while the value of $M$ is
fixed by the relation $\tilde \mpl^2 = M^3/K$ to be
$M\gta  10^9$ GeV.

The emission of KK modes with masses
$1/r_c \gta m \gta K$ is unsuppressed on the brane. Their contribution to
various processes, such as star cooling or high-energy experiments,
is analogous to those in standard toroidal compactifications,
\cite{sav2} with one extra dimension and a compactification radius $\sim 1/K$.
The strongest constraints arise from star cooling through the emission of
KK modes.

The cosmology of this scenario has several novel features, as was indicated
in the previous section.
For densities $\rho \lta M^3 K$ one expects the standard cosmological
evolution with $H^2 \sim \tilde\rho/{\tilde \mpl^2}$ (RS regime).
For $K$, $M$ near the lower
bound set by observations $K \sim 10^{-11}$, $M\sim 10^9$ GeV, this
regime extends up to densities $\tilde\rho \sim (10\,{\rm TeV})^4$.
However, for
$M^3 K \lta \tilde\rho \lta M^3/r_c $ the Hubble parameter behaves
$H^2 \sim \tilde\rho^2/M^6$ ($5d$ regime), while for $\tilde\rho \gta M^3/r_c$ we have
$H^2 \sim \tilde\rho/\mpl^2$ (IG regime).

Unsuppressed emission of single KK gravitons
can take place for the mass range $1/r_c> m> K$ as shown in section \ref{energyloss}.
For a brane with energy density $\rho \gta K^4$ it is
possible to produce such unsuppressed KK gravitons
that escape into the bulk.
We concentrate on the case of a radiation-dominated brane-universe
($\tilde\rho\sim T^4$), which is the most relevant for the energy scales of
interest. The scale
$K^4$ is smaller than $M^3K$ because we assume
$K\ll M$ (otherwise the
whole energy regime above $K$ is strongly coupled).
We also assume $Mr_c \gg 1$ (otherwise
induced-gravity effects are masked by strong five-dimensional gravity).

%\vspace{0.5cm}
\begin{figure}[htb]
\centering
\epsfxsize=5.7in
\epsfysize=1.5in
%\hspace{1.5cm}
\epsffile{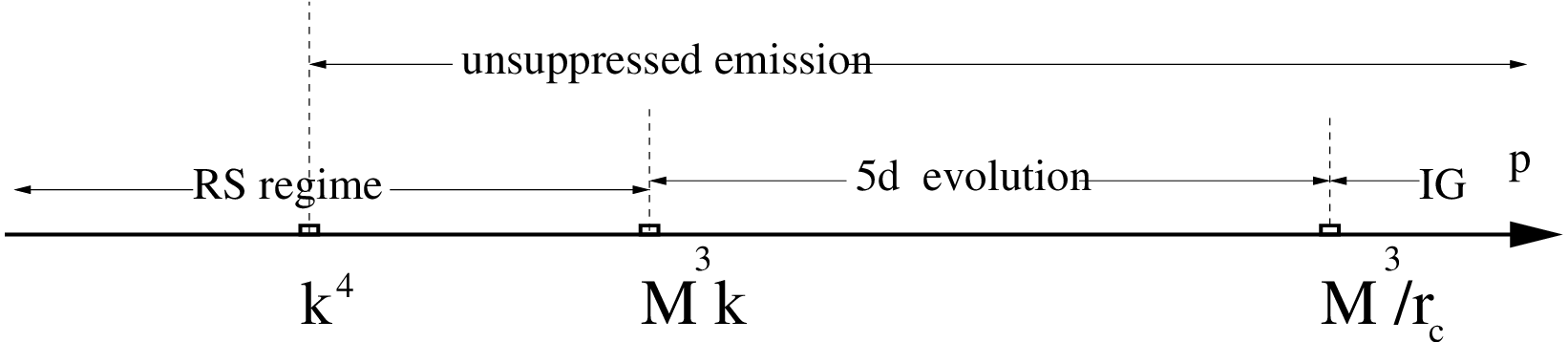}
\vspace{0.0cm}
\caption{Different regions in density, during the evolution of the universe for $Kr_c<<1$.
}
\label{sin1}
\end{figure}
%\vspace{0.5cm}

The energy lost through emission of unsuppressed KK gravitons is given by
eq. (\ref{energyloss1}). We estimate
\be
\left\langle \sigma_{a+b \to c + KK}\, v \right\rangle
\sim \frac{1}{M^3} \int_k^{\min(T,1/r_c)} dm = \frac{\min(T,1/r_c)}{M^3}.
\label{sigmavt2} \ee
For a given temperature $T$, the energy loss is maximized if $1/r_c > T$.
We concentrate on this case in the following and find
\be
\left( \frac{d\rho}{dt} \right)_{\rm lost}
\sim - \frac{T^8}{M^3} \sim - \frac{\rho^2}{M^3}.
\label{energyloss} \ee
The change in energy density because of the expansion is
\be
\left( \frac{d\rho}{dt} \right)_{\rm exp}
= -4 H \rho.
\label{expansion} \ee

In the RS regime
\be
{(d\rho/dt)_{\rm expansion}\over (d\rho/dt)_{\rm lost}} \sim \sqrt{M^3k\over \rho}\gg 1\ee

Thus, the energy loss during this period is negligible compared
to the dilution due to the expansion.
In particular, for $M^3 K \sim (10\,{\rm TeV})^4$, the energy loss during
nucleosynthesis is smaller by a factor $10^{14}$ than the rate of decrease
of the
energy density because of the expansion. Thus, during this period
the standard cosmological
evolution  is not affected by energy loss.
A differed way of saying this is that graviton emission is frozen-out
during the RS period.

The energy loss is substantial during the $5d$ regime, when
\be
(d\rho/dt)_{\rm exp} \sim -\rho^2/M^3\ee

Both $(d\rho/dt)_{\rm exp}$ and $(d\rho/dt)_{\rm lost}$ are of the same
order of magnitude and both lead to a decrease of the energy density for an expanding universe.
Thus, it must be included in the cosmological evolution. This can
be done using the metric of a radiating brane (Vadya metric)
\cite{lsorbo2},\cite{vadya}. We will come back to it in section \ref{va}.
The energy exchange between the brane and the bulk has been
studied in \cite{outf1}-\cite{lsorbo2}, \cite{kkttz1}-\cite{kkttz4}.

\subsection{Phenomenology of brane-bulk energy exchange and its
 impact on brane cosmology}
\setcounter{equation}{0}

We have seen in the previous section that energy exchange between the brane
and the bulk, sometimes can modify significantly the Friedman
cosmological evolution on the brane.
It is interesting to try to investigate various patterns of such
brane bulk energy exchange and its impact on the brane-world
cosmology.

We will thus reconsider the general cosmological evolution
equations derived in section \ref{crs}.

Rescaling the variables in the basic equations (\ref{aa}),
(\ref{chi}) as in (\ref{rescale}) as well as
\be
T={\tilde T\over 72 M^3K^2}\sp T_5={{ T^5}_5\over 3M^3K^2}
\ee
we obtain the general equations that describe the cosmological
evolution of the brane-world and its interaction with the bulk
\be
\dot{\rho}+3(1+w)\,{{\dot a}\over a} \, \rho = -T
\label{rho1}
\ee
\be
{{{\dot a}^2}\over {a^2}}=\rho^2+\rho+\chi -
{k\over{a^2}}+\lambda
\label{a1}
\ee
\be
\dot\chi+4\,{{\dot a}\over
a}\,\chi=\left(2\rho+1\right) T-{{\dot a}\over
a}T_5,
\label{chi1}
\ee
\be
q={\ddot a\over a}=-(2+3w) \rho^2-{3w+1\over
2}\rho-\chi-{1\over 2}T_5+\lambda\label{acce}
\ee
where as usual we may assume an equation of state $p=w\rho$ for the
brane matter.

In (\ref{rho1}),  $T>0$ is the rate of energy outflow from the brane to the
bulk, while it corresponds to inflow when negative.
$T_5$ can be considered as the bulk back-reaction.

In order to derive solutions
that are largely independent of the bulk dynamics,
the $T_{~5}$ term on the
right hand side of the same equation must be
negligible relative to the second one.
This is possible if we assume that the diagonal elements of the various
contributions to the energy-momentum tensor satisfy the schematic
inequality \cite{kkttz1}-\cite{kkttz4}
\be
\left|
\frac{\left. T\right|^{\rm diag}_{{\rm m},B}}{
\left. T\right|^{\rm diag}_{{\rm v},B}}
\right|
\ll
\left|
\frac{\left. T\right|^{\rm diag}_{{\rm m},b}}{
\left. T\right|^{\rm diag}_{{\rm v},b}}
\right|.
\label{vacdom} \ee
Our assumption is that the bulk matter
relative to the bulk vacuum energy is much less important than
the brane matter relative to the brane vacuum energy, {\it at the position of the brane}. In this
case,
the bulk is largely unperturbed by the exchange of energy with the brane.
When the off-diagonal term $T^0_{~5}$ is of the same order of magnitude or
smaller than the diagonal ones, the inequality (\ref{vacdom}) implies
$T\ll\rho $. Thus, under this assumption, we may neglect $T_5$ from
(\ref{chi1},\ref{acce}).

Even by neglecting $T_5$ from the equations, we do not obtain a
closed system. We still have $T$ to worry about.
In the previous section, we have calculated $T$ for the RS case
and graviton radiation in the bulk. We have found that for periods
of the evolution, $T$ is a power  of the driving energy density $\rho$ on the brane.
This is true for a large class of processes. Imagine that
particles of the driving energy density of the brane universe (the
one that dominates and drives the cosmological evolution) scatter
among themselves and radiate some bulk particles via some
interaction. The rate of energy loss will depend on the energy
density, on cosmological temperature T and several particle physics parameters.
using the cosmological evolution one can trade T with $\rho$:
$\rho\sim T^{3(1+w)}$ and eventually write $T$ as a function of
$\rho$ and other fundamental constants. If the theory on the
brane is conformally invariant then $T(\rho)$ behaves like a
power. Otherwise there will be scaling violations.

Motivated by this , we will parameterize $T$ as a function of the driving energy density $\rho$.
The precise form depends on the details of the interaction between
brane and bulk. In the scaling regime, $T$ will typically have a
power dependence $T\sim \rho^{\nu}$ up to  small corrections. We will use this to
investigate various physical instances in the sequel.

\renewcommand{\theequation}{\arabic{section}.\arabic{subsection}.\arabic{subsubsection}.\arabic{equation}}
\subsubsection{The cosmological equations for brane-bulk energy exchange}.

\setcounter{equation}{0}

With $T(\rho)$ a known function of $\rho$, and the approximation $T_5\to negligible$ is valid, the system of equations

\be
\dot{\rho}+3(1+w)\,{{\dot a}\over a} \, \rho = -T(\rho)\sp \dot\chi+4\,{{\dot a}\over
a}\,\chi=\left(2\rho+1\right) T(\rho)\sp {{{\dot a}^2}\over {a^2}}=\rho^2+\rho+\chi -
{k\over{a^2}}+\lambda
\label{eqs}
\ee
is self-contained and we can attempt to find solutions.
An interesting parameter of the cosmological evolution is the
acceleration that can be evaluated as
\be
q={\ddot a\over a}=-(2+3w) \rho^2-{3w+1\over
2}\rho-\chi+\lambda
\ee

Combining equations (\ref{eqs})  we obtain
\be
a{{d\rho}\over da}=-3(1+w)\rho-
\epsilon\,T(\rho) \left(\rho^2+\rho+\chi -
{ k\over{a^2}}+\lambda\right)^{-1/2}.
\label{rho(a)}
\ee
Similarly,
\be
a{{d\chi}\over da}=-4\chi+\epsilon \left(2\rho+1\right)
T(\rho)
\left(\rho^2+\rho+\chi -
{k\over{a^2}}+\lambda\right)^{-1/2},
\label{chi(a)}
\ee
where $\epsilon =1$ refers to expansion, while $\epsilon=-1$ to contraction.
These two equations form a two-dimensional dynamical system. The function
$\chi(\rho)$ is obtained from the equation
\begin{eqnarray}
\Biggl(3(1+w)\rho
\sqrt{\rho^2+\rho+\chi -
{k\over{a^2}}+\lambda}&+&\epsilon\,T(\rho)
\Biggr)\,
{{d\chi}\over{d\rho}}
\nonumber \\
=4\chi\sqrt{\rho^2+\rho+\chi -
{k\over{a^2}}+\lambda}
&-&\epsilon \left(2\rho+1\right) T(\rho).
\label{chi(rho)}
\end{eqnarray}
Note that the equations of contraction are those of expansion with the
roles of outflow and influx interchanged.

We will  now set  the spatial curvature on the brane-world $k=0$ and define

\be\hat T=
\epsilon \rho^{-{3/2}} T\sp \zeta=\sqrt{\rho
+{\chi\over\rho}+1+{\lambda\over \rho}}\ee
Then, the equation (\ref{chi(rho)}) becomes

\be
2\rho\zeta '={(1-3w)\zeta^2-\hat
T\zeta+\left[3w-1+2(1+3w)\rho-4{\lambda\over
\rho}\right]\over 3(1+w)\zeta+\hat T}
\ee
while (\ref{rho(a)4}) becomes
\be
a{d\log \rho\over da}=-3(1+w)-{\hat T\over \zeta}
\ee

We can also derive an equation for the acceleration with the observable energy density $\rho$ as the independent
variable:
\be
-{dq\over d\rho}={\left[2(1+3w)(2+3w)\rho^2+{9w^2-1\over
2}\rho+4\lambda-4q\right]\over
3(1+w)\rho R+\epsilon T}R+
\label{ac}\ee
$$
+{2(3w+1)\rho+{3w-1\over 2}+{1\over 2}{\partial T_5\over \partial \rho}\over
3(1+w)\rho R+\epsilon T}\epsilon T
$$
where
\be
R=\sqrt{{1-3w\over 2}\rho-(1+3w)\rho^2+2\lambda-q-{k\over a^2}-{T_5\over 2}}
\ee

For $k=0$, equation (\ref{ac}) simplifies to

\be
2R'={4R^2-4\lambda+(3w-1)\rho+2(3w+1)\rho^2\over
3(1+w)\rho R+\epsilon T}
\ee

\subsubsection{The four-dimensional regime}
In this regime $\rho<<1$ and the cosmological evolution in the
absence of energy exchange is given by the  standard Friedman
equation.
Equations (\ref{eqs}) linearize to
\be
\dot{\rho}+3(1+w)\,{{\dot a}\over a} \, \rho = -T\sp
{{{\dot a}^2}\over {a^2}}=\rho+\chi -
{k\over{a^2}}+\lambda\sp
\dot\chi+4\,{{\dot a}\over
a}\,\chi=T
\label{chi11}
\ee
while the acceleration is
\be
q={\ddot a\over a}=-{3w+1\over
2}\rho-\chi+\lambda
\ee

An inspection of these equations reveals a simple interpretation.
There two relevant energy densities appearing in the Friedman
equation (apart from a potential vacuum energy and curvature): the
energy $\rho$ of brane matter, and the mirage energy density (or
dark radiation)
$\chi$ reflecting the bulk dynamics.
The function $T$ is responsible for the conversion of energy from
$\rho$ to $\chi$ or vice-versa while the total energy $\rho+\chi$ is
conserved:
\be
\dot\rho+\dot\chi+3H\left[(1+w)\rho+{4\over 3}\chi\right]=0
\ee

We can manipulate further the linearized equations to
\be
a{{d\rho}\over da}=-3(1+w)\rho-
\epsilon\,T(\rho) \left(\rho+\chi -
{ k\over{a^2}}+\lambda\right)^{-1/2}.
\label{rho(a)4}
\ee
\be
a{{d\chi}\over da}=-4\chi+\epsilon
T(\rho)
\left(\rho+\chi -
{k\over{a^2}}+\lambda\right)^{-1/2},
\label{chi(a)4}
\ee

\begin{eqnarray}
\Biggl(3(1+w)\rho
\sqrt{\rho+\chi -
{k\over{a^2}}+\lambda}&+&\epsilon\,T(\rho)
\Biggr)\,
{{d\chi}\over{d\rho}}
\nonumber \\
=4\chi\sqrt{\rho+\chi -
{k\over{a^2}}+\lambda}
&-&\epsilon  T(\rho).
\label{chi(rho)4}
\end{eqnarray}

If we now $k=0$ and define $\hat T=
\rho^{-{3/2}} T$ and $\zeta=\sqrt{{\chi\over\rho}+1+{\lambda\over \rho}}$
then
\be
2\rho\zeta'={(1-3w)(\zeta^2-1)-\hat
T\zeta-4{\lambda\over \rho}\over
3(1+w)\zeta+\hat T}
\label{zeta}\ee

\be
a{d\log \rho\over da}=-3(1+w)-{\hat T\over \zeta}
\label{rr}\ee

\subsubsection{The exact solution for RS outflow \label{va}}
\setcounter{equation}{0}

Before we move on to find interesting solutions to our
(approximate) cosmological equations describing energy exchange
between the brane and the bulk, we will investigate an exact
solution for the energy outflow from the brane in the simple RS
case, that we treated perturbatively in section \ref{exchange}.

This was given in \cite{lsorbo2,vadya} and corresponds to setting in our general equations
$T=a\rho^2$ and $T_5=2a\rho^2$ where $a$  is a positive dimensionless constant .

We will thus consider the full system of equations (we choose
$w=1/3$ for concreteness, so that the leading energy density on
the brane is radiation, and a flat brane, $k=0$ without left-over cosmological constant, $\lambda=0$).
We also allow (for more generality)
$T=a\rho^2$ and $T_5=b\rho^2$. $b=2a$ corresponds to a radiating
RS brane.

The full system of equations are:

\be
\dot \rho+4H\rho=-a\rho^2\sp
\dot\chi+4H\chi=a(2\rho+1)\rho^2-bH\rho^2\sp H^2=\rho^2+\rho+\chi
\ee

The linear combination $z=H+\kappa \rho$
with $4\kappa=a-\sqrt{16+4b+a^2}$ satisfies the simple equation
\be
\dot z=-2 z^2\to  z={1\over 2(t-t_0)}
\ee

The general solution is given by
\be
\rho={1\over A(t-t_0)(t-t_1)}\sp t_1=t_0+{a-4\kappa\over A}
\label{solv1}\ee
\be
\chi={(\kappa^2-1)+{A^2\over 4}(t-t_1)^2-A\kappa(t-t_1)-A(t-t_0)(t-t_1)\over A^2(t-t_0)^2(t-t_1)^2}
\ee
\be
a(t)\sim (t-t_0)^{{a+\sqrt{16+4b+a^2}\over 4\sqrt{16+4b+a^2}}}(t-t_1)^{{\sqrt{16+4b+a^2}-a\over 4\sqrt{16+4b+a^2}}}
\label{solv3}\ee

From this exact solution we can study the approximation we have
made earlier when we dropped the $T_5$ term:  $HT_5<<(2\r+1)T$.
To do this we define the ratio

\be
\eta\equiv {HT_5\over (2\r+1)T}={b\over a}{H\over (2\r+1)}={b\over a}{A(t-t_1)-2\kappa\over 2(2+A(t-t_0)(t-t_1))}
\ee
As long as $\eta$ remains much smaller than one, our approximation
is justified.

The general solution obtained above \cite{vadya} described  three types of brane-cosmological
evolution:

\bigskip
{\bf
(I)} $A>0$ and $t<t_0$ describes a contracting universe with a big crunch at $t=t_0$ ($t_0<t_1$).
Here $\eta$ varies monotonically as

\be
0=\eta(-\infty)\geq \eta \geq \eta(t_0)=-{b(a+\sqrt{16+4b+a^2})\over 8a}
\ee
Here the mirage density $\chi$ is always positive if $A>4$ and switches sign at $t=a/(A-4)$ if $0<A<4$.

\bigskip

{\bf (II)} $A>0$ and $t>t_1$  describes an expanding universe with a big bang at $t=t_1$ ($t_0<t_1$).

\be
0=\eta(\infty)\leq \eta \leq \eta(t_1)={b(-a+\sqrt{16+4b+a^2})\over 8a}
\ee
Here $\chi$ can be always negative or switch sign .

\bigskip

{\bf (III)} $A<0$ and $t_1<t<t_0$ which describes an expanding and recontracting universe ($t_0> t_1$).

Here $\eta$ varies between $\eta(t_0)$ and $\eta(t_1)$. Since
$\eta(t_0)<0$ and $\eta(t_1)>0$, near the big bag or big crunch
$\eta$ is of order one while all the intermediate time is small

The overall message is that $\eta$ is always very small and can become of order unity
or larger near a big bang or a big crunch.
Moreover its value there is bounded by the coefficient of brane bulk energy exchange.

As mentioned above,  the realistic  case of RS radiation $b=2a$.
The solution (\ref{solv1})-(\ref{solv3}) simplifies to
\be
{a(t)\over a_0}= (t-t_0)^{{a+2\over 2(a+4)}}(t-t_1)^{1\over a+4}
\ee
\be
\rho={1\over A(t-t_0)(t-t_1)}\sp t_1=t_0+{a+4\over A}\sp H=\rho+{1\over 2(t-t_0)}
\ee

\be
\chi=H^2-\r-\rho^2={A(t-t_1)-4(t-t_0)+4\over
4A(t-t_0)^2(t-t_1)}
\ee

In the asymptotic 4d regime $t\to \infty$ we have
\be
\r\sim {1\over At^2}\sp \chi\sim {A-4\over 4A t^2}\sp H\sim {1\over 2t}
\ee
Thus A characterizes the asymptotic ratio of energy densities.

In the 5d regime $t\to t_1$ we have
\be
\rho\simeq H\sim {1\over (a+4)(t-t_1)}\sp \chi\sim {A-a-4\over (a+4)^2(t-t_1)}
\ee
We conclude that:

 In the late (4d) epoch $a\sim t^{1/2}$ while in the early (5d) one $a\sim t^{1/(a+4)}$.
Thus in the early epoch, outflow modifies the  $t^{1/4}$ behavior as expected from our arguments in the previous section.
Moreover, our approximation of dropping $T_5$ is good everywhere
except close to  a big crunch or a big bag.

Generally, if during the cosmological  evolution  $T_5$ remains of the same order of magnitude as $T$
for the whole period of evolution, then we have the following
estimate
\be
\eta\equiv {HT_5\over (2\r+1)T}\simeq {H\over (2\r+1)}={\sqrt{\r^2+\r+\chi+\lambda}\over 2\rho+1}
\ee

$\bullet$ If the vacuum energy is dominating the expansion then $\eta\to \sqrt{\lambda}$. If the vacuum energy is thus small
in natural units, $\eta<<1$

$\bullet$ In regions where $\chi$ can be neglected, $\eta$ is small when $\r<<1$ and can become at most $1/2$ for $\r$ large.

\subsubsection{Inflating fixed points}
\setcounter{equation}{0}

An interesting feature of the cosmological equations is the possible presence of
accelerating cosmological solutions. We may look for
exponential expansion with a constant Hubble parameter $H$, even if the
brane content is not pure vacuum energy.
We will restrict ourselves for simplicity to the 4d regime. For
the non-linear analysis we refer the reader to \cite{kkttz1}-\cite{kkttz4}.

For a  fixed point,
equations (\ref{chi11}) must have a time-independent
solution, without necessarily requiring $w=-1$ ($\lambda=k=0$).
The possible fixed points (denoted by $*$) of these equations satisfy
\begin{eqnarray}
3H_*(1+w)\rho_* = -T(\rho_*)
\sp
H^2_* =  \rho_* + \chi_*
\sp
4 H_* \chi_* =  T(\rho_*).
\label{fp}
\end{eqnarray}
It is clear from equation (\ref{fp}) that, for positive matter density
on the brane ($\rho >0$), flow of energy into the brane
($T(\rho)<0$) is necessary.

The accretion of energy from the bulk depends on the
dynamical mechanism that localizes particles on the brane.
Its details are outside the scope of our discussion. However, it is
not difficult to imagine scenaria that would lead to accretion.
If the brane initially has very low energy density,
energy can by transferred onto it by
bulk particles such as gravitons.
An equilibrium is expected to set in
if the brane energy density reaches a limiting value. As a result,
a physically  motivated behavior for the function
$T(\rho)$ is to be negative for small $\rho$ and cross zero towards positive
values for larger densities.
In the case of accretion it is also natural to expect that the energy
transfer approaches a negative constant value for $\rho \to 0$.

The solution of equations (\ref{fp})
satisfies
\be
T(\rho_*) = -\frac{3}{2} (1+w)\sqrt{1-3w}~\rho_*^{3/2}
\sp
H_*^2 = \frac{1-3w}{4}  \rho_*
\sp
\chi_* = - \frac{3(1+w)}{4}  \rho_*.
\label{sol1}\ee
For a general form of
$T(\rho)$ equation (\ref{sol1}) is an algebraic equation with
a discrete number of roots.
For any value of $w$ in the region
$-1<w<1/3$ a solution is possible.
The corresponding cosmological model has a scale factor that
grows exponentially with time. The energy density on the brane remains
constant due to the energy flow from the bulk.
This is very similar to the steady state model of cosmology
\cite{steady}. The main
differences are that the energy density is not spontaneously
generated, and the Hubble parameter receives an additional contribution from
the ``mirage'' density $\chi$ (see equation (\ref{fp})).

The stability of the fixed point
(\ref{fp}) determines whether the exponentially expanding
solution is an attractor of neighboring cosmological flows. If we
consider a small homogeneous perturbation around the fixed point
($\rho=\rho_*+\drho$, $\chi=\chi_*+\dchi$) we find that $\drho,\dchi$ satisfy
\be
\frac{d}{dt}
\left(
\begin{array}{c}
\drho \\ \dchi
\end{array}
\right)
=
\frac{T(\rho_*)}{\rho_*}
{\cal M}
\left(
\begin{array}{c}
\drho \\ \dchi
\end{array}
\right),
\label{pert} \ee
where
\begin{eqnarray}
{\cal M}&=&
\left(
\begin{array}{cc}
-\nu +3(1-w)/(1-3w)&~~~~~ 2(1-3w) \\
\nu-2/(1-3w) &~~~~~
-2(1+9w)/[3(1+w)(1-3w)]  \end{array}
\right)
\label{mat} \\
\nu &=& \frac{d\ln |T|}{d\ln\rho} \left(\rho_* \right),
\label{aaa}
\end{eqnarray}
and we have employed the relations (\ref{fp}) and $T(\rho) \propto \rho^\nu$.
The eigenvalues of the matrix $\cal M$ are
\be
M_{1,2}=\frac{
7+3w-3\nu (1+w)\pm\sqrt{
24(-3+2\nu)(1+w)+\left[7+3w-3\nu (1+w)\right]^2
}}{6(1+w)}.
\label{eigen} \ee
For $-1<w<1/3$, $0\leq\nu < 3/2$ they both have a positive real part. As we
have assumed  $T(\rho)<0$, the fixed point is stable in this case.
The approach to the fixed-point values depends on the sign of the quantity
under the square root. If this is negative the energy density oscillates
with diminishing amplitude around its fixed-point value.

\subsubsection{Tracking solutions}
\setcounter{equation}{0}

We will now analyze the case $\nu=3/2$ which lies at the boundary of the stability region discussed above.
We will thus assume that $T=A~\rho^{3/2}$, and that the universe expands and is dominated by non-relativistic matter
(w=0). Then, in the 4d regime,  equation (\ref{zeta}) becomes ($\lambda =0$)

\be
2\rho\zeta'={\zeta^2-A\zeta-1\over 3\zeta+A}
\label{track}\ee

We will parameterize the dimension-less coefficient $A$ as $A=\m-{1\over \m}$.
$A$ is determined by the details of the microscopic cross
section that gives rise to this type of energy exchange.
$\m$ running on non-negative real numbers parameterizes all
possible values of $A$.
When we have expansion, $A>0$ means outflow. When we have
contraction $A<0$ means outflow.

The general solution of equation (\ref{track}) is

\be
(\zeta-\m)^{-{2\over \mu}+8\m}~\left(\zeta+{1\over
\mu}\right)^{-2\m+{8\over \m}}=C~\rho^{\m+{1\over \m}}
\label{hr}\ee
where $C$ is a constant

Since $H^2=\rho+\chi$, the equation above can be re-written,
 in terms of
$\zeta=H/\sqrt{\rho}$.

Then equation (\ref{rr}) becomes
\be
{a\over \rho}{d\rho\over da}=-{3\zeta+A\over \zeta}
\ee
and
can be integrated as a function of $a$ with the result
\be
(\zeta-\m)^{2\m^2}~\left(\zeta+{1\over
\mu}\right)^{2}=C'~a^{-(\m^2+1)}
\label{ha}\ee

$\rho(a)$ can be obtained by solving (\ref{hr}) and substituting
into (\ref{ha}).
Finally $\chi(a)=\rho(\zeta^2-1)$.

We will first study a few special cases :

{\bf (i)} $\mu=1$.
Here we obtain \be
\zeta^2=1+C\rho^{1/3}\Rightarrow  H^2=\rho+C~\rho^{4/3}
\sp
 \rho={C'^3\over C^3}{1\over
a^3}\sp \chi={C'^4\over C^3}{1\over
a^4}
\ee
compatible with the absence of energy exchange in this case and consequent
independence of the evolution of $\rho,\chi$.

{\bf (ii)} $1/2<\mu$. It corresponds to $-{3\over 2}<A$.
Asymptotically ($a\to \infty$) we obtain the tracking solution

\be
\zeta=\m+\tilde C~\rho^{{\m^2+1\over 2(4\mu^2-1)}}+...\sp
H^2=\m^2 \rho+2\m\tilde C~\rho^{{\m^2+1\over 2(4\m^2-1)}+1}+...
\ee
\be
\rho\sim \tilde C'~ a^{{1\over \m^2}-4}+...\sp
\chi=(\m^2-1)\rho+...
\ee
Here, although the initial conditions for the real $\rho$ and
mirage $\chi$ energy density are arbitrarily different
(parameterized by the independent integration constants $C,C'$),
at late times they scale similarly with the scale factor
\be
\rho\sim \tilde C'~ a^{{1\over \m^2}-4}+...\sp
{\chi\over \rho}=(\m^2-1)+...
\ee
Thus, the dark energy behaves as the visible energy, and such a
mechanism could be used so that bulk energy simulates dark matter.

The case $-2\leq \mu<0$ has qualitatively similar behavior.
All other ranges have asymptotic $\zeta$ which is negative and thus unphysical.

Finally, in the case of outflow, there is a fixed point in the 5d regime, when  $A^2<9/4$ with
\be
\rho_*={9-4A^2\over 8.81}\sp H_*={A\over 108}\sqrt{9-4A^2\over 2}
\ee
This  is a saddle point

It is interesting to note that a similar tracking behavior has been
observed in matter interacting with the dilaton in \cite{venez}.

\subsubsection{Fixed points in the non-linear regime}
\setcounter{equation}{0}

\def\rs{\rho_*}
\def\r{\rho}
\def\rt{\tilde \rho}
\def\cs{\chi_*}
\def\hs{H_*}
\def\qs{q_*}
\def\ts{T_*}
\def\l{\lambda}
\def\r{\rho}

We will consider solutions to the non-linear system  (\ref{eqs}) with $H=\hs$ constant. The equations
also imply that $\rho=\rs$, $T=\ts$, $\chi=\cs$ are also constant.
We will see that although there may be a leftover cosmological
constant $\lambda$ on the brane, the cosmological acceleration
because of energy inflow, may be much smaller than
$\sqrt{\lambda}$.

From the equations we obtain
\be
\rs^{\pm}={1\over
144(1+3w)}\left[(1-3w)\pm\sqrt{(1-3w)^2+1152(1+3w)(\l-\hs^2)}\right]
\ee
\be
\cs=-{3\over 4}(1+w){\rs}\left[72\rs+1\right]\sp
\ts=-3(1+w)\hs\rs\sp \qs=\hs^2
\ee

Assuming the rate of expansion $\hs$ to be small compared to the
cosmological constant $\l$,  we have the following two possibilities

\bigskip
(i) $\l$ dominates in the square root.  In this case $\rs\simeq \sqrt{\lambda\over 18(1+3w)}$.
There is still space for this approximation to be correct and
$\rs<<1$ so that we are in the 4d period.

\bigskip
(ii) In the opposite case $\rs\simeq {1-3w\over 72(1+3w)}$ and we
can be either in the 4d or the 5d regime.

In either case , energy exchange can mask a leftover brane
cosmological constant

\subsubsection{Other accelerating solutions}

We will present here two different families of solutions that are
characteristic in their classes.

\begin{figure}[htb]
\begin{center}
\epsfig{figure=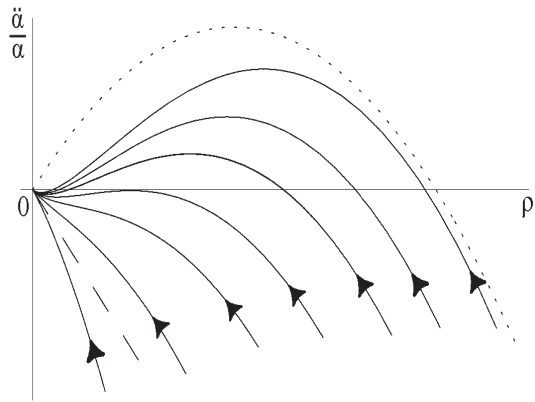, width=200pt,height=150pt} \caption{\label{e1}
\small
\small Outflow, $k=0$, $w=0$,
$\nu=1$. The arrows show the direction of increasing scale factor}
\end{center}
\end{figure}

A global phase portrait of $q\equiv\ddot{a}/a$ with respect to
$\rho$ during expansion in the outflow case for $k=0,\,w=0, \,\nu=1$ is shown
in Figure \ref{e1}.
All solutions are below the limiting parabola $q<{1-3w\over
2}\rho-36(1+3w)\rho^2$.

One recognizes two families of solutions:
The first have $q<0$ for all values of $\rho$, while the  second
start with a deceleration era for large $\rho$, enter an acceleration era and then
return to deceleration for small enough values of $\rho$.

Solutions corresponding to initial conditions with positive $q$
(always under the limiting parabola shown with the dotted line),
necessarily had a deceleration era in the past,
and are going to end with an eternal deceleration era also.
The straight dashed line represents the standard FRW solution without the effects of
energy exchange.

\begin{figure}[htb]
\begin{center}
\epsfig{figure=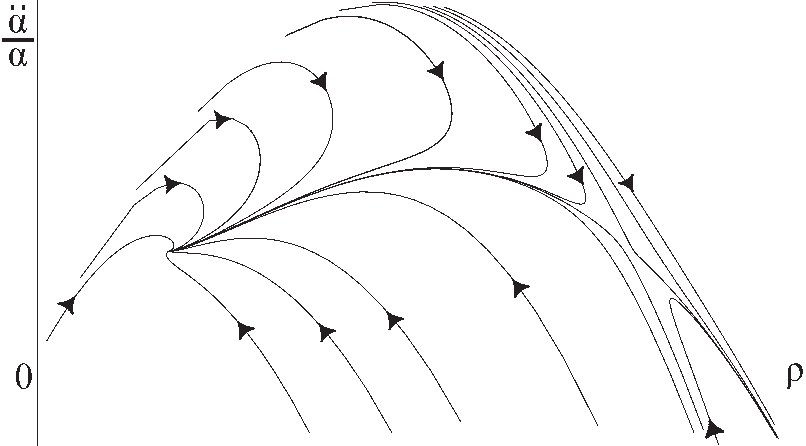, width=200pt,height=150pt} \caption{\label{e2}
\small
\small Influx, $k=0$, $w=0$,
$\nu=1$.}
\end{center}
\end{figure}

The global phase portrait of $q\equiv \ddot{a}/a$ with respect to $\rho$ during
expansion for the case $k=0$, $w=0,\nu=1$ is shown in Figure
\ref{e2}. The presence of the limiting parabola as in the outflow case is apparent.
However, new
characteristics appear. For example, $\rho_*^{(-)}$ attracts to eternal
acceleration a whole family of solutions which start their evolution at either
very low or very high densities. There is another family of solutions which
are attracted to acceleration by $\rho_{*}^{(+)}$ and which eventually exit to
a deceleration era.
Finally, there is a family of solutions, near the limiting parabola, which
start with acceleration at very low
densities, and eventually exit to eternal
deceleration, while their density increases
monotonically with time because of the influx.

For $\nu\neq 1$ one expects a different set of fixed points with
varying behaviors around them.
Such accelerating attractors may be relevant for the present
acceleration of the universe

\subsection{Mirage Cosmology}

We have already seen that vacua of string theory are promising to
describe the standard model, involve D-branes that localize the
standard model particles, embedded in a ten-dimensional bulk space
in which gravity propagates.
This notion of brane-world, embedded in the full space-time
indicates a different view (and dynamics) for cosmology.
Some aspects of this
have been already explored in section \ref{rsbigcosm} in the context of
a RS realization of gravity on a three-brane.

Here, we would like to advocate a more general approach to
brane-world cosmology that
is very powerful, since it is generally and easily applicable.
This is the approach of mirage cosmology \cite{mirage}.
It studies the motion of the brane world in the bulk which is
affected by the energy density (and other data) on the brane-world
as well as bulk fields due to other branes or just bulk geometry.
An important observation in this approach \cite{real1,real2,mirage} is
that the motion of the brane world, via its interactions with the
bulk fields, induces a cosmological evolution on itself, driven by
brane energy densities (as in standard cosmology) as well as bulk
fields (which from the brane-world point of view are interpreted as mirage energy
densities).
Coupled with a probe approximation for the brane-world it is
powerful enough to be able to handle many non-trivial
cosmological contexts.

 We can thus consider our ``universe"
(standard model  collection of almost coincident branes)
in potential motion in the field of other branes carrying hidden
gauge groups. Gravity as well as other universal interactions are living in the bulk space.

There are two different regimes that may be considered.

$\bullet$ The weak bulk field regime. In this case the
matter density on the brane alone drives the cosmological expansion, in the
traditional fashion.

$\bullet$ The strong bulk field regime. Here, the cosmological
evolution of the brane is of the mirage type driven by the bulk
fields and may be interpreted by the inhabitants of the
brane-world as real energy, living on the brane. Detailed
measurements can identify it though as mirage energy, sourced in
the bulk.

An important ingredient in the physics is the induced metric on the brane
which depends on the brane positions but also on brane-localized fields or energy.
This dependence is at the heart of new phenomena, like variable
speed of light \cite{sym} and  induced cosmological evolution
\cite{mirage}.

Mirage cosmology, by mapping cosmological evolution of the brane to
geodesic motion, geometrizes the Friedman-like cosmology and gives
a simple and generic pictures of accelerating, bouncing and cyclic
cosmological evolution as we will explain in the sequel.

The picture of AdS/CFT correspondence and its avatars, provides a
resolution of initial singularities in this context \cite{mirage} which make
use of the duality between gravity and gauge theory.

In order to put in context different approaches of study of
time-dependent solutions of brane-worlds we will compare them
in the following problem: Relative motion of two
branes.

$\bullet$ The effective-field theory approach. One needs to
calculate the effective interactions of brane-fluctuations. The
solutions for the scalars parameterizing the inter-brane distance
describe the relative motion of branes \cite{lecture}.

$\bullet$ The probe approach. One considers the
geodesics of one of the branes moving in the fields of the other.
This approach is in general approximate. In  this case however,
(two sets of branes ) it is exact.

$\bullet$ Full solution of the supergravity/string equations with
a time dependent solution that describes two (solitonic) branes in relative
motion. This is a straight-forward approach albeit very difficult
in general.

Here we will focus on the second approach.
The transverse space may be compact or non-compact.
In the first case, when the size of the compact directions is
large with respect to the string scale, we can treat most regions
of brane geodesics as being embedded in non-compact space.
There is also the possibility of non-compact transverse
dimensions. In such cases, the brane-induced-gravity mechanism,
introduced in previous sections,
must be invoked to implement four-dimensional gravity on the
branes.

Thus, the central idea is that the universe brane is moving into the bulk
background fields of other branes of the theory.
The motion of the brane follows, a classical geodesic
in the bulk geometry. This geodesic is also affected by the matter
densities' associated dynamics, localized on the brane.
The prototype branes we are using here are Dp-branes.
However, as it will become obvious, the results are
valid for general branes, since the IR relevant parts of their
world-volume actions are essentially universal.

There are two steps in the procedure:
\begin{itemize}
\item
Determine the brane motion by solving the world-volume field equations
for the scalar  fields determining the position of the brane in the
bulk. Other brane fields may also be excited.

\item Determine the induced metric on the brane which now becomes an
implicit function  of time. This gives a cosmological evolution in the induced
brane metric. This cosmological evolution can be reinterpreted in terms of
cosmological ``mirage" energy densities on the brane via a Friedman-like
equation.
The induced metric on the brane is the natural metric felt by the observers
on the brane. We assume that our universe (SM fields) live on the brane and are made off
open string fluctuations.
\end{itemize}

An important reminder here is that mirage cosmological evolution is
$not$ driven by four-dimensional gravity on the brane but by higher-dimensional gravity.
One can however include the effects of induced four-dimensional
gravity on the brane as we have done earlier in the RS case.

We will analyze various combinations of branes and simple background
fields.
They correspond to a stack of Dp-branes on and out of extremality (black
Dp-branes). They provide a cosmological evolution on the probe
brane
that can be simulated by various types of mirage matter \cite{mirage}.
Most prominent is radiation-types ($w$=1/3) or massless scalars ($w$=1).
It should be stressed however that at small scale factor size, there are
many exotic types of mirage matter including $w$ values that are outside the
range $|w|\leq 1$ required by four-dimensional causality \cite{mirage}.
Such matter, coined phantom matter \cite{gib1}-\cite{gib3}  appears naturally here.
We interpret the presence of such mirage densities
 as an indication that super-luminal (from that
four-dimensional points of view)``shocks" are possible in such
cosmologies. Super luminal signal propagation in a brane-world context
have been recently pursued  in \cite{free1}-\cite{sol5}.

It has been also pointed out in \cite{sym} that the effective
speed of light on the brane, in the presence of non-trivial  bulk fields is in general
different (lower) than that of gravitational waves. This has been
shown in general in \cite{gh}.
Moreover, when the brane approaches horizons, the effective speed
of light vanishes \cite{sym}. This was also observed later, during
tachyon condensation  and was coined a Carolean limit
\cite{gibbons}. This effect indicates a similarity between unstable
D-brane decay and the emergence of closed strings  on one hand and
confinement in the context of AdS/CFT.

Another peculiarity is that ``individual" densities of mirage dilute-matter can be negative
(without spoiling the overall positivity at late times).

Inflation can be produced in simple backgrounds, at the cost of
breaking supersymmetry. As it was shown in \cite{mir1,mir12} this
can be obtained from type-0 D-branes.

In mirage
cosmology, the initial singularity is an artifact of the low energy description.
This can be seen by studying brane motion in simple spaces like
$AdS_5\times M$ which are globally non-singular.
   The induced cosmological evolution of a brane moving in such a space
has a typical expansion profile due to radiation and an initial singularity
(from the four-dimensional point of view).
However, this singularity is an artifact.
At the point of the initial singularity the universe brane joins a
collection of
parallel similar branes and there is (non-abelian) symmetry enhancement.
The effective field theory breaks down and this gives rise to the
singularity. The non-singular description is that of a non-abelian
gauge theory.

An obvious question is how ``real" matter/energy densities on the
brane affect its geodesic motion and consequently the induced cosmological
evolution.
This can be studied by turning on such energy densities on the brane.  We
will consider as an example electromagnetic energy density and find a solution of the moving brane
with a covariantly constant electric field.
This gives the expected  additional   effect  on the
cosmological evolution
similar to the analogous problem of radiation density in four-dimensions \cite{mirage}.
Although  an electric field is an unrealistic cosmological background
the solution we obtain is also valid when the electric energy density is
thermal (and thus isotropic) in nature.
This indicates that the formalism  is capable of handling the
most general situation possible, namely cosmological evolution driven by bulk
background fields (mirage matter) as well as world-volume energy
densities (real matter).
Such issued have been developed further in
\cite{mir1}-\cite{mir2,sb,kach}.

\subsubsection{Probe-brane geodesics and induced cosmology\label{geod}}

\setcounter{equation}{0}

According to our previous discussion, we will consider a brane-world
moving in the field of a localized collection of p-branes. Here, we
first review the motion of the (probe) brane-world moving in a
generic static, spherically symmetric background. The brane will
move in a brane-geodesic. We
 assume the brane to be light compared to the background so that
we can neglect the back-reaction. The simplest case corresponds to
the background of a  (black) Dp-brane and we will focus mostly on
this case.

The metric of a gravitating Dp-brane may be parameterized as \be
ds^2_{10}=g_{00}(r)dt^2+g(r)(d\vec
x)^2+g_{rr}(r)dr^2+g_S(r)d\Omega_{8-p} \label{mm} \ee and we may
also generically have a dilaton $\phi(r)$ as well as a RR
background $C(r)=C_{0...,p}(r)$. For black Dp-branes we have \be
g_{00}(r)=-{f(r)\over \sqrt{H_p(r)}}\sp g(r)={1\over
\sqrt{H_p(r)}} \sp g_{rr}(r)={\sqrt{H_p(r)}\over f(r)} \ee \be
g_S(r)=r^2\sqrt{H_p(r)}\sp e^{\phi}=H_p^{(3-p)/4} \sp C_{0,...,p}=
\sqrt{1+{r_0^{7-p}\over L^{7-p}}}\left(1-{1\over H_p}\right) \label{c4}\ee
and \be H_p=1+{L^{(7-p)}\over r^{(7-p)}}\sp f=1-{r_0^{(7-p)}\over
r^{(7-p)}} \ee

The probe 3-brane will in general  move in this background along a
geodesic. Its dynamics is governed by the DBI action. In the case
of maximal supersymmetry it is given by \be S=T_3\int d^{4}\xi
e^{-\phi}\sqrt{-det(\hat G_{\alpha\beta} +(2\pi
\alpha')F_{\alpha\beta}-\hat B_{\alpha\beta})}+ T_3\int d^{4}\xi  ~\hat
C_4 +{\rm anomaly~~ terms} \label{action} \ee where we have
ignored the world-volume fermions.

The embedded data are given by
\be
 \hat G_{\alpha\beta}=G_{\mu\nu}{\partial x^{\mu}\over
\partial\xi^{\alpha}}{\partial x^{\nu}\over
\partial\xi^{\beta}}\sp \hat B_{\alpha\beta}=B_{\mu\nu}{\partial x^{\mu}\over
\partial\xi^{\alpha}}{\partial x^{\nu}\over
\partial\xi^{\beta}}
\ee
Due to reparametrization invariance, there is a gauge
freedom which may be fixed by choosing the static gauge,
$x^{\alpha}=\xi^{\alpha}$,  $\alpha=0,1,2,3$. A generic motion of
the probe D3-brane will  have a non-trivial angular momentum in
the transverse directions as well as non-trivial momentum in the
directions transverse to the probe brane but longitudinal to the
background brane if $p>3$ (also known as  Neumann-Dirichlet (ND)
directions) In the static gauge, the relevant  (bosonic) part of
the brane Lagrangian reads \be
L=e^{-\phi}\sqrt{g(r)^3[|g_{00}|-g_{rr}\dr^2-g(r)\sum_{i=4}^{p}(\dot
x^i)^2-g_S(r)h_{ij}\dt^i\dt^j]}-C(r)\delta_{p,3} \label{the}\ee
where $h_{ij}(\varphi)d\varphi^i\varphi^j$ is the line element of
the unit (8-p)-sphere. For future purposes (generality) we will
parameterize the Lagrangian as \be {\cal L}=\sqrt{A(r)-B(r)\dot r
^2-\sum_I D_I(r)h^I_{ij}\dt_I^i \dt_I^j}-C(r)\delta_{p,3}
\label{l} \ee with \be A(r)=g^3(r)|g_{00}(r)|e^{-2\phi}\sp
B(r)=g^3(r)g_{rr}(r)e^{-2\phi} \sp D_I(r)=g^3(r)g_I(r)e^{-2\phi}
\label{sub}\ee and $C(r)$ is the RR background. The RR background
only appears if the dimensions of the background and probe brane
are equal. The metric $h^I_{ij}$ is that of $S^{d_I}$ while
$\phi^I_i$ are the coordinates of $S^{d_I}$. $d_I$ can be one
(relevant for compact ND directions). Several spheres can appear
if the central brane is a collection of intersecting elementary
branes.

In spherically symmetric backgrounds, radial motion entails
non-trivial cosmological evolution on the brane  with
a scale factor that satisfies the mirage cosmology equations
\cite{mirage}.

Roughly speaking the brane-world motion depends on both its
charges (energy density on the brane, and RR and dilaton  charge)
as well as on the bulk fields, generated by the
central branes. This motion induces a time-dependent effective
metric on the brane and thus a cosmological evolution. In the
absence of world-volume energy distributions, homogeneity and
isotropy on the brane is ``explained" by the spherical symmetry of
the central fields.

In the absence of 2-index antisymmetric tensor fields the
effective metric on the brane-world is the induced metric. Thus,
for bulk metrics of the form advocated in (\ref{mm}) the scale
factor is $a^2=g(r)$ and its time dependence is due to the
changing distance $r(t)$ from the central source.

We will now proceed further and solve for the brane-universe
motion. The problem is effectively one-dimensional and can be
solved by quadratures.The momenta are given by
\begin{eqnarray}
&& p_r=-{B(r) \dot r\over \sqrt{A(r)-B(r)\dot r ^2-
\sum_I D_I(r)h^I_{ij}\dt_I^i \dt_I^j}}\\ \nonumber &&
p^I_i=-{D_I(r) h^I_{ij}\dt_I^j\over
\sqrt{A(r)-B(r)\dot r ^2-\sum_I D_I(r)h^I_{ij}\dt_I^i \dt_I^j}}
\end{eqnarray}

The angular momenta
as well as the  Hamiltonian
\be
H=-E=C-{A(r)\over
\sqrt{A(r)-B(r)\dot r ^2-\sum_I D_I(r)h^I_{ij}\dt_I^i \dt_I^j}}
\label{the1}\ee
are conserved.
The conserved total angular momenta in direction I are
$h_I^{ij}p^I_ip^I_j=\ell_I^2$ and
\be
h^I_{ij}\dt_I^i\dt_I^j={\ell_I^2\over D_I^2}(A(r)-B(r)
\dot r ^2-\sum_I D_I(r)h^I_{ij}\dt_I^i \dt_I^j)\ee
\be             \sum_I D_I h^I_{ij}\dt_I^i\dt_I^j=
(A(r)-B(r)\dot r^2){\sum_I {\ell_I^2\over D_I}
\over 1+\sum_I {\ell_I^2\over D_I}}
\ee
The final equation for the radial variable is
\be
\sqrt{{A(r)-B(r)\dr^2\over 1+\sum_I {\ell_I^2\over D_I}}}=
{A(r)\over E+C(r)}\Rightarrow
\dr^2={A\over B}\left(1-{A\over (C+E)^2}\left[1+\sum_I
{\ell_I^2\over D_I}\right]\right)
\label{solu}\ee
and using it we can rewrite
\be
h^I_{ij}\dt_I^i\dt_I^j={\ell_I^2\over D_I^2}{A^2\over (C+E)^2}
\ee

The induced four-dimensional metric on the 3-brane universe
is\footnote{We assume here that $B_{\mu\nu}=0$ for D-branes. In
the opposite case there are modifications to the effective metric
that are interesting but will not be further discussed here. An
example of mirage cosmology in the presence of nontrivial
$B_{\mu\nu}=0$ was presented in \cite{mirage}
.}

 \be d\hat
s^2=(g_{00}+g_{rr}\dr^2+ \sum_Ig_Ih^I_{ij}\dt_I^i
\dt_I^j)dt^2+g(d\vec x)^2 \label{met} \ee

This can be brought to the standard RW form by defining cosmic
time $\tau$ (coinciding with the proper brane time) as \be
d\tau^2=-(g_{00}+g_{rr}\dr^2+ \sum_Ig_Ih^I_{ij}\dt_I^i
\dt_I^j)dt^2={A^2\over (C+E)^2}{e^{2\phi}\over
g^3}dt^2\label{cosmic}\ee and the scale factor as \be
a^2=g(r)\label{scale}\ee

Definitions (\ref{cosmic},\ref{scale}) and the dynamical equation
(\ref{solu}) imply that

\be {\dot a\over a}={1\over 2}{1\over
g}{dg\over dr}{dr\over dt}{dt\over d\tau}\ee which can be written as  a
Friedman-like equation
\be \left({\dot a (\tau)\over
a(\tau)}\right)^2=\rho_{\rm eff}(a(\tau))\equiv {1\over
4}{g'^2\over g^2}\left((C+E)^2-A\left[1+\sum_I {\ell_I^2\over
D_I}\right]\right){{e^{-2\phi}~ g^3}\over AB}
\label{frie}\ee
(\ref{frie})  describes the effective cosmological
evolution on the brane-universe.  As
advocated in \cite{kach}, more detailed information about the
(local) physics of the brane can be recovered from the brane
action and its coupling to the bulk fields.

\subsubsection{Orbits, bounces and static universes}

At this point and without further calculation we can qualitatively
describe the connection between different brane-universe geodesics
and the  (effective) cosmological type of expansion or
contraction perceived on the brane-universe. The function $g(r)$ linking the
distance to the effective scale-factor of the brane-universe is
typically a monotonic function of $r$. We
ignore here the potential effect of brane-energy densities over
and above the (BPS) negative tension and RR charges. Their effects
will be discussed later on.

The eventual type of cosmology (behavior of the scale factor)
depends on two distinct ingredients \cite{unp}-\cite{branonium}:

(a) The dependence of the scale factor on the radial distance $r$
of the universe brane from the center of the bulk distribution,
encapsulated in the relation $a^2=g(r)$. There are two distinct
possibilities here: In asymptotically flat bulk configurations,
$a(r)$ is a monotonic function with a maximum. For
non-asymptotically flat configurations (like AdS) $a(r)$ is
monotonic without a bound.

(b) The type of orbit in transverse space. Such orbits are of the
following form:

 (i) {\tt Unbounded (hyperbolic) orbits}. These
are ``scattering-type" orbits where the brane starts at infinite
distance approaches to a minimal distance and finally retreats
back to infinity. The associated behavior of the brane cosmology
is then as follows: The universe is contracting until it reaches a
minimum size. It subsequently (and smoothly) bounces back to an
ever expanding phase. No cosmological singularity will be ever
perceived on the brane-universe in this case. In the
asymptotically flat case, the ever expanding phase can saturate
asymptotically due to the bound in $a(r)$.

(ii) {\tt Bounded (parabolic) orbits}. These are bound orbits
where the distance to the central point (hyper-galaxy)
 oscillates between a
minimum and a maximum value. The associated cosmology is cyclic
alternating between expanding and contracting phases with no
cosmological singularity. A very special case is that of circular
orbits that corresponds to a static brane-universe. As it will
be expanded upon later, the speed of light on the brane is smaller
than the one in the bulk \cite{sym,alex}. At large distances, one
deals with Newtonian potentials that behave like $r^{2-d}$ where
$d$ is the number of transverse directions. It is well known that
bound orbits exist only when the centrifugal potential (always
behaving like $r^{-2}$ independent of dimension) can balance the
Newtonian one. This can only happen  when the Newtonian attractive
potential behaves as $r^{-1}$ or $r^{-2}$.

(iii) {\tt ``absorption" or ``emission" orbits}. These are orbits
where the universe brane finally enters the horizon of the  black
brane (central hyper-galactic black brane) or it is ``emitted" and
moves outwards. We should distinguish here the following two
subcases.

(iiia) {\tt Bound absorption or emission orbits}. These obits
originate or end at the horizon but they never reach infinity.
They correspond to a cosmology of an expanding universe (outward
motion) which reaches a maximum size, then recontracts  and has a
final singularity associated to the brane falling at the center
$r=0$. It has been argued \cite{mirage} that this singularity is
resolved in the D-brane description by passing to the enhanced
non-abelian theory of the full system. Nonetheless, this encounter
is catastrophic for the brane-matter (and its inhabitants) since
it entails a complete rearrangement of the effective matter theory
similar to that of phase transition of symmetry restoration. It
has been argued that already at the crossing of the horizon the
brane thermalizes with the Hawking heat-bath associated with the
central black brane \cite{kt1,kt2} provided it is colder than the black
brane\footnote{See also \cite{verlinde}
for a related view.}. The
time reversed situation involves an expanding universe with an
apparent (initial) singularity in its past.

(iiib) {\tt Unbounded absorption or emission orbits}. Such orbits
start above the  horizon and reach infinity, or start at infinity
and fall through the horizon. They correspond to always expanding
cosmologies with an initial singularity or always contracting
cosmologies with a final singularity. The scale factor can also
saturate here in asymptotically flat bulk geometries. The same
remarks as above apply to the resolution of the cosmological
singularity.

\subsubsection{A partial survey of  various induced mirage cosmologies}
\setcounter{equation}{0}

Here we will take a closer look at the various orbits. We consider
a 3-brane moving in the background of a black p-brane. The
internal charges that are relevant are the angular momentum $l_s$
on $S^{8-p}$ transverse to the p-brane as well as the momenta
$l_i$, $i=1,2,\cdots ,p-3$ along the longitudinal directions of
the black p-brane that are transverse to the 3-brane universe. The
relevant functions are \be A=f~H_p^{p-7\over 2}\sp B={1\over
f}H_p^{p-5\over 2}\sp D_i=H_p^{p-7\over 2}\sp D_s=r^2
H_p^{p-5\over 2}\ee In the special case $p=3$ there is also a RR
potential $C=\sqrt{1+{r_0^4\over L^4}}{L^4\over L^4+r^4}$. We will
first describe the case $p>3$.

 The
orbit is given by the solution of \be \dr^2={f^2\over
H_p}\left[1-{f\over E^2}\left({1\over H_p^{7-p\over 2}}+{l_s^2
\over  r^2 H_p}+{\sum_i l_i^2}\right)\right]\ee where $l_s$ is the
angular momentum on the $S^{8-p}$ and $l_i$ are angular momenta on
the Neumann-Dirichlet  $S^1$s. We will scale $r$, $t$ and $l_s$ appropriately so
as to set the length $L$ to one.

The radius-scale factor relation is \be r=\left({a^4\over
1-a^4}\right)^{1\over 7-p} \ee Note that here the scale factor is
bounded above by 1. This can be rescaled to any desirable finite
value by a choice of units.
 The associated Friedman equations take
the form
\be
\left({\dot a\over a}\right)^2={(p-7)^2\over 16}a^{2{3-p\over
7-p}}(1-a^4)^{2{8-p\over 7-p}}\left[{(C(a)\delta_{p,3}+E)^2\over
a^{2(7-p)}}-\right.
\ee
$$
-\left. f_p(a)\left(1+{\sum_i l_i^2\over
a^{2(7-p)}}+{l_s^2(1-a^4)^{2\over 7-p}\over a^{2{(p-6)^2+3\over
7-p}}}\right)\right]
$$
with
\be
C(a)=\sqrt{1+r_0^4}(1-a^4)\sp f_p(a)=1+r_0^{7-p}-{r_0^{7-p}\over
a^4}
\ee

 To discuss the
behavior of orbits we must study the effective potential
\be
V_{\rm eff}={f^2\over H_p}\left[1-{f\over E^2}\left({1\over
H_p^{7-p\over 2}}+{l_s^2 \over  r^2 H_p}+{\sum_i
l_i^2}\right)\right]
\ee
Zeros of the effective potential signal a turning point in the
radial motion.  At the horizon $r=r_0$ the effective potential
always vanishes
\be
V_{\rm eff}={(7-p)^2\over r_0^{p-5}(r_0^{7-p}+1)}(r-r_0)^2+{\cal
O}\left((r-r_0)^3\right) \label{horizon}\ee

This zero is not a turning point. If no other zero $r_*>r_0$
exists, then the brane crosses eventually the horizon. Moreover
$V_{\rm eff}$ remains positive for $r$ approaching $r_0$ from
above.

The asymptotic form of the effective potential at large $r$ is
\be
V_{\rm eff}=1-{\sum_i l_i^2+1\over E^2}+{\cal O}\left({1\over
r}\right) \ee where $3<p<7$.

We distinguish the following cases
\begin{itemize}
\item{($\bf A$)} $E^2>\sum_i l_i^2+1$. In this case $V_{\rm eff}(\infty)>0$ and
there is always the possibility of unbounded motion. If there are
no non-trivial zeros, then the possible motion is of type (iiib)
in the nomenclature of the previous section (unbounded emission or
absorbtion orbits).

If there are non-trivial zeros they must appear in pairs. In this
case there  are type (i) unbounded orbits  that bounce at the
largest non-trivial zero. There is always also type (iiia) orbits
between the lowest non-trivial zero and the horizon. Finally, if
there are more than two non-trivial zeros, there are type (ii)
bound orbits corresponding to oscillating universes.

\item{($\bf B$)} $E^2<\sum_i l_i^2+1$. In this case $V_{\rm eff}(\infty)<0$ and
there is no unbounded motion. In this case the number of
non-trivial zeros is odd. There is always a type (iiia) motion
between the lowest non-trivial zero and the horizon.

If there are more than one non-trivial zeros then there are always
type (ii) orbits leading to oscillating universes.

\item{($\bf C$)} $E^2=\sum_i l_i^2+1$. This is a marginal case with  $V_{\rm eff}(\infty)=0$.
We must   investigate whether $V_{\rm eff}$ vanishes from
positive or negative values.

This crucially depends on $p$. For $p=4$
\be
V_{\rm eff}=-{l_s^2\over E^2}{1\over r^2}+{\cal O}\left({1\over
r^3}\right)\ee When $l_s\not= 0$ the orbits are as in case ($\bf
B$). When $l_s=0$
\be
V_{\rm eff}={3+2E^2r_0^3\over 2E^2}{1\over r^3}+{\cal
O}\left({1\over r^6}\right)
\ee
and the orbits are those of case ($\bf A$)

 For $p=5$
\be
V_{\rm eff}={r_0^2E^2-l_s^2+1\over E^2}{1\over r^2}+{\cal
O}\left({1\over r^4}\right)\ee If $r_0^2E^2+1>l_s^2$ we have
orbits similar to case ($\bf A$) while in the opposite case
$r_0^2E^2+1<l_s^2$ we have orbits like case ($\bf B$) In the
marginal case $l_s^2=r_0^2E^2+1$
\be
V_{\rm eff}=r_0^2(1+r_0^2)~{1\over r^4}+{\cal O}\left({1\over
r^6}\right)\ee and the orbits are of case  ($\bf A$).

 For $p=6$

\be V_{\rm eff}={2r_0 E^2+1\over 2E^2}{1\over r}+{\cal
O}\left({1\over r^2}\right)\ee and we have obits like case ($\bf
A$).
\end{itemize}

The case of the black-D3-brane background is described by the
following effective potential.
\be
V_{\rm eff;D3}={f^2\over H_3}\left[1-{f\over (E+C)^2}\left({1\over
H_3^2}+\sum_il_i^2+{l_s^2\over r^2H_3}\right)\right]=1-{\sum_i
l_i^2+1\over E^2}-{l_s^2\over E^2}+{\cal O}\left({1\over
r^4}\right)
\ee
The presence of the RR 4-form appears first at order $r^{-4}$. The
general structure of orbits is similar to the $p=4$ case.

 Note that in all the above situations, unbounded orbits in
 $r-space$ correspond to universes with a scale factor $a\to 1$ in
 the asymptotic past and future.

\subsubsection{Near-horizon region}
\renewcommand{\theequation}{\arabic{section}.\arabic{subsection}.\arabic{subsubsection}.\arabic{equation}}
\setcounter{equation}{0}

We will investigate here the near-horizon region of black
Dp-branes. The effective potential is given by

\be
V_{\rm eff}={f^2~r^{7-p}}\left[1-{f\over
(E+\delta_{p,3}r^4)^2}\left( r^{(7-p)^2\over 2}+l_s^2
~r^{5-p}+{\sum_i l_i^2}\right)\right]
\ee

The radius-scale relation is given by
\be
r=a^{4\over 7-p}
\ee
and the effective Friedmann equation reads \cite{mirage}
\be
\left({\dot a\over a}\right)^2={(p-7)^2\over 16}a^{2{3-p\over
7-p}}\left[{(C(a)\delta_{p,3}+E)^2\over
a^{2(7-p)}}-f_p(a)\left(1+{\sum_i l_i^2\over
a^{2(7-p)}}+{l_s^2\over a^{2{(p-6)^2+3\over 7-p}}}\right)\right]
\label{met4}\ee
with
\be
C(a)=a^4\sp f_p(a)=1-{r_0^{7-p}\over a^4}
\ee

The behavior of the effective potential at the horizon is the same
as in (\ref{horizon}) and similar remarks apply. At large $r$ we
obtain that for all $3<p<7$ the effective potential reaches
$-\infty$. Consequently we have only bound orbits of (ii) and
(iiia) type. The maximal radial distance of such orbits grows with
the energy $E$ of the motion roughly as $r_{max}^{7-p}\simeq E^4$.

The case $p=3$ is special. The asymptotic form of the effective
potential is here
\be
V_{\rm eff}=2E+r_0^4-{l_s^2\over r^2}+{\cal O}\left({1\over
r^4}\right)
\ee

\begin{itemize}
\item $2E+r_0^4>0$. Here we have generically unbounded orbits.

\item $2E+r_0^4<0$. Here we have bound orbits.

\item $2E+r_0^4=0$. The orbits are bound.

\end{itemize}

\subsubsection{Static universe solutions}
\setcounter{equation}{0}

Fixed $r$ orbits correspond to static universes with a constant speed of light.
Using $\dot r=0$ and the equations of motion we obtain \be d\hat
s^2={g_{00}(r)\over 1+\sum_I {\ell_I^2\over D_I}}dt^2+g(r)(d\vec
x)^2 \label{met1} \ee from which we can read the effective light
velocity as \be c^2_{\rm eff}={g_{00}\over g\left(1+\sum_I
{\ell_I^2\over D_I}\right)} ={f\over 1+\sum_I {\ell_I^2\over
D_I}}\leq 1 \ee where the second equality holds for black
D-branes. The numerator comes from the presence of the horizon in
the background brane \cite{sym} and the denominator is due to the
circular motion in transverse space \footnote{This term is missing
in
\cite{alex}
.}.

We will now discuss  orbits for $D_{p>3}$ backgrounds. In this
case
\be
A=fH_p^{(p-7)/2}\sp B={H_p^{(p-5)/2}\over f}\sp D_{ND}=H_p^{(p-7)/2}
\sp D_T=r^2H_p^{(p-5)/2}
\ee
where $D_{ND}$ corresponds to the Neumann-Dirichlet directions and $D_T$ corresponds to the (8-p)-sphere.
The energy condition for circular orbits at $r=r_*$ implies
\be
{E^2\over A}=1+\sum_I {\ell_I^2\over D_I}
\ee
while the no-force condition
\be
{\partial\over \partial r}\log\left[ A \left( 1+\sum_I {\ell_I^2\over D_I}\right)\right]={\partial\over \partial r}\log\left[fH_p^{(p-7)/2}\left( 1+\ell_{ND}^2
H_p^{(7-p)/2}+{\ell_{T}^2\over r^2}H_p^{(5-p)/2}\right)\right]=0
\ee
Using the above, the effective light velocity becomes
\be
c^2_{\rm eff}={f^2H_{p}^{(p-7)/2}\over E^2}={f\over 1+\sum_I {\ell_I^2\over D_I}}
\ee
It is implied that as $r_*\to \infty$ $c_{eff}\to 1/|E|$.

It is easy to argue that the existence of stable circular orbits for a range of parameters is possible only if the ``gravitational" attraction
can be balanced by the centrifugal force which always falls like $r^{-2}$.
This can happen only if the attraction behaves as $r^{-1}$ or $r^{-2}$ and this singles out the cases of $p=5,6$

In these cases we always have stable circular orbits at any radius
by appropriately adjusting the energy and angular momenta.
Analyzing the dynamics of small perturbations around the circular
motion we obtain the condition for stability: \be {1\over
2}\left(A''-\sum_I{D_I''{\ell_I^2\over D_I^2}{A^2\over
E^2}}\right)+\sum_{I,J}B_IM^{-1}_{IJ} B_J\geq 0 \ee where \be
M_{IJ}=\left(D_I\delta_{IJ}+\ell_I\ell_J\right)\sp B_I={D_I' \over
D_I}\ell_I \ee

We will now examine the two relevant cases.

$p=5$.
In this case there are stable circular orbits for a wide range of parameters.
In the special case $r_0=0$, $L={\ell_T}$ and there is a stable orbit for any r.
Here
\be
c^2_{\rm eff}={1\over (1+\ell_{ND}^2)\left(1+{L^2\over r^2}\right)}
\ee
and varies between zero and $1/ (1+\ell_{ND}^2)$ for various orbits.
In the general case, $r_0>0$, there are stable circular orbits provided
$\ell_{T}^2 >L^2+r_0^2(1+\ell_{ND}^2)$.

$p=6$. This is similar to the $p=5$ case. In the extremal case ($r_0=0$),
the radius of the circular orbit is $r_{*}=(\ell_T^2+\ell\sqrt{\ell_T^2+L^2})/L$
and it is stable.
Similarly
\be
c^2_{\rm eff}={1\over 1+\ell_{ND}^2\sqrt{H_6}+{\ell_{T}^2\over r_*^2\sqrt{H_6}}}
\ee
This situation generalizes to $r_0>0$.

\subsubsection{The Hyper-universe}
\setcounter{equation}{0}

We will advocate here a picture for the universe that is motivated by our previous
discussion.
It is the Copernican approach to the brane-universe idea and contains two
levels.

1) Our conventional 4-dimensional universe is a p-brane embedded
in a D-dimensional space. One possibility is $p=3$. However, we
may have $p>3$ with the extra $p-3$ dimensions compactified. D
maybe ten or eleven if the underlying theory is assumed to be
string (M)-theory\footnote{A reduced version of a universe
containing branes has been advocated in
\cite{brande}.
The full
D-dimensional space is assumed to be compact and the wrapped
branes are treated like a gas in a finite box}.

2) The hyper-universe is populated by many brane-worlds in the
same way that our universe is populated by stars. Brane-worlds
gravitate (and in general  interact via other related
interactions) and can form ``brane-galaxies" whose long range fields are
those of ``black-branes"

There are several fundamental questions/problems associated with
such a picture.

(i) ``Who ordered that?"  It seems natural that if the fundamental
theory contains brane-like objects then it is expected that the
vacuum will be populated by them. The simplest examples of branes
have to be infinite in size. We understand however contexts where
compact branes can be stabilized by external fields \cite{myers}
or a compact background geometry. Whether the energy of the
hyper-universe is finite or not will depend on the type of branes
(compact or not), the type of extra -dimensions (compact or not)
and the potential presence of ``negative tension branes".

Such branes appear in string theory (known as orientifold or
orbifold planes) and are crucial for generating flat vacua in
string theory. They cannot fluctuate, in order to preserve
unitarity (their potential fluctuations have necessarily negative
norm). They can ``effectively move" via a changing background
geometry. Their number in simple ground-states of string theory seems
to be bounded above
($2^5$ in the superstring and $2^{13}$ in the bosonic string,
constrained by the tadpole of the Klein-bottle projection). Up to
now, no fundamental reason is known for this fact. It is
known however, that when their charge is large, they develop an
``enhancon-like" singularity. Whether such a behavior is
forbidding is not known. It seems however, than in
compactifications with fluxes such upper bounds can be quite high.
Macroscopic negative tension defects can provide interesting
cosmological models \cite{negative1}-\cite{negative3}.

\bigskip

(ii) There is no known reason so far, that a special dimensionality
of branes is preferred. Thus, one may assume that originally any
possible dimensionality  of branes is produced. After a long
evolution and process (that remains to be understood) the
hyper-universe ends-up with a collection of hyper-galaxies,
(gravitating collections of brane-worlds). Such aggregations with
different dimension branes and  arbitrary orientations of their
world volumes run counter to the (strongly favored by data)
homogeneity and isotropy of our own brane world.

A dynamical mechanism \cite{quevedo1,quevedo2,watson} can be argued to provide
alignment of world-volumes. However, a mechanism  for the
uniformisation of world-volume dimension needs to be found.

(iii) An important question is whether six (or D-4) extra
dimensions are compact or (effectively) non-compact. An important
experimental ingredient is that  gravity at distances
$10^{-5}m<l<10^{26}m$ has been measured to be of the $r^{-2}$ type
(4-dimensional).

(iiia) If the extra space-time dimensions are compact, there
should be no phenomenological problems provided the compactness
scale is small enough. If however the branes carry gauge charges (as D-branes do)
 the total charge
of the branes is constrained to be zero. Thus, a hyper-universe requires an equal
number of branes and anti-branes and orbifold/orientifold planes.
The presence of anti-branes seems to be an important ingredient of
some cosmological approaches to the early universe
\cite{ekp,dvali-in,alex2,quevedo1,quevedo2}.
However, the branes might not have gauge charges but are stable
because of dynamical reasons.

(iiib) An interesting alternative is that (some of) the extra
dimensions are non-compact. A mechanism is needed to guarantee
that gravity on our brane universe is four-dimensional in the
required range of distances. There are three such mechanisms known
as explained earlier:

(1) The internal space is non-compact with its Laplacian having a
sufficiently large gap above zero \cite{gw1}-\cite{gw3}.

(2) The internal space is non-compact with a gap-less Laplacian
with an appropriate density of eigenvalues close to zero, and
finite volume \cite{rs2}. It has been argued that this can be
implemented in string theory \cite{hv}. In the case of codimension
higher than one, an effective UV cutoff is needed in AdS. This
maybe provided by angular momentum in the internal dimensions.

Both of the approaches (1) and (2) guarantee four dimensional
gravity in the IR. In case (2) gravity becomes higher dimensional
in the UV \cite{Kakushadze:2001rz,irs}.

(3) Quantum corrections of matter fields induce a localized
Einstein term  on the brane \cite{big1}-\cite{big5}. This term induces
4-dimensional gravity on the brane in the UV while gravity is
higher dimensional in the IR. Obviously the cross-over scale has
to be larger than the observable size of the universe or it has to
be combined with one of the mechanisms described above
\cite{dvali,irs}. Some model building in string theory has been
done along these lines \cite{orient1,orient2,fat,amv,kohlprath}.

(iv) The cosmological evolution  on our brane world is driven by
matter localized on the brane and also bulk fields (``mirage
cosmology" \cite{mirage}). Such a picture ,
allows for novel cosmological mechanisms, namely inflow or
outflow of matter from the bulk space \cite{irs}, \cite{kkttz1}-\cite{kkttz4}, that can
trigger undesirable (constraining) effects \cite{sav2}, but
can also provide natural mechanisms for inflation or late-time
acceleration \cite{kkttz1}-\cite{kkttz4}.

\renewcommand{\theequation}{\arabic{section}.\arabic{subsection}.\arabic{equation}}
\subsection{Brane/antibrane systems and inflation}
\setcounter{equation}{0}

Inflation is by now a mechanism favored for explaining the recent
cosmological data as well as providing conceptual solutions to the
flatness, horizon, and defects problems of cosmology.
Although there many phenomenological models that implement
inflation, there are two important problems remaining to be
solved.

$\bullet$ {\it The fine-tuning problem}: Fine tuning is required in the
phenomenological scalar models in order (i) to produce inflation
(slow roll) (ii) to produce the experimentally observable size of
CMB fluctuations \cite{lecture,review1}.

$\bullet$ The phenomenological model must be {\it incorporated} in the theory
of fundamental interactions.

Here, we will describe the framework of brane-world inflation
\cite{dvt}, and in particular, the one generated by the interaction
of branes with anti-branes
\cite{quevedo1,quevedo2,dvali-in,alex2}.
There are also attempts to generate inflation for tachyon
condensation \cite{tachyon1}-\cite{tachyon28} but we will not
discuss it here.

We will consider a pair of a $D3/\bar{D3}$ brane and their static
interaction potential
\be
V_{D\bar D}(r)=2T_3\left(1-{1\over 2\pi^3}{T_3\over M_{10}^8
r^4}\right)
\ee
where $T_3$ is the tension of the branes, $M_{10}$ the
ten-dimensional Planck scale, and r is the distance between the
brane and anti-brane.
The kinetic term of $r$ induced on the branes (in the low-velocity regime) is multiplied by
$T_3$. Thus, the canonically normalized scalar is $\phi=\sqrt{T_3}~
r$ and the potential becomes
\be
V_{D\bar D}(\phi)=2T_3\left(1-{1\over 2\pi^3}{T_3^3\over M_{10}^8
\phi^4}\right)
\ee

Such potentials can in principle give rise to inflation for large
$\phi$. To investigate this we must investigate the slow-roll
parameters \cite{lecture,review1}
\be
\epsilon ={M_P^2\over 2}\left({V'\over V}\right)^2\sp
\eta=M_P^2{V''\over V}
\ee

To get sufficient inflation we must arrange that $\epsilon<<1$,
$\eta<<1$. Taking  into account that if six dimensions transverse
to the 3-branes are compact $M_{P}^{2}=M_{10}^8V_6$ where $V_6$ is
the volume of the compact manifold we obtain
\be
\eta\sim {V_6\over r^6}
\ee
In order for the interaction potential to be valid , we must take
$r<<V_6^{1/6}$. In that case $\eta >>1$ and inflation is not
possible \cite{quevedo1,quevedo2}. This result is valid also in the
anisotropic case \cite{quevedo1,quevedo2,kklmmt}.

Things can get better if we consider a warped background geometry.
This is equivalent to consider a large number of $D3$ branes,
which in this case gravitate. In the near horizon limit they generate (locally)
 an $AdS_5$ geometry. Of
course, this is part of the compact manifold, so the $AdS_5$ must be
cutoff at some point.
The metric of $AdS_5$ in Poincar\'e coordinates is
\be
ds^2={r^2\over R^2}({-dt^2+d\vec x^2})+{R^2\over R^2}dr^2
\ee
and there is the standard 4-form background.
The AdS length-scale is given by \cite{mald1}
\be
R^4\sim g_s N\alpha'^2
\ee
The exact constant of proportionality depends on the embedding of
the AdS slice in the six-dimensional compact manifold.

We will the assume that the range of $r$ is $r_0<r<r_{\rm max}$
for $r>r_{\rm max}$ (the UV region) the manifold is glued to the
internal compact manifold. The manifold is also cut-off in the IR
$(r<r_0)$ smoothly as in solutions dual to confining gauge
theories (see for example \cite{non-ads1})
Thus, this region is approximate to the simple RS setup with an UV and an IR
brane \cite{rs1}.
In particular, there is an effective four-dimensional
gravitational interaction, with a finite Planck mass due to the UV
cutoff $r_{\rm max}$.

We would like now to consider the effective world-volume action of
a $D_3$ brane in this background.
The $C_4$ form is the near horizon limit of (\ref{c4})
\be
C_4={r^4\over R^4}
\ee
and the induced action reads
\be
S_{D_3}=-T_3\int d^4 x\sqrt{-g}{r^4\over R^4}\sqrt{1-{R^4\over
r^4}g^{\m\n}\p_{\m}r\p_{\n}r}+T_3\int d^4 x {r^4\over R^4}
\ee
For a $\bar D_3$, the sign of the second term is reversed.

For a $D_3$ brane in this background the static potential
cancels
(no force condition, for a BPS configuration).
However, in the presence of $D_3$ brane at r and a $\bar D_3$ at
$r_0$ there is a non-trivial potential that can be estimated as
follows:
The harmonic function in the metric due to N $D_3$ branes is
\be
h(r)={R^4\over r^4}
\ee
The analogous function for an extra $D_3$ brane at $ r=r_*$ is
\be
h(r)+\delta h(r)=R^4\left[{1\over r^4}+{1\over N}{1\over
r_*^4}\right]
\ee
Thus, the potential for a $\bar D_3$ at $r=r_0$ and a $ D_3$ brane
at r, for large r , from the DBI action is
\be
V=2T_3{r_0^4\over R^4}\left[1-{1\over N}{r_0^4\over r^4}\right]
\ee
There are two observations on this interaction. Th first term is
there but does not affect the motion of the $D_3$ brane.
the second term is responsible for the attractive interaction
between the brane and anti-brane. At large distance, this is a
slowly varying potential with an "effective" tension
\be
T_{eff}=T_3
{r_0^4\over R^4}<<T_3
\ee
Thus in this regime we expect that
\be
S_{D_3\bar D_3}=\int d^4x\left[{T_3\over
2}g^{\m\n}\p_{\m}r\p_{\n}r-2T_3{r_0^4\over R^4}\left(1-{1\over N}{r_0^4\over
r^4}\right)\right]
\ee
will give rise to slow roll and inflation.
When eventually the branes collide, there will be brane-anti-brane
annihilation. This could be the starting point of reheating.
There are, however, several problems with this type of inflationary
scenarios \cite{kklmmt}. They have to do with additional couplings
on the D-branes that might destroy inflations.
There is active research currently on this realization of
inflation,\cite{dud}-\cite{db3}, but it seems that the last word has not yet
been said.

\subsection{The cosmology of massive gravity and late time acceleration}
\setcounter{equation}{0}

It has been argued in  \ref{energyloss} that sometimes four-dimensional gravity in the UV
can be mediated by a massive-like graviton.
It turns out that in all case where branes and induced gravity are
involved, four-dimensional gravitons are massive in some regime.
We will investigate here the
cosmological evolution of a universe where gravity is massive.
We will show that a very light graviton produces a late time
acceleration of the universe that has the right properties to
explain today's data.

The cosmological equations have been derived in \cite{gri} for
quadratic potential and in \cite{kogan1,kogan2} for more general
potentials. For simplicity I will follow the treatment in
\cite{gri}, although some of the conclusions will turn out to be
different.

We will consider a four-dimensional massive graviton field
$h_{\m\n}$ in a background Minkowski metric $\eta_{\m\n}\sim (1,-1,-1,-1)$.
The case of a general background constant metric is equivalent to
the above, upon appropriate constant rescalings of the scale
factors in $g_{\m\n}$.
The full metric is defined as
\be
\sqrt{-g}g^{\m\n}=\sqrt{-\eta}(\eta^{\m\n}+h^{\m\n})
\label{metr}\ee
and $g_{\m\n}$ is the inverse of $g^{\m\n}$.
Matter couples to gravity via a minimal coupling to $g_{\m\n}$

The action is given by
\be
L=L_{GR}+L_{mass}+L_{matter}
\ee
where
\be
L_{GR}=-{1\over 2\kappa^2}\sqrt{-g}~R\sp L_{mass}=-{1\over
2\kappa^2}\sqrt{-\eta}\left[k_1 h^{\mu\nu}h_{\mu\nu}+k_2
(h^{\mu\nu}\eta_{\mu\nu})^2\right]\ee

Here we have added only a quadratic mass term. We will comment on
a more general potential later.

A study of the quadratic fluctuations of the graviton field
indicates the presence of a massive spin-two filed with mass
$m_{g}$ and a scalar component with mass $m_0$.
The masses are related to $k_{1,2}$ as \cite{gri}
\be
k_1={m_g^2\over 4}\sp k_2 =-{m_g^2\over 8}~~{m_g^2+2m_0^2\over
2m_g^2+m_0^2}
\ee
We define for further convenience
\be
\z={m_0^2\over m_g^2}
\ee

The field equations obtained are
\be
G_{\mu\nu}+M_{\mu\nu}=T_{\mu\nu}\sp G_{\mu\nu}=R_{\mu\nu}-{1\over
2}g_{\mu\nu}R
\label{eqq}\ee
where
\be
M_{\mu\nu}=\left(\delta^a_{\mu}\delta^b_{\nu}-{1\over
2}g^{ab}g_{\mu\nu}\right)(2k_1h_{ab}+2k_2
(h^{cd}\eta_{cd})\eta_{ab})
\ee
and $T_{\m\n}$ is the matter stress tensor.

The homogeneous and isotropic cosmological ansatz is
\be
h^{00}={a^3\over b}-1\sp h^{11}=h^{22}=h^{33}=1-ab
\label{ans}\ee
where $a,b$ are functions of time.
Then, the components of the full metric, defined in (\ref{metr}), are
\be
g_{00}=b^2\sp g_{11}=g_{22}=g_{33}=-a^2
\ee
We also assume a dilute homogeneous matter distribution
\be
T_{00}=\rho\sp T_{11}=T_{22}=T_{33}=-p=-w~\rho
\ee
which is conserved
\be
\dot \rho+3(1+w){\dot a \over a}\rho=0
\ee
where the dot implies differentiation with respect to cosmological
time $\tau$ defined as
\be
{d\over d\tau}={1\over b}{d\over dt}
\ee
Substituting the cosmological ansatz (\ref{ans}) into the
gravitational equations (\ref{eqq}) we obtain

\be
3\left({\dot a\over a}\right)^2+M^0_0=\kappa^2 ~\rho\sp 2{d \over
d\tau}\left({\dot a\over a}\right)+3\left({\dot a\over
a}\right)^2+M^1_1=-\kappa^2 ~p
\label{final}\ee
where
\be
M^0_0={3m_g^2\over 8(\zeta+2)}{a^5+(4\zeta
-1)ab^4+2 \zeta ~a^2 b-6 \zeta ~b^3\over a^2b^3}
\ee
\be
M^1_1=-{m_g^2\over 8(\zeta+2)}{3a^5-4(1+2\zeta) a^3b^2+(1-4\zeta
)ab^4+6\zeta ~a^2 b+6\zeta ~b^3\over a^2b^3}
\ee

The integrability condition on $M$ is
\be
\dot M^0_0+3\left({\dot a\over a}\right)\left[M^0_0-M^1_1\right]=0
\label{inte}\ee
Thus, the first of the equations in (\ref{final}) and (\ref{inte})
are an equivalent set of equations, from which the second
equation
in (\ref{final}) follows.

Equation (\ref{inte}) can be integrated explicitly to the following
algebraic equation
\be
3{a^6\over b^2}+(4\zeta-1)
~a^2b^2-2(2\zeta+1)a^4+8\zeta {a^3\over b}-8\zeta=0
\label{alg}\ee
We have chosen the constant of integration such that Minkowski
space (a=b=1) is a solution.

In the phenomenologically interesting case, $m_0>>m_g$ so that
$\z\to \infty$. (in fact $\zeta\sim 10^{30}$).
The $\zeta\to\infty$ limit corresponds to $k_2=-k_1$ and the mass
term becomes of the Pauli-Fierz type. As it was shown in
\cite{ags} in this case the validity of the effective action is
enhanced up to scales $\sim (M_Pm_g^4)^{1\over 5}$.

The algebraic condition  (\ref{alg}) in the $\zeta\to\infty$ limit becomes
\be
a^2 ~b^3-(a^4+2)~b+2a^3=0
\label{alg1}\ee
while the Friedman equation is
\be
\left({\dot a\over a}\right)^2={\kappa\over 3}\rho+\rho_m\equiv {\kappa\over 3}\rho-{m_g^2\over
4}\left(2{b\over a}+{1\over b^2}- 3{1\over a^2}\right)
\label{fried}\ee

The algebraic equation (\ref{alg1}) can be solved exactly for b as a function
of a, which can then be substituted in (\ref{fried}) to give an
equation for $a(b)$.
However, we are interested in the large $a$ behavior of the
evolution. For $a>>1$ the three solutions of (\ref{alg1}) are
\be
b_{\pm}=\pm a -{1\over a}\mp{1\over 2a^3}+\cdots\sp b_3={2\over
a}+{4\over a^5}+\cdots
\ee
while the effective energy density due to the massive graviton becomes

\be
\left.{\rho_{m}\over m_g^2}\right|_{\pm}=\mp{1\over
2}+{1\over a^2} \mp{1\over 4a^4}+\cdots
\sp
\left.{\rho_{m}\over m_g^2}\right|_{3}=-{a^2\over 16}-{1\over
4a^6}+\cdots
\label{solut}\ee

Out of the three solutions, only the minus one is real for all
values of a.
The other two are complex when $1<a<(1+\sqrt{3})^{3\over 4}$.
Moreover,
$\rho_m$ is a monotonically decreasing positive function of $a$.

Thus, the physical solution is the minus one, and as can be seen
from (\ref{solut}) it gives a positive effective cosmological
constant at late times, $\Lambda_{\rm eff}={m_g^2\over 2}$.
It should also be remarked that the subleading contribution to the
energy density behaves as an effective curvature term with $k_{\rm
eff}=-1$.

It is intriguing that the value of the graviton mass needed to be
consistent with standard massless gravity inside today's horizon,
$m_{g}\sim H^{-1}_{\rm today}$ gives a contribution to the vacuum
energy which is of the same order of magnitude as that coming from
cosmological observations.

As it was argued in \cite{ags} massive gravity with a Fierz-Pauli
term becomes strongly coupled at energies of the order $E_5\sim (m_g^4~
M_P)^{1/5}$. Adding higher potential terms in $h$ and fine-tuning
such interactions can increase this cutoff to $E_3\sim (m_g^2~
M_P)^{1/3}$.
It is conceivable that the inclusion of  interactions with other fields (namely the dilaton
and axion)  may
increase this cutoff even further to $E_2\sim (m_g~
M_P)^{1/2}$.
To obtain some numbers we set the graviton mass to the inverse
horizon size today $m_g\sim (10^{26} m)^{-1}$ and we obtain
\be
E_5^{-1}\sim 10^{19}~m\sp
E_3^{-1}\sim 10^{6}~m
\sp
E_2^{-1}
\sim 10^{-4}~m
\ee
It is interesting that a massive gravity with a $E_2$ cutoff
breaks down precisely at the boundary of today's short distance  experimental tests.
The lower cutoffs indicate that such theories are not viable for
describing four-dimensional gravity.

\subsubsection{Cubic interactions}
\renewcommand{\theequation}{\arabic{section}.\arabic{subsection}.\arabic{subsubsection}.\arabic{equation}}
\setcounter{equation}{0}

In general there will be higher order terms in the potential of
the massive graviton. To see how such terms affect the late
cosmological evolution we will study the cubic terms.

The most general cubic interaction is
\be
L_3=-{1\over 2\kappa}\sqrt{-\eta}(r_1 ({h^{\mu}}_{\mu})^3+r_2~
({h^{\mu}}_{\mu})h^{\nu\rho}h_{\nu\rho}+r_3~h^{\mu\nu}h_{\mu\rho}{h^{\rho}}_{\nu})
\ee
where $r_i$ have dimensions of (mass)$^2$.

The  correction to the energy density is
\be
\rho_3=-{r_1\over 2}{(a^2-3b^2)(a^3-4b+3ab^2)^2\over a^2
b^4}-
\ee
$$-
{r_2\over 2}{a^8-4a^5b+4a^2b^2+a^6b^2-12b^4-a^4b^4+20ab^5-9a^2b^6\over a^2
b^4}-
$$
$$
-{r_3\over 2} {a^8 - 2 a^5 b + a^2 b^2 - 3 b^4 + 6 a b^5 - 3 a^2
b^6\over a^2 b^4}
$$

The algebraic condition, generalizing (\ref{alg}) is
\be
I_2+I_3=0
\ee
where
\be
b^2~I_2=3(k_1+k_2)a^6+4(k_1+4k_2)(b^2-a^3b)+6k_2a^4b^2-3(k_1+3k_2)a^2b^4
\ee
\be
b^3~I_3=-6(r_3+3r_2+9r_1)a^3b^6+3(3r_3+10r_2+36r_1)a^2b^5-4(r_3+4r_2 +16r_1+3(r_2+6r_1)a^4)b^3+
\ee
$$
+6(r_3+4r_2+16r_1+(r_2+3r_1)a^4)a^3b^2
-9(r_3+2r_2+4r_1)a^6b+4(r_1+r_2+r_3)a^9
$$

The asymptotic form of the energy density after solving for $b$ is
of the form
\be
\rho=m_1^2+m_2^2~a^2 +{\cal O}\left({1\over a^2}\right)
\label{eff3}\ee
where the masses $m_{1,2}$ are functions of $k_{1,2}$,
$r_{1,2,3}$.
This gives again eternal acceleration.
In particular, $m_2=0$ when $r_3+4r_2+16r_1=0$ (and $m_1^2=-2k_1-5k_2$ as without the cubic terms)
or when $r_2=-{3\over 7}r_1$, $r_3=-{4\over 7}r_1$ (and $m_1^2={14k_1-35k_2-780 r_1\over
7\sqrt{15}}$).

At late times, ignoring the subleading terms, the effective energy
density (\ref{eff3}) provides super-acceleration for $m_2^2>0$
\be
a(t)=2{m_1\over m_2}{C~e^{m_1 t}\over 1-C^2~e^{2m_1 t}}
\ee
For $m_2^2<0$, we have initially acceleration and later
exponential deceleration.

Thus, higher order terms in the potential do not substantially
affect the nature of late cosmology (although they may affect the details).

\renewcommand{\theequation}{\arabic{section}.\arabic{subsection}.\arabic{equation}}
\subsection{Massive gravity and gauge theory}
\setcounter{equation}{0}

In this section we will present yet another idea on the
realization of 4-d gravity.
There are several motivations for the approach advocated here:

$\bullet$ Closed string theory generically predicts gravity. Fundamental string theories
provide a consistent (perturbative) quantization of gravity.
Despite its successes, string theory, although  well defined at
energies below or at the string
scale,
breaks down at energies close to the Planck scale. In particular,
the perturbation theory breaks down due to the strong effective
gravitational coupling. Despite speculations, the nature of the
extreme UV degrees of freedom of
the theory is still obscure \cite{wittat1}-\cite{ooguri}.
Perturbative string theory is essentially a cutoff theory of
gravity and other interactions. This is obvious at the one-loop
level of closed string theory, where the theory has a (smart indeed) cutoff at the string
scale, implemented by Schwinger parameters confined to the
fundamental domain of the torus. A similar structure persists at
higher orders in perturbation theory.
In perturbative string theory, the string scale is much lower than
the Planck scale. Taking this at face value, we do not expect the theory to give
useful information about physics at energies hierarchically  higher than the
string scale, namely around or above the Planck scale without running at a singularity/strong coupling.

It would seem that non-perturbative dualities might give a way
out, since they provide information about strong coupling physics.
Indeed non-perturbative dualities relate theories with different
(dimensionless) couplings and string scales.
This is however not the case for gravity, since any
non-perturbative duality we know, leaves the Planck scale fixed,
and thus cannot address questions on physics at or beyond the
Planck scale.

$\bullet$ Since the early work of 't Hooft \cite{hoof} it was understood that
the low energy limit of large N-gauge theories is described by some string theory.
The gauge theory versus string theory/gravity
correspondence \cite{mald1,mald2} is a more precise indication that gravity can be realized as an
effective theory of a four-dimensional gauge theory. The inverse
is also true: fundamental string theory in some backgrounds
describes the physics of theories that at low energy are standard
gauge theories. Although bulk-boundary duality is a concept
transcending that of four-dimensional gauge theories, it is most
powerful in the four-dimensional cases.

We can claim that the lesson of AdS/CFT correspondence is that any
gauge theory has a dual gravity/string theory.
The idea of 't Hooft that gravity must be holographic \cite{holog1,holog2}
indicates that a gravity theory must have a dual gauge theory
description.

$\bullet$ A standard gauge theory realization of four-dimensional  gravity generically
predicts massive composite gravitons. The graviton is the spin-two glueball generated out of the vacuum by the
stress-tensor of the theory. Confinement typically comes together with a mass gap.
A graviton mass is severely constrained by observations. Its presence may have two potential advantages.
It predicts an intrinsic  cosmological constant that may be of the
order of magnitude observed today, if the graviton mass modifies
gravity at or beyond the horizon today. Also, the fact that the
graviton is a bound state, provides  mechanism
for suppressing the cosmological constant.
In particular, the graviton {\sl does not} directly couple
to the standard ``vacuum energy" of the SM
fields.

Thus, the idea is that the building blocks of a theory of
all interactions are four-dimensional  gauge theories. Such theories
are special in many respects both
nature-wise and mathematics-wise. Four-dimensional  gravity is  an
effective, almost classical theory, emanating from a large-N sector of the gauge
theory.

The approach advocated here has similarities with ideas in
\cite{sundrum1,sundrum2} and \cite{zee}.
The qualitative model of \cite{sundrum1}, is somewhat different since the SM
particles are not charged under the strong gauge group. Gravity here is
mediated by (heavy) messager matter charged both under the SM
group and the strong gauge group, as suggested by
gauge-theory/string theory correspondence.
In fact, a light scalar graviton can be a meson but not a spin-two
one.
There is also some similarity with the idea of deconstruction
\cite{deconstr}, but here it is gravity rather than higher
dimensional matter theories that is realized by the gauge theory.

There are  direct similarities with attempts to describe
fundamental string theory in terms of matrix models
\cite{c=11}-\cite{matrix2}. Here, however, the gauge physics is four
dimensional and provides a wider class of gravity theories.
Moreover, a four-dimensional large N gauge theory, although more
complicated than a standard Matrix model gives a better intuitive
handle on the physics.

Consider a large N-gauge theory with gauge group $\gn$ and large-N matter (scalars and
fermions) that we
will not specify at the moment.
We would like the theory to be asymptotically free or conformal,
so that it is a well-defined theory at all scales. This
will put constraints on the type of large-N matter content.

At low energy, the effective degrees of freedom are colorless
glueballs as well as mesons (baryons are heavy at large N,
\cite{wittenn}).
Among the effective low-energy degrees of freedom there is always
a spin-two particle (that is generated from the gauge theory vacuum by the
total stress tensor of theory). Typically this theory, being confining,
will have a mass gap, and the spin-two particle will be massive.
However, on general principles (conservation of the gauge theory
stress-tensor) we expect to have a spin-two gauge invariance (that
may be spontaneously broken by the gauge theory vacuum).
Thus, the interactions of this particle, are those of a massive
graviton.

There are, however, other universal composites.
Let us consider for simplicity an $SU(N)$ pure gauge theory.
The leading operators, that are expected to create glueballs
out of the vacuum are a scalar (the ``dilaton")
$\phi\to Tr[F_{\m\n}F^{\m\n}]$, a spin-2 ``graviton" $g_{\m\n}\to
tr[F_{\m\r}{F^{\r}}_{\n}-{1\over 4}\delta_{\m\n}F_{\r\s}F^{\r\s}]$
and a pseudoscalar ``axion" $a\to
\e^{\m\\r\s}Tr[F_{\m\n}F_{\r\s}]$ \cite{bag}.
These particles will be massive, and their interactions at low energy are non-perturbative from the
point of view of the gauge theory.
In this relatively simple theory, with two parameters, the mass scale $\Lambda$ and
$N$, the masses are expected to be of order $\Lambda$ and the
interactions controlled at large N by $g_s\sim {1\over
\sqrt{N}}$.
In  particular, three point couplings scale as $1/\sqrt{N}$.

As first advocated by 't Hooft \cite{hoof}, the effective interactions
of the glueballs are expected to be described by an effective
string theory in the low energy regime. Most probably, the
world-sheets of this string theory are discrete, and only a tuning
of the gauge theory (double scaling limit?) might give rise to a
continuous string.

If the large-N gauge theory contains fermions in the fundamental,
$\psi^i_a$ ($i$ is a flavor index while $a$ is color),
then there are other generic low energy bound states.
In particular, the conserved fermionic currents $J^a_{\mu}\sim \bar \psi_a^iTâ_{ij}\gamma^{\mu}\psi_a^j$
generate (generically massive) vector particles.
The scalar or pseudoscalar densities $\bar \psi^i_a\psi^j_a$, $\bar \psi^i_a\gamma^5\psi^j_a$
correspond to scalar or pseudoscalar mesons, (and give rise to
open strings).
The antisymmetric tensor composites $\bar \psi^i_a\gamma^{\m\nu}\psi^j_a$,
$\bar \psi^i_a\gamma^{\m\nu}\gamma^5\psi^j_a$
should correspond to higher oscillator string states, and must
consequently be heavier.

Finally, adjoint scalars give rise to extra scalar bound-states that
appear as extra internal dimensions, as AdS/CFT indicates.

Unlike fundamental string theory, the graviton here is a
bound state of glue, and in the UV, the proper description of its
interactions are in terms of gluons. Thus, in this theory, the low
energy theory is string-like.
But the hard scattering of ``gravitons" is described by
perturbative gauge theory, while their soft scattering by an
effective (massive) gravity/string-theory.
In particular, gravitational interactions turn off at high energy
due to asymptotic freedom.

There are several immediate questions that beg to be answered in such a scheme.

\vskip .7cm

{\bf (A)} The effective graviton must have a mass that is very small (probably
of the order of the inverse horizon size) in order not to be upset by current data.
Just lowering the scale $\Lambda$ of the gauge theory is not
enough. A simple gauge theory with an ultra-low $\Lambda$ (of the order of the inverse horizon size today), has
light gravitons that are on the other hand very loosely bound states
with a size comparable to that of the universe.
We need that their size is hierarchically larger than
their mass.
An important issue is whether a small mass for the graviton is
technically natural. It is conceivable that coordinate invariance
protects the graviton mass as gauge invariance does for the photon
mass.

Moreover, the other generic low lying scalars (dilaton and axion as
well as the spin-0 component of the graviton) must be substantially
heavier so that we are not again upset by data.

What types of large N-gauge theories have a
small or no mass gap? What determines the mass gap?
What determines the hierarchy of
masses of $\phi,g_{\mu\nu}$ and $a$?
Although, there has been considerable efforts to answer such
questions for several gauge theories, no unifying picture exists \footnote{For example, in SU(2) and SU(3) gauge theory
there is an inverted hierarchy for the $0^{++}$ and $2^{++}$ glueballs, \cite{lattice}.}.
This is due to the fact that these questions involve
non-perturbative gauge theory physics. It is also due to the fact that glueballs
have been conspicuously absent from particle physics experiments.
Such questions may be studied using the general ideas of AdS/CFT
correspondence and its generalizations.

An important lesson from AdS/CFT correspondence is that the
gravity dual to four-dimensional gauge theory is five-dimensional
(with additional compact dimensions if extra (adjoint) scalar matter appears
in the gauge theory).
Polyakov has advocated general reasons why this is expected
\cite{polyakov}.
Indeed counting the degrees of freedom of massive
$g_{\mu\nu},\phi,a$ we could expect that their effective
interaction can be described by a five-dimensional massless graviton
as well as five-dimensional scalars $\phi$ and $a$.
The non-trivial gauge theory vacuum should correspond to a
nontrivial background of the five-dimensional theory (as
$AdS_5\times S^5$ describes N=4 super Yang-Mills via AdS/CFT
duality).

We would also like the full theory to be
asymptotically free. In that case the short distance physics
will be well defined. In the five-dimensional picture this will
imply an $AdS_5$ asymptotic region.

\vskip .7cm

 {\bf (B)} At low energy in the gauge theory
(if it is confining), the effective physics is described by some
string theory (at large N). Also non-confining
theories have a string description as AdS/CFT indicates but only
for the gauge singlet sector.
The important question is: what are the scales of  the string theory/gravity in terms
of the fundamental scales of the gauge theory?
The AdS/CFT paradigm is suggestive.
 Here, on the string theory side there
are three parameters: The AdS radius $R$, the string scale $l_s$
and the string coupling $1/N$. On the gauge theory side there are
only two: N and the 't Hooft coupling $\lambda=g^2N$. N=4 super
Yang-Mills is scale invariant in the symmetric vacuum. This
implies that only the ratio $R/l_s$ is observable:
$R/l_s=\l^{1/4}$.

When a mass gap $\Lambda$ is generated because of temperature
effects, it corresponds to a long distance (far away from
boundary) cutoff in AdS, namely the position $r_0$ of the horizon of the AdS
black-hole,
$r_0=\Lambda~ R^2$. The energy on the AdS side is given by
$E=r/R^2$. The cutoff implies a non-trivial effective string
length for the gauge theory, obtained by red-shifting the AdS
string scale at the horizon \cite{polchinski}:
\be
l_s^{\rm eff}\sim l_s{r_0\over
R}\sim {1\over \l^{1/4}\L}
\ee

Finally there could be masses and/or Yukawa couplings in the
large-N gauge theory. They modify the higher-dimensional geometry
by turning on fluxes \cite{non-ads1,non-ads2}

\vskip .7cm

{\bf (C)} Another important question is: how is the SM accommodated in such a
picture?  The expectation is that the SM gauge group is a separate factor from
 the large-N group. It may be  so by fiat, or it may be connected to
 the  large-N gauge group $G_N$ by symmetry breaking. It could
 also be enlarged to a unified group (SU(5), SO(10) etc).
 The standard model particles are neutral under  $\gn$.
 There should be new massive particles charged under both the
 $\gn$ and the SM gauge group.
 Integrating out these particles, the gravitational interaction is
 generated for the standard model particles. Thus, such particles
 are messagers of the gravitational interaction.
 This is analogous to the picture we have of probe D-branes in
 AdS/CFT \cite{probe1}-\cite{probe4,sym,kt1}

One could also advocate a certain ``unification" in this context:
The theory starts from a simple large-N gauge group which is
broken to a large-N subgroup generating gravity, as well as
``splinters" (the SM or the conventional unified group). The
massive states communicate gravity to the SM particles.

\vskip .7cm

{\bf (D)} The issue of the cosmological constant is qualitatively
different here. The standard matter loop diagrams that contribute
to the cosmological constant do not couple to gravity here.
Matter loops induce a potential for the graviton. Since the
graviton is composite, its form factors cut-off the matter
contributions at much lower energies (hopefully at $10^{-3}$ eV)
than the matter theory cutoff.
This is similar to the mechanism advocated in \cite{sundrum1}.

\vskip .7cm

 {\bf (E)} As we have learned from AdS/CFT, and expected on more general principles \cite{polyakov}
 the low energy gravitational theory of a large N-gauge theory
 have at least five non-compact dimensions. The obvious  question is:  how is this compatible with the observed
 4-d gravity. Here there are  two complementary ideas that might resolve this puzzle: RS localization and
 brane induced gravity. The RS solution can be implemented if the
 ''vacuum" of the gauge theory imposes an  effective UV cutoff at the position of
 the SM branes. Brane induced gravity is always present, however
its strength is crucial for it to effectively turn gravity
four-dimensional in agreement with data.

\vskip .7cm

{\bf (F)} Some of these questions can be put in perspective by
utilizing the essentials of the D-brane picture which underlies gauge-theory/gravity correspondence.
They provide a close link between gauge theory and gravity.

The large N gauge group $\gn$ can be represented as a heavy (large
N) black brane. The SM gauge group can be viewed as some collection
of a few probe branes in the background of the black-hole.
Ideally,
integrating out the strings that connect the probe branes with the
central stack (massive matter charged under both $G_N$ and SM)
induce effective gravitational interactions for the SM fields.
If more than five dimensions are present, the SM branes may be in
a nearly stable orbit around the central black-hole.
Thus, the  picture with a single large-N component of the gauge group
gives a hyper-planetary model. The SM branes must be very close to
the central black hole (near horizon region). Falling on the naked
(or not) singularity is catastrophic since it implies fusion of
the SM gauge group with $G_N$ and a radical rearrangement of the
IR field theory (gauge group enhancement) . There could be other
simple components of the gauge group, giving presumably rise to
other hyper-planetary systems and eventually to hyper-galaxies and
a hyper-universe.

Here, the picture of the gauge theory representation for gravity we advocate,
 matches smoothly to
the cosmological evolution of D-branes we discussed in previous
sections.

\vskip .7cm
{\bf (G)} There is an extra issue of divergences. We usually assume that a 4-d
large-N gauge theory has 4d-type divergences. However this need
not be true. As a counterexample, consider a 5-D theory, and compactify one direction
on a lattice. This is a large-N 4-d theory, (where N is the number of lattice sites)
but in the continuum
(large N) limit should reproduce the 5-D divergencies for certain energies.
Such divergencies will thus show up as large-N divergences, and
understanding them is central in this context.

\vskip .7cm
{\bf (H)} The approach described here has a potentially serious problem:
it relies on non-perturbative physics. Typically such a problem
proves fatal. However here we would like to advocate a
5-dimensional gravitational approach to the problem.
Several of the questions described above can be attacked in this
fashion, namely determining the 5-d action and its vacuum solution
and tuning it to achieve small graviton mass and correct
gravitational interactions for standard model particles.
In the next subsections we start a preliminary investigation of
some simple issues in this context. Whether a fully workable model
can emerge remains to be seen.

\vskip 2cm

%%%%%%%%%%%%%%%%%%%%%%%%%%%%%%%%%%%%%%%%%%%
\section{ Acknowledgements}

 I would like to thank my collaborators, P. Anastasopoulos, I.
Antoniadis, A. Kehagias, G. Kofinas, J. Rizos, N. Tetradis, T. Tomaras, V.
Zarikas who have contributed to some of the physics presented
here. I would like to thank S. Alexander, C. Bachas, T. Banks, R. Blumenhagen, R. Brandeberger,
 R. Brustein, C. Charmousis, J. Cline, L. Cornalba, M. Costa,
 A. Davis, N. Deruelle, G. Dvali,
L. Iba\~nez,  N. Kaloper, J. Khoury, C. Kounnas, D. Langlois, D. L\"ust, D. Marolf, R.
 Myers, A. Petkou, M. Porrati, F. Quevedo, H. Real, C. Skenderis,  D. Steer,     S. Trivedi, A.
Uranga C. Van der Bruck, G.
Veneziano and A. Zaffaroni for discussions and correspondence and M. Gra\~na, A. Hammou and H. Partouche
 for comments and a careful reading of the manuscript.
This review is an expanded version of lectures given at the RTN school in Torino, the Ahrenshoop Symposium
and the graduate school of Roma I University.
I would like to thank the organizers for giving me the opportunity
to give these lectures.

The author acknowledges partial support from  RTN contracts HPRN--CT--2000--00122,  --00131, --00148 and INTAS contract
N~2000-254,.
%%%%%%%%%%%%%%%%%%%%%%

\end{document}